\documentclass[twocolumn,prl,showpacs,superscriptaddress]{revtex4-1}
\pdfoutput=1

\usepackage{euscript,amssymb,amsmath}
\usepackage[hyperindex,breaklinks]{hyperref}
\usepackage{enumitem}

\newcommand{\const}{\mbox{\rm const}\,}

\usepackage{graphicx}

\newcommand{\etal}{{\it et al }}
\newcommand{\rf}[1]{(\ref{#1})}
\newcommand{\nn}{\nonumber}

\renewcommand{\Im}{\text{Im }}
\renewcommand{\Re}{\text{Re }}

\newcommand{\dd}{\dagger}

\def\RR{\mathbb{R}}
\def\NN{\mathbb{N}}
\def\CC{\mathbb{C}}

\def\ZZ{\mathbb{Z}}

\def\p{\partial}
\def\a{\alpha}

\def\d{\delta}

\def\sg{\sigma}
\def\e{\varepsilon}

\def\k{\kappa}

\def\D{\Delta}

\def\Om{\Omega}

\newcommand{\cE}{\mathcal{E}}

\newcommand{\cG}{\mathcal{G}}

\newcommand{\cK}{\mathcal{K}}
\newcommand{\cM}{\mathcal{M}}

\newcommand{\cO}{\mathcal{O}}
\newcommand{\cP}{\mathcal{P}}
\newcommand{\cT}{\mathcal{T}}

\newcommand{\cR}{\mathcal{R}}
\newcommand{\cS}{\mathcal{S}}

\newcommand{\cY}{\mathcal{Y}}

\newcommand{\ra}{\rangle}
\newcommand{\la}{\langle}
\newcommand{\Ai}{\mbox{\rm Ai\,}}

\begin{document}

\title{IR-truncated $\cP\cT-$symmetric $ix^3$ model and its asymptotic spectral scaling graph}

\author{Uwe G\"unther}
\email{u.guenther@hzdr.de}
\author{Frank Stefani}
\affiliation{Helmholtz-Zentrum
Dresden-Rossendorf, Bautzner Landstrasse 400, D-01328 Dresden, Germany}

\date{January 24, 2019}

\begin{abstract}The $\mathcal{PT}-$symmetric quantum mechanical $V=ix^3$ model over the real line, $x\in\mathbb{R}$, is infrared (IR) truncated and considered as Sturm-Liouville problem over a finite interval $x\in\left[-L,L\right]\subset\mathbb{R}$. Via WKB and Stokes graph analysis, the location of the complex spectral branches of the $V=ix^3$ model and those of more general $V=-(ix)^{2n+1}$ models over $x\in\left[-L,L\right]\subset\mathbb{R}$ are obtained. The corresponding eigenvalues are mapped onto  $L-$invariant asymptotic spectral scaling graphs $\mathcal{R}\subset \mathbb{C}$. These scaling graphs are geometrically invariant and cutoff-independent so that the IR limit $L\to \infty $ can be formally taken. Moreover an increasing $L$ can be associated with
an $\mathcal{R}-$constrained spectral UV$\to$IR renormalization group
flow on $\mathcal{R}$. The existence of a scale-invariant $\cP\cT$ symmetry breaking region on each of these graphs allows to conclude that the unbounded eigenvalue sequence of the $ix^3$ Hamiltonian over $x\in\RR$ can be considered as tending toward a mapped version of such a $\cP\cT$ symmetry breaking region at spectral infinity. This provides a simple heuristic explanation for the specific eigenfunction properties described in the literature so far and clear complementary evidence that the $\mathcal{PT}-$symmetric $V=-(ix)^{2n+1}$ models over the real line $x\in\RR$ are not equivalent to Hermitian models, but that they rather form a separate model class with purely real spectra. Our findings allow us to hypothesize a possible physical interpretation of the non-Rieszian mode behavior as a related mode condensation process.
\end{abstract}
 \maketitle

{\em Introduction}\quad
The past 20 years witnessed a strongly increasing interest in physical models with parity-time ($\cP\cT$) symmetry. Starting from the realization \cite{cmb-pt-prl-1998} that $\cP\cT$ symmetry is an inherent property of the class of non-Hermitian quantum mechanical Hamiltonians of the type $H=p^2 +mx^2 -(ix)^N$, $N\in\RR$, Bender and Boettcher related the numerically observed \cite{Bessis-JZ-1995} reality and positivity of the eigenvalues for Hamiltonians with imaginary cubic potentials,  $N=3$, to an unbroken $\cP\cT$ symmetry of these Hamiltonians.

The first years of mathematical and conceptual explorations of such and related model classes (summarized, e.g., in \cite{cmb-rev,ali-rev-2010,ali-phys-sc-2010,croke-pra-2015,pt-math-book-2015})  were followed, since ca 2007, by still ongoing investigations into $\cP\cT$ symmetric classical and quantum effective models with balanced gain and loss components and their various implementations  \cite{christodoul-nature-phys-2018,KGM-prl-2008,darmstadt-prl-2012,konotop-rmp-2016}.

More recent investigations revealed new types of $\cP\cT-$symmetry related phase transition regimes  in strongly  correlated many-body systems  \cite{PT-nature-phase-trans-2017}, an extension of Zamolodchikov's c-theorem to non-unitary, but $\cP\cT-$symmetric QFTs \cite{PT-c-theorem-jpa-2017}, and a reversability/irreversabiliy transition for information flows between $\cP\cT-$symmetric systems and  environments \cite{unit-ext-japan-2017prl} and options for
modeling non-unitary dynamics by a larger unitary operator
\cite{unit-ext-joglekar2018}, with \cite{unit-ext-japan-2017prl,unit-ext-joglekar2018} using unitary extensions of non-Hermitian systems as suggested, e.g., in \cite{unit-ext-GS-2008prl} for a complementary understanding of $\cP\cT$ quantum brachistochrones \cite{cmb-brach}. Relaxing unitarity requirements allowed to exactly compute Schur indices for $D_2[SU(2N+1)]$ superconformal field theories \cite{nonunitary-cft-2018prl}.

Since relaxing hitherto fundamental requirements like unitarity might help in deriving new deep structural relations, it appears highly desirable to understand simple nontrivial $\cP\cT-$symmetric setups most comprehensively.
One of the simplest nontrivial models \cite{DDT-pramana} remains the $\cP\cT-$symmetric quantum mechanical (QM) Hamiltonian with cubic potential $H=p^2 +igx^3$
defined over the real line $x\in\RR$ \cite{cmb-pt-prl-1998}. The setup is the one-dimensional (1D) toy model analogue of a Landau-Ginzburg theory with $i\phi^3$ interaction in $D=6-\e$ dimensions  introduced in \cite{fisher-prl-1978} for a field theoretic description of the phase-transition behavior of the 2D Ising model at a Yang-Lee edge singularity \cite{yang-lee-1952,lee-yang-1952}.  The corresponding 2D field theoretic model at criticality is a non-unitary conformal field theory (CFT) of minimal type $\cM_{2,5}$ \cite{cardy-prl-1985,mussardo-book} with algebraically growing correlation function  \cite{mussardo-book}, $\la\phi(x)\phi(0)\ra\sim |x|^{4/5}$,
in contrast to algebraically decaying behavior as for unitary CFTs. One of the interesting question is which of the specific peculiarities of non-unitary CFTs will be reflected, in one or another way, as nonstandard behavior in the 1D QM toy model analogues, such as the $\cP\cT-$symmetric $ix^3$ model.

A first clear indication for nonstandard behavior in the $ix^3$ model are the recent results on its eigenfunctions: apart from the completeness proof \cite{siegl-krej-pra-2012} and the lack of the Riesz property of the complete set of these eigenfunctions \cite{siegl-krej-pra-2012,siegl-krej-jmp-2015}, it was shown in \cite{henry-2014} that these eigenfunctions have diverging projector norms and cannot form a basis in $L^2(\RR)$. Diverging projector norms of eigenfunctions (and, for finite-dimensional matrices, of eigenvectors) typically occur \cite{GRS-jpa-ep-2007} when the operator or matrix is close (in parameter space) to a spectral branch point (exceptional point (EP)), for $\cP\cT-$symmetric setups close to a $\cP\cT$ phase transition regime. The diverging projector norm is just a reformulation of the concept of self-orthogonality of bi-orthogonal eigenvector pairs \cite{GRS-jpa-ep-2007,GS-pra-2008,moiseyev-book}, which in turn is closely related to vector isotropy in Krein spaces \cite{azizov,L2,AGK-jpa2009,GK-jpa2010}, a generalization of the concept of light-like vectors to arbitrary dimensions.

In the present Letter, we combine the knowledge about this eigenfunction property with analytic WKB (Wentzel-Kramers-Brillouin) results of Bender and Jones \cite{bender-jones-1,bender-jones-2} for the real eigenvalues of spectral problems with Hamiltonians
\begin{equation}\label{1}
H=p^2+V(x),\ V(x)=-g(ix)^{2n+1},\  n=0,1,2,\ldots
\end{equation}
defined over the  finite real interval $x\in\left[-1,1\right]\in\RR$. For varying coupling $g\in\RR_+$ these $\left[-g(ix)^{2n+1},x\in[-1,1]\right]$ models show a rich $\cP\cT$ phase transition behavior with pairwise passing of real eigenvalue branches into branches of complex conjugate eigenvalues (numerically described in \cite{bender-jones-2,gsz-jmp2005}). Here, we extend the Bender-Jones results \cite{bender-jones-2} by using the length scale $L$ of the interval $x\in[-L,L]\subset \RR$ as additional scaling parameter. This allows us to link the results over finite intervals via infra-red (IR) completion $L\to\infty$ to the results for models defined over the full line $x\in\RR$, i.e. to consider the $\left[-g(ix)^{2n+1},x\in[-L,L]\right]$ models as IR truncated versions of $\left[-g(ix)^{2n+1},x\in\RR\right]$ models. In this way, we demonstrate that $\left[-g(ix)^{2n+1},x\in\RR\right]$  models are not equivalent to Hermitian models, but that they form a separate class of physical models with purely real spectra.

{\em Relevant properties of the $(igx^3,x\in\RR)$ model}\quad
Hamiltonians \rf{1} defined over the real line $x\in\RR$ with wave functions vanishing at infinity, $\psi(x\to \pm\infty)\to 0$, are non-selfadjoint in a usual Hilbert space $L^2(\RR)$. Rather they are selfadjoint in a Krein space $\cK_\cP$ \cite{azizov,L2,lt-czech2004}, a Hilbert space with an indefinite inner product (metric) induced by the parity operator $\cP$: $\left[H\phi,\psi\right]_\cP=\left[\phi,H\psi\right]_\cP$, $\left[\phi,\psi\right]_\cP:=\la\cP\phi,\psi\ra=\int_\RR [\phi(-x)]^*\psi(x)dx$, i.e. the $\cP\cT$ inner product of \cite{BBJ-prl-2002,trinh-jpa-2005}.  $\la \phi, \psi\ra:=\int_\RR [\phi(x)]^*\psi(x)dx$ is the usual $L^2(\RR)$  inner product. With initial WKB results given in \cite{cmb-pt-prl-1998,cmb-berry-jpa-2001}, $H$ with $V=igx^3$ is  known to have a countably infinite set of discrete purely real, positive \cite{DDT-PT-jpa-2001,shin-cmp-2002,DDT-review-jpa-2007}, non-degenerate (geometrically simple) \cite{shin-cmp-2002} as well as algebraically simple (no Jordan blocks in the spectral decomposition) \cite{trinh-jpa-2005,tai-jdiff-eqs-2006} eigenvalues with accumulation at infinity.
This implies that the $\cP\cT$ symmetry of the Hamiltonian is exact \cite{cmb-pt-prl-1998}, i.e. that not only $[\cP\cT,H]=0$, but also the corresponding eigenfunctions $\psi_j$, with $H\psi_j=E_j\psi_j$, can be chosen $\cP\cT-$symmetric, $\cP\cT\psi_j(x)=[\psi_j(-x)]^*=\psi_j(x)$. These properties, clarified till ca 2005, led to the conjecture that the Hamiltonian $H$ for $x\in\RR$  might be quasi-Hermitian \cite{cmb-rev,ali-rev-2010,ali-jmp2002-2} and equivalent to some special type of non-local Dirac Hermitian Hamiltonian \cite{nonlocal-cmb-2004,nonlocal-mostafazadeh-2006,nonlocal-cmb-2009}. More recent operator-theoretic results \cite{siegl-krej-pra-2012} on the set of eigenfunctions $\left\{\psi_j\right\}_{j=1}^\infty$ provided strong evidence that the currently known quasi-Hermiticity mappings will not work for the $(igx^3,x\in\RR)$ model, because the eigenfunctions, although complete in $L^2(\RR)$, do neither form a Riesz basis (and therefore cannot be mapped by nonsingular similarity transformations into standard Euclidean type orthogonal Hilbert space bases \cite{gohberg-krein-book}) nor can they form any basis \cite{siegl-krej-jmp-2015}.  Rather it had been demonstrated in \cite{henry-2014} that the projector norms $\k_j$ of the eigenfunctions $\psi_j$ diverge as $\k_j\propto \exp[\frac\pi{\sqrt{3}}j]$ for $j\to\infty$. Exact $\cP\cT-$symmetry implies for the projector norm \cite[sect.~A]{suppl}
\begin{equation}\label{2}
\k_j=\frac{\la\psi_j,\psi_j\ra}{|[\psi_j,\psi_j]_\cP|}
=\frac{\int_\RR|\psi_j|^2dx}{\left|\int_\RR\psi_j^2dx\right|}
\end{equation}
so that $\k_j\to\infty $, for $\la\psi_j,\psi_j\ra=1$, means $[\psi_j,\psi_j]_\cP=\int_\RR \psi_j^2(x)dx\to 0$, i.e. a limiting process directed toward isotropic Krein space states \cite{GRS-jpa-ep-2007,GS-pra-2008} and self-orthogonality regimes \cite{GRS-jpa-ep-2007,GS-pra-2008,moiseyev-book}. This is a behavior typical for approaching an exceptional point (a $\cP\cT$ phase transition regime with a spectral branch point) \cite{GRS-jpa-ep-2007,GS-pra-2008,moiseyev-book}.

\begin{figure}[htb]
	\includegraphics[width=1\columnwidth]{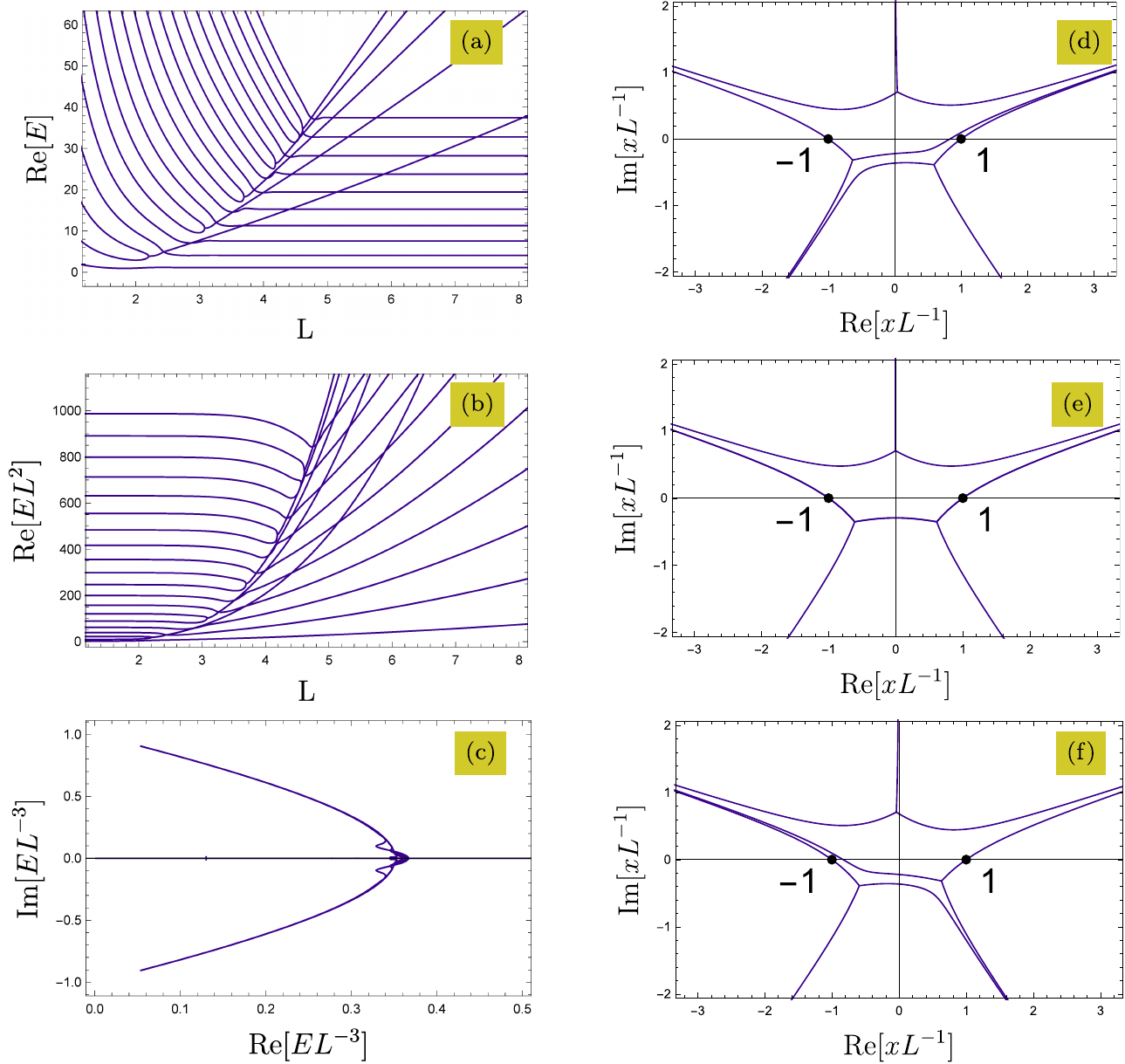}
	\caption{$(ix^3,x\in[-L,L])$ model: Numerically (shooting method) obtained lowest eigenvalue branches $\{E_j(L)\}_{j=1}^{20}$: (a) $E_j$ and (b) $E_j L^2$ against $L$, (c) location of $\cE_j=E_j L^{-3}$ in the complex $\cE-$plane. Clearly visible: $L-$independence of low-lying BS type (a), high-lying $E_jL^2\approx \pi^2 j^2/4$ BT (b) eigenvalues and (c) coincidence of the $E_jL^{-3}$ branches signalling the existence of a spectral scaling graph $\cR$. Sample Stokes graphs for $[\cE(0.9\,\tau_c)]^*\not\in\RR$ (d),  $\cE(\tau_c)\in\RR$ (e) and  $\cE(0.9\,\tau_c)$ (f) derived via Eq. \rf{8}.
	\label{fig1}}
\end{figure}

{\em WKB for $\left[-g(ix)^{2n+1},x\in[-L,L]\right]$}\quad
We provide complementary evidence for this observation by analyzing  $\left[-g(ix)^{2n+1},x\in[-L,L]\right]$ models, i.e. eigenvalue problems for Hamiltonians \rf{1} defined over  finite intervals $x\in[-L,L]\subset\RR$ with Dirichlet boundary conditions (BCs) imposed at the end points $\psi(x=\pm L)=0$ (box-type problems). We start from recent analytical results of  Bender and Jones \cite{bender-jones-2} on the real eigenvalues of $\left[-g(ix)^{2n+1},x\in[-1,1]\right]$ models. These authors  used a lowest-order WKB ansatz for the wave functions
\begin{eqnarray}\label{3}
\psi(x)&=&A_+ \psi_+(x)+A_-\psi_-(x),\ Q(x):=E+g(ix)^{2n+1}\nn\\
\psi_\pm(x)&:=&Q^{-1/4}(x)\exp\left[\pm i\hbar^{-1}\int_{x_0}^x \sqrt{Q(x')}dx'\right],
\end{eqnarray}
$E\in \RR$ and deformed the interval $x\in [-1,1]\subset \RR$ into $\cP\cT-$symmetric paths $\cG$ in the complex coordinate plane $\CC\supset \cG\ni x$ such that $\cG$ pass through those two WKB turning points $x=a_{1,2}$, $Q(a_k)=0$ which are located closest to the real $x-$axis and are connected one with the  other by anti-Stokes line segments \cite{remark-1-airy-paths}. This allowed for the derivation of a secular equation \cite{bender-jones-2}.  With slightly more general end points $x=\pm L$ and allowing for a second solution class \cite[sect.~B]{suppl}, the pair of these secular equations takes the form
\begin{equation}\label{4}
\pm\sin(I_T)+e^\Delta \cos(I_M)=0,
\end{equation}
where $\D:=(-1)^{n+1}2\hbar^{-1}\Im \int_{-L}^{a_1}\sqrt{Q(x')}dx'$,
$I_T:=\hbar^{-1}\int_{-L}^L\sqrt{Q(x')}dx'$, and $I_M:=\hbar^{-1}$ $\int_{a_1}^{a_2}\sqrt{Q(x')}dx'$. For $E\in\RR$, the integrals $I_T$ and $I_M$ are real \cite[sect.~B]{suppl}.
The secular equation pair \rf{4} not only defines the real eigenvalues, rather it also  encodes a smooth transition between two qualitatively different real spectral regimes [Fig.~\ref{fig1}(a,b)]. To make the transition mechanism transparent we analyze the Stokes graphs of the $\left[-g(ix)^{2n+1},x\in[-L,L]\right]$ models [\cite[sects.~E,G]{suppl} and Fig.~\ref{fig1}(d-f)].

{\em Stokes graph analysis}\quad
Stokes graphs $\cS\subset \CC\ni x$ are formed by the sets of anti-Stokes lines \cite{child-book}. Along these lines in the complex $x-$plane, defined by the conditions $\Im \int_{a_k}^{x}\sqrt{Q(x')}dx'=0$, the wave functions show oscillatory behavior \cite[sect.~C]{suppl}, \cite{remark-1a-stokes-lines}.
Off $\cS$, the action functions $I_k(x)=\hbar^{-1}\int^x_{a_k}\sqrt{Q(x')}dx'$ have non-vanishing imaginary components so that the corresponding locally defined WKB wave functions $\psi_{k,\pm}:=Q^{-1/4} \exp[\pm i I_k]$ either exponentially decay (subdominant Stokes sectors) or blow up (dominant Stokes sectors).
For the secular equation pair \rf{4} this means that only along $\cS$ it holds $\D=0$, whereas away from $\cS$ one can assume $|\D|\gg 0$. Depending on the sign of $\D$, Eq. \rf{4} comprises the two real spectral regimes: $e^\D\ll1: \ \sin(I_T)\approx 0$ and $e^\D\gg1: \ \cos (I_M)\approx 0$, i.e. a box type (BT) quantization $I_T\approx N_1 \pi $ and a Bohr-Sommerfeld (BS) quantization $I_M\approx (N_2+\frac12)\pi$, \ $N_{1,2}\in\ZZ$. The BS type eigenvalues are defined by the turning point positions and do not depend on the box size $L$ [Fig.~\ref{fig1}(a)], whereas for the high lying $(j\to\infty)$ BT eigenvalues the standard $E_j\approx \frac {\pi^2}{4L^2} j^2$ behavior is reproduced \cite{lt-czech2004} [Fig.~\ref{fig1}(b)].  The transition between the two real spectral regimes, $BT\rightleftarrows BS$, occurs when the box end points $x=\pm L$ can be connected by anti-Stokes lines to the BS turning points $x=a_{1,2}$, i.e. for $\pm L\in \cS$ and $\D=0$. In this case, the secular equation pair \rf{4} factorizes, $2\sin\left[\frac12\left(\pm I_T-I_M+\frac\pi2\right)\right]
\cos\left[\frac12\left(\pm I_T+I_M-\frac\pi2\right)\right]=0$, so that
\begin{equation}\label{4q}
\pm I_{L}+\frac\pi4=N_3\pi\ \cup \ \pm (I_L+I_M)+\frac\pi4=N_4\pi
\end{equation}
with $N_{3,4}\in\ZZ$ and $I_L:=\hbar^{-1}\Re \int_{-L}^{a_1}\sqrt{Q(x')}dx'=\hbar^{-1} \int_{-L}^{a_1}\sqrt{Q(x')}dx'$. Geometrically, this signals the possibility for a break-up of the Stokes graphs [Fig.~\ref{1}(d-f)] with connection to  a third type of quantization condition --- for complex eigenvalues. The various quantization regimes for arbitrary finite scales $L$ and couplings $g$ can be compared by splitting the action terms into purely real scale factors and scale invariant integrals with boundary positions fixed at $y:=x/L=\pm 1$:
\begin{eqnarray}
 I_k(x)&=&\hbar^{-1}\int_{a_k}^x\sqrt{E+g(ix')^{2n+1}}dx',\quad \cE:=\frac E{gL^{2n+1}}\nn\\
 &=&\hbar^{-1}g^{1/2}L^{(2n+3)/2}\!\!\int_{\a_k}^y\!\!\sqrt{\cE+i(-1)^n y'^{2n+1}}dy'\label{5a}
\end{eqnarray}
where $\a_k:=a_k/L$.
\begin{figure}[htb]
	\includegraphics[width=1\columnwidth]{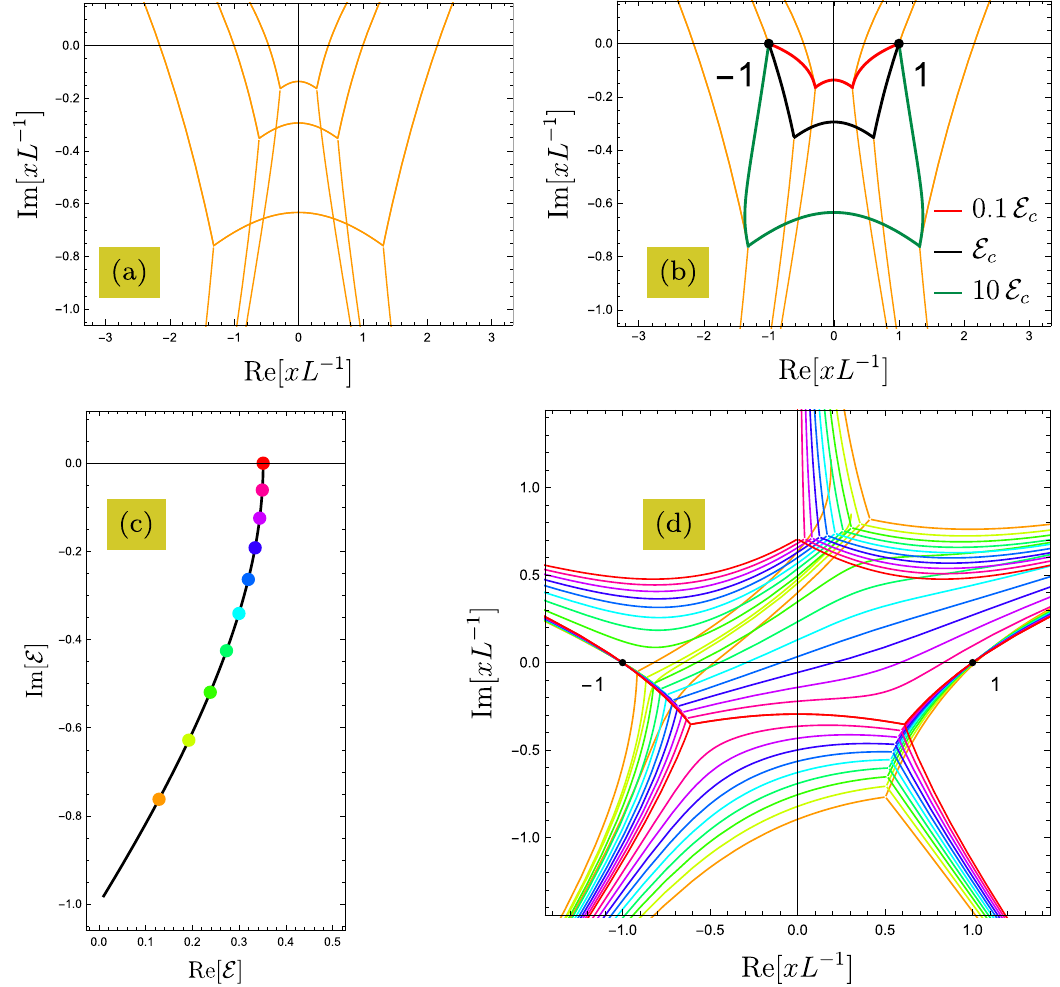}
	\caption{$(ix^3,x\in[-L,L])$ model: Form-invariant Stokes graph family for real sample values $10^{-1}\,\cE_c$, $\cE_c$, $10\,\cE_c$ (a) and corresponding paths $\cG$ (b) as deformations of the interval $x/L\in [-1,1]\in\RR$. Spectral scaling graph segment $\cR_{CO}(\Im \cE<0)\cup\cE_c$ with 10 marked sample values $\cE_j$ (c) and their associated Stokes graphs (d).}
	\label{fig2}
\end{figure}
The form (and topology) of the Stokes graphs depends only on the argument $\arg \cE$ \cite{voros-1983}. For fixed $\arg\cE$ these graphs are form invariant and they merely scale with $|\cE|$.  For real $\cE\in \RR_+$ this results in form invariant Stokes graph families like in Fig.~\ref{fig2}(a) with the BT and BS spectral regimes identified in the complex $y-$plane and in terms of $\cE$ [Fig.~\ref{fig2}(b)] via the asymptotic behaviors: fixed $E$, $L$ yield in the (empty box) limit $g\to 0$ diverging $\cE\to \infty$ (BT), whereas fixed $E$, $g$ for $L\to \infty$ should reproduce the BS-type WKB results for $\left[-g(ix)^{2n+1},x\in \RR\right]$ \cite{cmb-pt-prl-1998,cmb-berry-jpa-2001} and, hence, imply $\cE\to 0$  for BS. The single critical $\cE_c\in\RR_+$, for which $\pm L\in\cS$ and the $BT\rightleftarrows BS$ transitions with possible Stokes graph break-ups occur, are derivable as limiting points from the complex spectral branches.

{\em Complex spectral branches}\quad  For complex conjugate eigenvalue pairs $E_i,E_j\in\CC, \ E_j=E_i^*$ the wave functions are mutually $\cP\cT-$symmetric $\psi_j=\cP\cT\psi_i$ and their Stokes graphs $\cS_i,\cS_j$ are mutually reflection symmetric with regard to the imaginary axis in the complex $x-$plane, $\cS_j=\cP\cT\cS_i$  \cite[sect.~F]{suppl}. Therefore,  it suffices to consider one of the complex spectral branches only. We provide evidence that the location of these branches in the complex plane is defined by the same general mechanism as for a Hamiltonian with complex linear potential $V=-igx$. For the latter Hamiltonian it is shown analytically \cite[sect.~D]{suppl} that its complex eigenvalues $E\in\CC$ are defined by a quantization condition over anti-Stokes-line segments $\cY\subset \cS$ connecting the relevant turning points with the nearest interval-end-points  $x=\pm L$ (with exact Dirichlet BC $\psi(x=\pm L)=0$ imposed)
\begin{equation}\label{5}
\sin\left[\hbar^{-1}\int_\cY\sqrt{Q(x')}dx'+\frac\pi4\right]=0.
\end{equation}
These quantization conditions \cite[sect.~D]{suppl} with additional reality constraints $\int_\cY\sqrt{Q(x')}dx'\in\RR_+$  follow from uniform WKB approximations  \cite[sect.~2.3]{child-book}, \cite{langer-pr-1937} of the wave functions $\psi$ in terms of  {\em single} Airy functions defined over paths $\cG\subset\CC$ (the deformed intervals $x\in[-L,L]\subset\RR$) in extended vicinities of these turning points. Such a path starting, e.g., at $x=-L$ with exact Dirichtlet BC, $\psi(x=-L)=0$, will follow  the anti-Stokes line segment $\cY\subset\cS$ (with opposite orientation) to the turning point and further \emph{inside} the subdominant Airy function Stokes sector up to $x=L$, where the Dirichlet BC is satisfied only asymptotically (Fig.~4 in \cite{suppl}). For a path end point deep in the subdominant sector this BC holds with precision higher than the $0$th-order WKB. When a complex eigenvalue enters the vicinity of the $\cP\cT$ phase transition region, toward a transition to exact $\cP\cT-$symmetric configurations with real energies, the distance between the path end point and the anti-Stokes line bounding the subdominant Stokes sector becomes smaller, the $0$th-order WKB breaks down \cite{remark-2-higher-order-wkb} and a smooth transition toward exact $\cP\cT$ symmetric regimes starts (Fig.~4 in \cite{suppl}).

\begin{figure}[htb]
	\includegraphics[width=1\columnwidth]{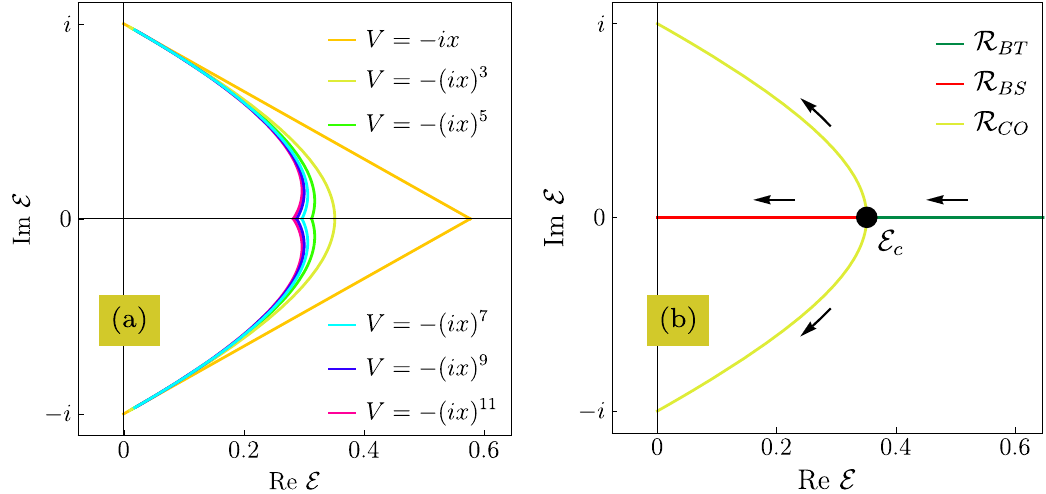}
	\caption{(a): Complex components $\cR_{CO}:=\left\{\cE(\tau),[\cE(\tau)]^*| \ \tau\in[0,\tau_c)\right\}$ of the asymptotic spectral scaling graphs $\cR=\cR_{CO}\cup\RR_+$ in the $\cE=\frac{E}{gL^{2n+1}}$ plane obtained via ODE \rf{8} for potentials $V=-(ix)^{2n+1}$, $n=0,\ldots,5$. (b): Spectral UV$\to$ IR RG flow directions (arrows) of the eigenvalues on $\cR\subset\CC$ for the $(ix^3,x\in[-L,L])$ model and increasing $L$.}
	\label{fig3}
\end{figure}

Making use of the splitting \rf{5a}
the locations of the complex eigenvalues can be described in terms of the rescaled energies $\cE$. The anti-Stokes-line related reality constraints (in terms of $y=x/L$) are fulfilled only for very special values of $\cE$ located on single curves in $\CC$. For $\cY$ between $x=\a_k L$ and $x=-L$ and with $\cE=-i(-1)^n\a_k^{2n+1}$ and a rescaling $y=\a_k \xi$ it should hold
\begin{eqnarray}\label{6}
\int_\cY\sqrt{Q(x')}dx'&=&g^{1/2}L^{\frac{2n+3}2}\int_{\a_k}^{-1}\sqrt{\cE+i(-1)^ny'^{2n+1}}dy'\nn\\
&=:&g^{1/2}L^{\frac{2n+3}2}\tau(\cE)\in\RR_+,\ \cE=\frac{E}{gL^{2n+1}}\\
\tau&=&\cE^{1/2}\a_k\int_1^{-\a_k^{-1}}\sqrt{1-\xi^{2n+1}}d\xi\in\RR_+.\nn
\end{eqnarray}
Passing from this integral constraint to its differential equivalent
one arrives at an ODE \cite[sects.~E,G]{suppl} for the complex rescaled energies as function $\cE(\tau)\in\CC$ of the real variable $\tau\in\RR_+$
\begin{equation}\label{8}
\frac{d\cE}{d\tau}=\frac{(2n+1)\cE}{\frac{2n+3}2\tau+\sqrt{\cE-(-1)^n i}}
\end{equation}
with initial condition $\tau_0=\d \tau\ll1, \qquad \cE_0=(-1)^n i+\left[3(2n+1)\d \tau/2\right]^{2/3}\exp\left[-(-1)^n i\frac{7\pi}3\right]$ and the solution graphs in Fig.~\ref{fig3}(a) \cite{remark-3-end-comment-complex-branches,remark-3a-end-comment-complex-branches}.
This allows for a visualization of the Stokes graph deformation mechanism [Fig.~\ref{fig2}(c,d)].

{\em $\cP\cT$ phase transition and spectral scaling graph}\quad
The numerical shooting-method results provide strong evidence \cite{suppl} that for a given potential $V(x)=-g(ix)^{2n+1}$, $n=1,2,\ldots$ the curve $\cE(\tau)\in\CC$, $\tau\in[0,\tau_c)$ together with its complex conjugate $[\cE(\tau)]^*$ and $\RR_+$ are the only possible locations of the eigenvalues $\cE$ in the complex plane. They constitute the corresponding spectral scaling graph $\cR$ so that $\cE\in\cR\subset\CC$. (For $V(x)=-igx$ and its $\cR$ see \cite{suppl,gsz-jmp2005}.)
Variations of $g$ and $L$ lead to variations of the positions of the eigenvalues on $\cR$, what for increasing $L$ can be understood as an $\cR$-constrained  spectral UV$\to$ IR renormalization group (RG) flow on $\cR$ [Fig.~\ref{fig3}(b)]  with $\cR$ itself scale (and flow) invariant \cite{remark-4-spectral-scaling-graph} (some rough 1D analogue of the QFT results of \cite{PT-c-theorem-jpa-2017}).
The coinciding BS quantization conditions for the $\left[-g(ix)^{2n+1},x\in\RR\right]$ model \cite{cmb-pt-prl-1998,cmb-berry-jpa-2001} and the $\cR_{BS}:=(0,\cE_c)$ segment of the  IR-truncated $\left[-g(ix)^{2n+1},x\in[-L,L]\right]$ model show that, under IR-completion $L\to\infty$, the $E_j$ corresponding to $\cR_{BS}$ will reproduce the full eigenvalue set  of $\left[-g(ix)^{2n+1},x\in\RR\right]$, i.e. the $L-$cutoff can be considered as equivalent $j-$cutoff in $\left[-g(ix)^{2n+1},x\in\RR\right]$. Once the bijection between $E_j$ and $\cE_j$ remains valid for arbitrarily large, but finite $L$, it suffices to interpret the $L\to\infty$ limit as synchronized $j\to\infty$ limit. In this way, $\cR_{BS}\ni\cE\to \cE_c$ as limit toward an EP justifies an interpretation of the $\left[-g(ix)^{2n+1},x\in\RR\right]$ eigenvalue sequence $\{E_j\}_{j=0}^\infty$ as tending asymptotically toward a $\cP\cT$ phase transition region (an EP) at spectral infinity $E_{j\to\infty}\to\infty$ associated with $\cE_{j\to\infty}\to\cE_c$. The fine-tuned mapping of the various spectral infinities to $\cE_c$ and $\cR_{CO}$ shows some rough analogy to Penrose diagrams \cite{penrose-diag} (with $\cE_c$ corresponding to some well defined spectral horizon position).

{\em Outlook}\quad
What is still lacking, is a simple \emph{physical} explanation scheme for the non-Rieszian behavior of the eigenfunction sets. For matrices approaching an EP it is known \cite{GRS-jpa-ep-2007} that the corresponding eigenvectors are tending to coalesce. For the $\infty-$dimensional Hilbert space (and Krein space) setup of the $(igx^3,x\in\RR)$ model, the eigenfunctions of the Hamiltonian having diverging projector norms and asymptotically approaching a $\cP\cT$ phase transition region at spectral infinity signal a possible tendency toward collinearity and isotropy of an infinite number of these eigenfunctions. Physically, this may find an explanation as a kind of condensation mechanism, possibly as a rough 1D toy model analogue to the formation process of a graviton BEC close to quantum criticality  \cite{dvali-grav-BEC,remark-5-condensation}.

The coincidence of the Stokes graph reconnection mechanism [Fig.~\ref{fig1}(d-f)] of $(igx^3, x\in[-L,L])$ at a $\cP\cT$ phase transition and that described in \cite{BPS-wall-crossing-wkb} for BPS setups and related wall-crossings signals structural and physical relations between these two model classes \cite{remark-6-Picard-Fuchs}.

\begin{acknowledgments}
U.G. thanks Gia Dvali, Sergii Kuzhel, Heinz Langer, and Boris Shapiro for useful discussions.
\end{acknowledgments}

\onecolumngrid

\medskip
\vspace{2cm}
\begin{center}
{\textbf{\Large Supplemental material}}
\end{center}

\medskip
\setcounter{equation}{0}
\setcounter{figure}{0}
\setcounter{table}{0}
\makeatletter
\renewcommand{\theequation}{S\arabic{equation}}
\renewcommand{\thefigure}{S\arabic{figure}}
\newcommand{\ba}[1]{\begin{eqnarray}\label{#1}}
\newcommand{\ea}{\end{eqnarray}}
\newcommand{\be}[1]{\begin{equation}\label{#1}}
\newcommand{\ee}{\end{equation}}

\makeatother

\section{A: $\cP\cT-$symmetric Hamiltonians:  projector norms of eigenfunctions}
For an operator $H$, in general non-self-adjoint $(H\neq H^\dd)$ with regard to the standard Euclidean type inner product $\la \psi,\xi\ra:=\int_\Omega [\psi(x)]^*\xi(x)dx$, \ $\Omega:=[-L,L]\subset\RR$, the projector norm $\k_j$ associated with an eigenvector $\psi_j$ of a  simple isolated eigenvalue $\lambda_j$ is defined as (see, e.g., \cite[sect. 3]{henry-2014})
\ba{p1}
\k_j&:=&||\Pi_j||=\frac{||\phi_j||\ ||\psi_j||}{|\la\phi_j,\psi_j\ra|}\label{p1a}\\
H\psi_j&=&\lambda_j\psi_j \qquad H^\dd\phi_j=\lambda_j^*\phi_j.\label{p1b}
\ea
For $\cP\cT-$symmetric Hamiltonians, $[\cP\cT, H]=0$, of type  $H=p^2 + V(x)$, $V(x)=V_+(x)+V_-(x)$, $V_\pm(-x)=\pm V_\pm(x)=[V_\pm(x)]^*$ with $[\cP,\cT]=0$ and $[\cP,p^2]=[\cT,p^2]=0$, \ $\cP^2=\cT^2=I$, \ $p=p^\dd$ it follows that these Hamiltonians $H$ are $\cP-$pseudo-Hermitian \cite{ali-jmp2002-1,GK-jpa2010}
\ba{p2}
H^\dd&=&\cP H\cP.
\ea
With $[\psi,\xi]_\cP:=\la\cP\psi,\xi\ra=\int_\Omega [\psi(-x)]^*\xi(x)dx$ this implies self-adjointness of $H$ in the Krein space $(\cK_\cP,[.,.]_\cP)$ \cite{GK-jpa2010,lt-czech2004}
\ba{p3}
[\psi,H\xi]_\cP=[H\psi,\xi]_\cP.
\ea
Furthermore, it follows from \rf{p1b} and \rf{p2} that
$
H^\dd\phi_k=\lambda_k^*\phi_k\ \Longrightarrow\ H\cP\phi_k=\lambda_k^*\cP\phi_k
$
and, hence, for some $j,k$
\be{p5}
\psi_j=\cP\phi_k,\qquad \lambda_j=\lambda_k^*.
\ee
For real eigenvalues, $\lambda_j=\lambda_j^*$, this gives $\phi_j=\cP\psi_j$, what together with the corresponding exact $\cP\cT$ symmetry of the eigenvectors $\psi_j=\cP\cT\psi_j$ implies $\phi_j=\cT\psi_j$ and
\ba{p6}
\k_j&=&\frac{||\phi_j||\ ||\psi_j||}{|\la\phi_j,\psi_j\ra|}=\frac{||\cP\psi_j||\ ||\psi_j||}{|\la\cP\psi_j,\psi_j\ra|}=\frac{||\psi_j||^2}{|[\psi_j,\psi_j]_\cP|}
=\frac{\la\psi_j,\psi_j\ra}{|\int_\Omega [\psi_j(-x)]^*\psi_j(x)dx|}\nn\\
&=&\frac{\int_\Omega|\psi_j|^2dx}{|\int_\Omega \left(\cP\cT\psi_j\right)\psi_j dx|}=\frac{\int_\Omega|\psi_j|^2dx}{|\int_\Omega \psi_j^2(x)dx|}.
\ea
One immediately concludes that a Hilbert space normalization $\la\psi_j,\psi_j\ra=1$ (or any other  normalization $0< c< \la\psi_j,\psi_j\ra< C< \infty$ with finite non-vanishing bounds $c,C$) yields for $\k_{j\to\infty}\to \infty$ a limit toward Krein space isotropy \cite{azizov,L2,AGK-jpa2009}, $|[\psi_j,\psi_j]_\cP|\to 0$, and self-orthogonality \cite{GRS-jpa-ep-2007,GS-pra-2008}, \cite[chapt. 9]{moiseyev-book}, $|\int_\Omega \psi_j^2(x)dx|\to 0$. This is a behavior typical for approaching a $\cP\cT$ phase transition regime, i.e., an exceptional point/spectral branch point \cite{GRS-jpa-ep-2007,GS-pra-2008}. Comparison with simple $2\times 2$ matrix models shows that the projector norm $\k$ is just the inverse of the phase rigidity $r=\k^{-1}$ as discussed, e.g., in \cite[sect. 6]{GRS-jpa-ep-2007}.

\section{B:  $\left[-g(ix)^{2n+1},x\in[-1,1]\right]$ models: solution classes of the secular equation}
The derivation of the secular equation for the real eigenvalues $E\in\RR$ of the $\left[-g(ix)^{2n+1},x\in[-1,1]\right]$ models in \cite{bender-jones-2} was based on a zeroth-order WKB approximation of the eigenfunctions over $\cP\cT-$symmetric paths $x\in\cG\subset\CC$ as deformations of the real interval $x\in[-1,1]\subset\RR$. A crucial role played Airy function approximations in the vicinities of the corresponding WKB turning points. For convenience, we summarize the relevant Airy function properties needed subsequently:
\begin{enumerate}[label=\emph{(\roman*)}]
\item Airy differential equation (DE) and its pairwise linearly independent solutions \cite[9.2.1]{dlmf}\label{en1}
\ba{s3}
w''(z)=zw(z),\qquad z\in\CC, \qquad w(z)\sim \Ai(z),\ \Ai(\mu z),\ \Ai(\mu^2 z),\qquad \mu:=e^{ i\frac{2\pi}3}
\ea
\item The Airy function $\Ai(z)$ is an entire function \cite{airy-encmath,sibuya1966} and, therefore, single-valued, $\Ai(e^{2\pi i}z)=\Ai(z), \quad z\in\CC, \ |z|\neq \infty$.\label{en1a}
\item connection formula \cite[9.2.12]{dlmf}\label{en2}
\ba{s4}
\Ai(z)+\mu\Ai(\mu z)+\mu^2\Ai(\mu^2 z)=0
\ea
\item zeroth-order WKB approximation \cite{suppl-rem-1}, \cite[9.7.5, 9.7.9]{dlmf}\label{en3}
\ba{s5}
\Ai(z)&\sim &\frac1{2\sqrt{\pi}}\,z^{-1/4}e^{-\frac23 z^{3/2}}\left[1+o(|z|^{-3/2})\right],\quad 1\ll |z|,\quad |\arg z|\le\pi-\d,\quad 0<\d \ll 1\label{s5a}\\
\Ai(-z)&\sim & \frac1{\sqrt{\pi}} \,z^{-1/4}\left[\cos\left(\frac23 z^{3/2}-\frac\pi4\right)+o(|z|^{-3/2})\right],\quad 1\ll |z|,\quad |\arg z|\le\frac{2\pi}3-\d,\quad 0<\d \ll 1\label{s5ab}
\ea
Below, approximation \rf{s5ab} over the real negative semi-axis is often used in the form
\be{s5b}
\Ai(-s)\sim  \frac1{\sqrt{\pi}} \,s^{-1/4}\left[\sin\left(\frac23 s^{3/2}+\frac\pi4\right)+o(s^{-3/2})\right],\qquad 1\ll s\in \RR_+, %\label{s5b}
\ee
where the  restriction on $|\arg z|$ in \rf{s5ab} implies  the branch selection $s^{3/2}\in\RR_+$.
\end{enumerate}

In the derivation of the secular equation (Eq. (17) in \cite{bender-jones-2} and Eq. \rf{4} in the main text of the present Letter), Airy function WKB asymptotics over both sectors \rf{s5a}, \rf{s5ab} in the complex plane  are used. For the explicit calculations in \cite{bender-jones-2} it turned out convenient to formally pass in \rf{s5ab} to the same complex variable $z=\rho e^{i\phi}$, $\rho\in\RR_+$ as for the sector (coordinate patch) \rf{s5a} (see Eq. (10) in \cite{bender-jones-2}) --- instead of using two different $z$ in the two sectors (coordinate patches) \rf{s5a} and \rf{s5ab}. Obviously, this passing can be done via matching in the sector overlaps either in the upper complex half-plane or in the lower one, i.e. with matching either over $\phi\in (\frac\pi3 +\d,\pi-\d)$ or over $\phi\in (-\pi+\d, -\frac\pi3 -\d)$. In terms of the same variable $z=\rho e^{i\phi}$ as in \rf{s5a}, this formally leads to $\RR_-\ni-s=z=e^{\pm i\pi}\rho$ with $s\in\RR_+$. Using the single-valuedness of $\Ai(z)$ (and $\Ai(-s)$), one, hence, has two cases $s=\rho=e^{\mp i \pi}z\in\RR_+$ to plug into \rf{s5b}, what results in \emph{two} formal extensions for \rf{s5ab}, \rf{s5b} in terms of $z$ from the sector (coordinate patch) \rf{s5a}
\ba{p7}
\Ai(-s)&\sim& \frac1{\sqrt{\pi}} \,s^{-1/4}\left[\sin\left(\frac23 s^{3/2}+\frac\pi4\right)+o(s^{-3/2})\right], \quad s=e^{\mp i \pi}z\in\RR_+, \quad s^{3/2}\in\RR_+\nn\\
&\sim& \frac1{2\sqrt{\pi}}z^{-1/4}\left[e^{-\frac23 z^{3/2}}\pm i e^{\frac23 z^{3/2}}+o(|z|^{-3/2})\right].
\ea
Using the \emph{two} representations \rf{p7} in the derivation of the secular equation (Eqs. (6) - (17) in \cite{bender-jones-2}) one arrives at a secular equation \emph{pair} (Eq. \rf{4} in the main text of the Letter)
\be{p8}
\pm\sin(I_T)+e^\Delta \cos(I_M)=0.
\ee
Here $I_T:=\hbar^{-1}\int_{-L}^L\sqrt{Q(x')}dx'$, $I_M:=\hbar^{-1}$ $\int_{a_1}^{a_2}\sqrt{Q(x')}dx'$, and
$\D:=(-1)^{n+1}2\hbar^{-1}\Im \int_{-L}^{a_1}\sqrt{Q(x')}dx'$ with $Q(x):=E-V(x)=E+g(ix)^{2n+1}$. For $E\in\RR$, the integrals $I_T$ and $I_M$ are real \cite{suppl-rem-2}.
That both equations of the secular equation pair \rf{p8} are relevant can be seen from the factorization behavior at the $BT\rightleftarrows BS$ transition point (where $\D=0$)
\ba{p9}
\pm\sin(I_T)+\cos(I_M)&=&2\sin\left[\frac12\left(\pm I_T-I_M+\frac\pi2\right)\right]
\cos\left[\frac12\left(\pm I_T+I_M-\frac\pi2\right)\right]=0.
\ea
With $I_T=I_M+2I_L$, $I_L:=\hbar^{-1}\Re \int_{-L}^{a_1}\sqrt{Q(x')}dx'=\hbar^{-1} \int_{-L}^{a_1}\sqrt{Q(x')}dx'$ this implies the Stokes graph break-up conditions
\ba{p10}
\pm I_{L}+\frac\pi4=N_3\pi\ \cup \ \pm (I_L+I_M)+\frac\pi4=N_4\pi,\qquad N_{3,4}\in\ZZ.
\ea
From comparison with the quantization condition for the complex energy branches (Eq. \rf{5} in the main text) it appears conceptually natural that the real endpoints of these complex spectral branches can match with the break-up conditions $\pm I_{L}+\frac\pi4=N_3\pi$ --- with both complex branches entering the picture on equal footing. Due to a possible matching of the condition $I_L+\frac\pi4=N_3\pi$  (following from $\sin(I_T)+\cos(I_M)=0$) with the real endpoint of the spectral branch with $\Im \frac{E}{gL^3}>0$ and matching incompatibility with the $\Im \frac{E}{gL^3}<0$ branch, the \emph{single} secular equation obtained in \cite{bender-jones-2} indicates a possibly missing second piece in the structural picture. The extension scheme \rf{p7} provides this piece in a natural way and hints on a selection rule \ $ \pm I_L+\frac\pi4=N_3\pi \quad\longrightarrow\quad \pm\Im \frac{E}{gL^3}>0$ \ in the $\cP\cT$ phase transition mechanism. Obviously, this phase transition mechanism  will relate the Stokes graph break-up with a restructuring of the Stokes sectors and a subtle underlying WKB Riemann surface structure still to be described in full detail. The origin of this interrelated behavior can be traced back to the approximation of entire (single-valued) wave functions (all solutions to Schr\"odinger equations with purely polynomial potentials are entire functions \cite{sibuya1966}) in terms of corresponding (multi-valued) WKB-type functions, a fact discussed already in \cite{langer-pr-1937}.

\section{C: Comments on Stokes lines and Stokes graphs}
There are different defining conventions for the terms \emph{Stokes line} and \emph{anti-Stokes line} in the literature. We follow \cite{berry-1989,child-book} associating oscillatory behavior with anti-Stokes lines, opposite to \cite{fedoryuk-book,cmb-orszag}. For simple turning points $a_k$ (as for all models studied in this Letter) three anti-Stokes lines are starting in each turning point \cite{strebel-book}. They either end in another turning point or tend toward infinity $|x|\to\infty$. Closed anti-Stokes lines are not possible for polynomial potentials $V(x)$, they may occur for potentials of rational type \cite{strebel-book}.

\section{D: The $(-igx,x\in[-L,L])$ model: asymptotic location of the spectrum}
\noindent The model is described by the Schr\"odinger type Hamiltonian
\ba{s1}
H=p^2 +V(x)=-\hbar^2\p_x^2-igx,\qquad g\in [0,\infty) =:\RR_+
\ea
considered over a finite real interval $x\in \left[-L,L\right]\subset \RR$. The corresponding eigenvalue problem
\ba{s2}
H\psi=E\psi,\qquad \psi(x=\pm L)=0
\ea
is a Sturm-Liouville problem. It can be considered as infrared truncated (IR-truncated) version of the eigenvalue problem over the whole real line $\RR$, i.e. an eigenvalue problem whose Dirichlet BCs at real infinity, $\psi(x\to \pm\infty)\to 0$, are replaced by Dirichlet boundary conditions (BCs) at the finite end points $\psi(x=\pm L)=0$.
Solutions $\psi(x)$ of this eigenvalue problem can be obtained in terms of Airy functions.
Below we will work solely in a zeroth-order WKB approximation. To go beyond this heuristically convincing lowest-order approximation, higher-order Poincar\'e type asymptotic approximations \cite[9.7.5, 9.7.9]{dlmf} or even super- and/or hyper-asymptotic techniques \cite{hyper-berry} should be used.

By a simple linear substitution $x=az+b$ the DE \rf{s1}, \rf{s2} can be brought into the form of an  Airy DE \rf{s3}. From the intermediate DE
\be{s6}
\p_z^2\psi=-\frac{iga^3}{\hbar^2}z\psi-\frac{a^2}{\hbar^2}(igb+E)\psi
\ee
it follows
\be{s6a}
a^3=i\hbar^2 g^{-1},\qquad b=iEg^{-1}
\ee
so that
\be{s7}
z=e^{-i\frac\pi6}e^{-i\frac{2\pi}3 k}\hbar^{-\frac23}g^{\frac13}(x-iEg^{-1})
\ee
with $k=0,1,2$ denoting the three cubic root branches induced by the condition \rf{s6a}. Without loss of generality one can fix $k=0$. The solutions of the eigenvalue problem \rf{s2} will then have the general form
\be{s8}
\psi(x)=c_1 w_1[z(x)]+c_2 w_2[z(x)]
\ee
with $w_{1,2}(z)$ any two of the three linearly independent Airy functions \rf{s3}. The constants $c_{1,2}$ should be chosen such that the BCs are fulfilled, $\psi(x=\pm L)=0$, what leads to the characteristic equation for the energy eigenvalues
\ba{s9}
w_1(z_+)w_2(z_-)-w_1(z_-)w_2(z_+)=0,\qquad z_\pm :=z(x=\pm L).
\ea
We will solve this characteristic equation in the semi-classical regime $|z_\pm|\gg1$, i.e. for $|g^{\frac13}(\pm L-iEg^{-1})|\gg \hbar^{\frac23}$ where the zeroth-order WKB approximations \rf{s5a}, \rf{s5b} will hold.
\begin{figure}[htb]
\begin{center}
 \includegraphics[scale=0.4]{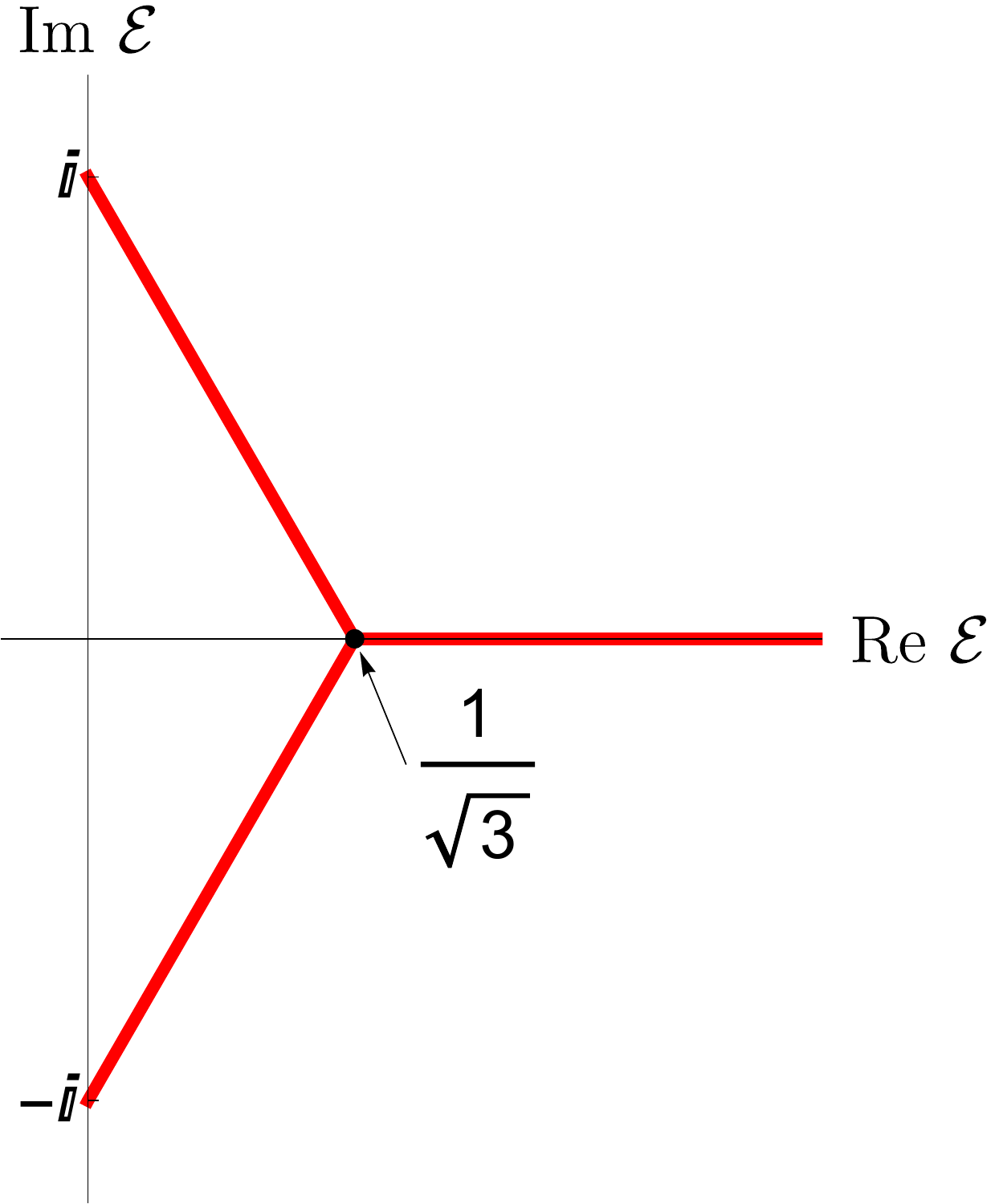}
\caption{\label{figsa1}Location of the rescaled asymptotic energy spectrum $\cE:=E(gL)^{-1}\in\CC$ in the complex plane (see Eqs \rf{s11a}, \rf{s16} below) --- which can be interpreted as asymptotic spectral scaling graph $\cR$ of the $(-igx,x\in[-L,L])$ model.}
\end{center}
\end{figure}
From the literature it is known \cite{RSH-1993,stepin1997,shkali2003,gsz-jmp2005} that in this semi-classical limit the rescaled energy spectrum $\cE:=E(gL)^{-1}$ is located on three straight-line segments in the complex $\cE$-plane (see Fig. \ref{figsa1}). This is an astonishingly simple spectral picture so that it appears natural to conjecture that this very simple picture is produced by a corresponding very simple underlying mechanism  --- despite the technically rather complicated approach in \cite{stepin1997}. As guiding hint for a possible working hypothesis we will check whether certain Airy-function combinations $w_{1,2}$ might lead to some strong simplification of the characteristic equation \rf{s9} in the sense that in the semi-classical limit it would hold, e.g.,
\be{s10}
w_1(z_+)=w_1(z_-)w_2(z_+)/w_2(z_-)\approx 0,\qquad |z_\pm|\gg 1.
\ee
As implication,  the roots of the characteristic equation would be simply given by the Airy function roots and, in zeroth-order approximation, by the sine-zeros in \rf{s5b}. (A similar behavior was recently discussed for other problems in \cite[5.2.25]{bauer-book}.) Using the connection formula \rf{s4} in \rf{s9} one finds that all three Airy-function pairings $w_{1,2}=\left\{\Ai(z),\Ai(\mu z)\right\}, \left\{\Ai(z),\Ai(\mu^2 z)\right\}, \left\{\Ai(\mu z),\Ai(\mu^2 z)\right\}$ will yield the same result in \rf{s9}, i.e., the same eigenvalues $E$. As we will see below, only in a few special cases the strong-simplification hypothesis (SSH) will be fulfilled so that the semi-classical spectral straight-line segments can be read-off directly. In the other cases, an extraction of the concrete solutions of the characteristic equation \rf{s9} would be much harder because there would be no clear control over the location of possible zeros from the interplay of the various diverging oscillating components in the terms of the characteristic equation.

\begin{figure}[htb]
\begin{center}
\includegraphics[scale=0.4]{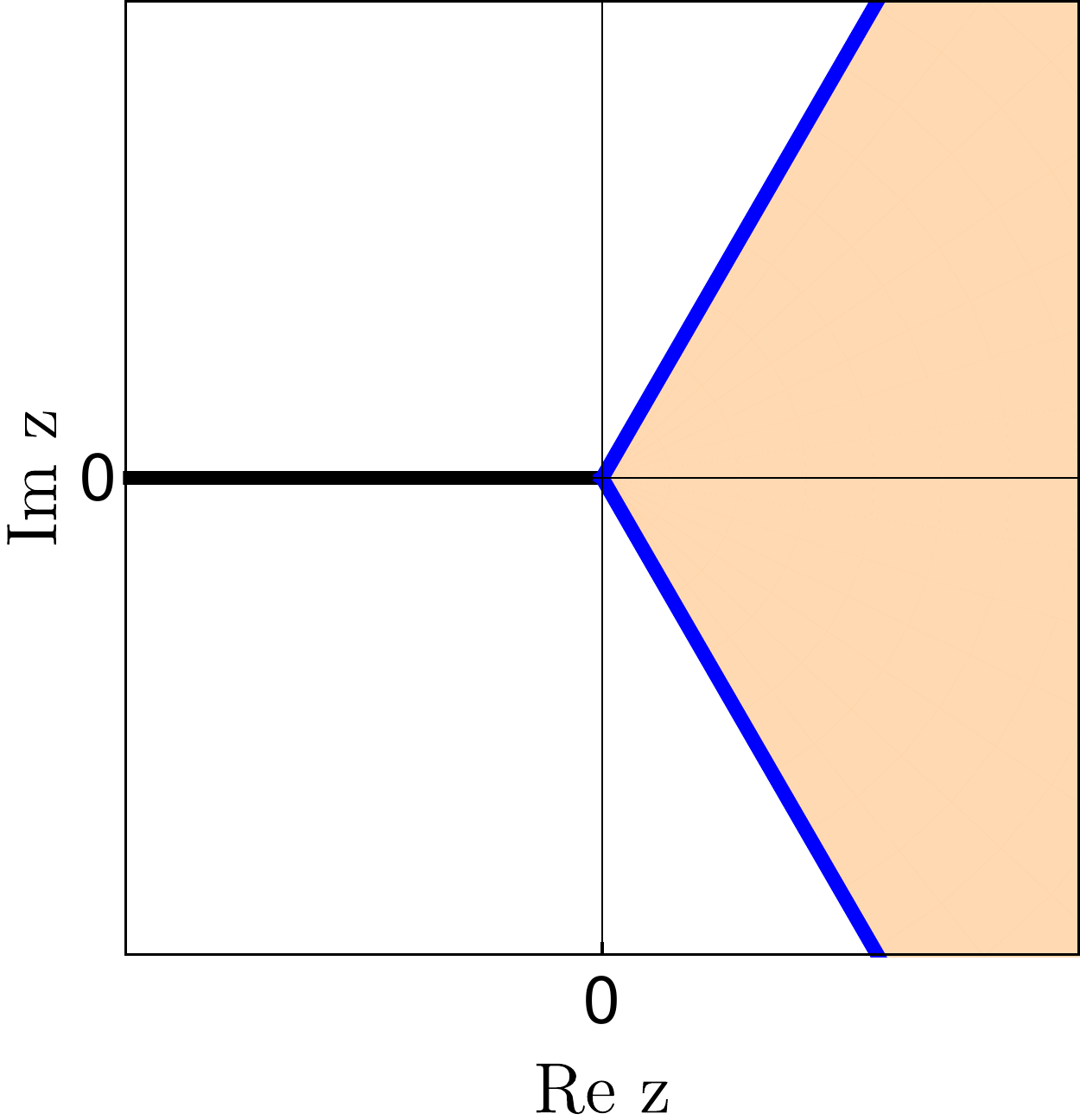}
\caption{\label{figa2} Sectors in the complex $z-$plane with dominant (white) and subdominant (shaded) asymptotic behavior of the Airy function $\Ai(z)$. On the (blue) anti-Stokes lines (along the rays $z\in e^{\pm \pi/3}\RR_+$) it holds $|e^{-\frac23 z^{3/2}}|=1$.  Along the negative semi-axis $z=-s\in (-\infty,0)$  (black line) the two oscillating contributions $|e^{-\frac23 z^{3/2}}|=1$ from the upper and lower edges of the dominant sectors superpose to produce the effective $\sin\left(\frac23 s^{3/2}+\frac\pi4\right)$ contribution in \rf{s5b}.}
\end{center}
\end{figure}
To identify the SSH cases we start from the specific asymptotic approximations \rf{s5a}, \rf{s5b} of the Airy functions on the complex plane. Obviously, \rf{s5a} is valid on the complex $z-$plane with a narrow $\d-$sector cut out along the negative real axis. Moreover, in the same zeroth-order WKB, the modulus of the relevant exponent $e^{-\frac23 z^{3/2}}$ in  $\Ai(z)$ for $|z|\gg1$ is growing in the two sectors $\arg (z)\in (-\pi,-\pi/3) \cup (\pi/3,\pi)$, decaying for $\arg(z)\in (-\pi/3,\pi/3)$ and one has $|e^{-\frac23 z^{3/2}}|=1$ along the rays $\arg (z)=\pm \pi/3$. Along $z=-s\in (-\infty,0)$ the two oscillating contributions $|e^{-\frac23 z^{3/2}}|=1$ from the upper and lower sector edges superpose \cite{suppl-rem-3}  to produce the effective $\sin\left(\frac23 s^{3/2}+\frac\pi4\right)$ contribution in \rf{s5b}. The behavior is schematically sketched  in Fig. \ref{figa2}.

\begin{figure}[htb]
\begin{center}
\includegraphics[scale=0.95]{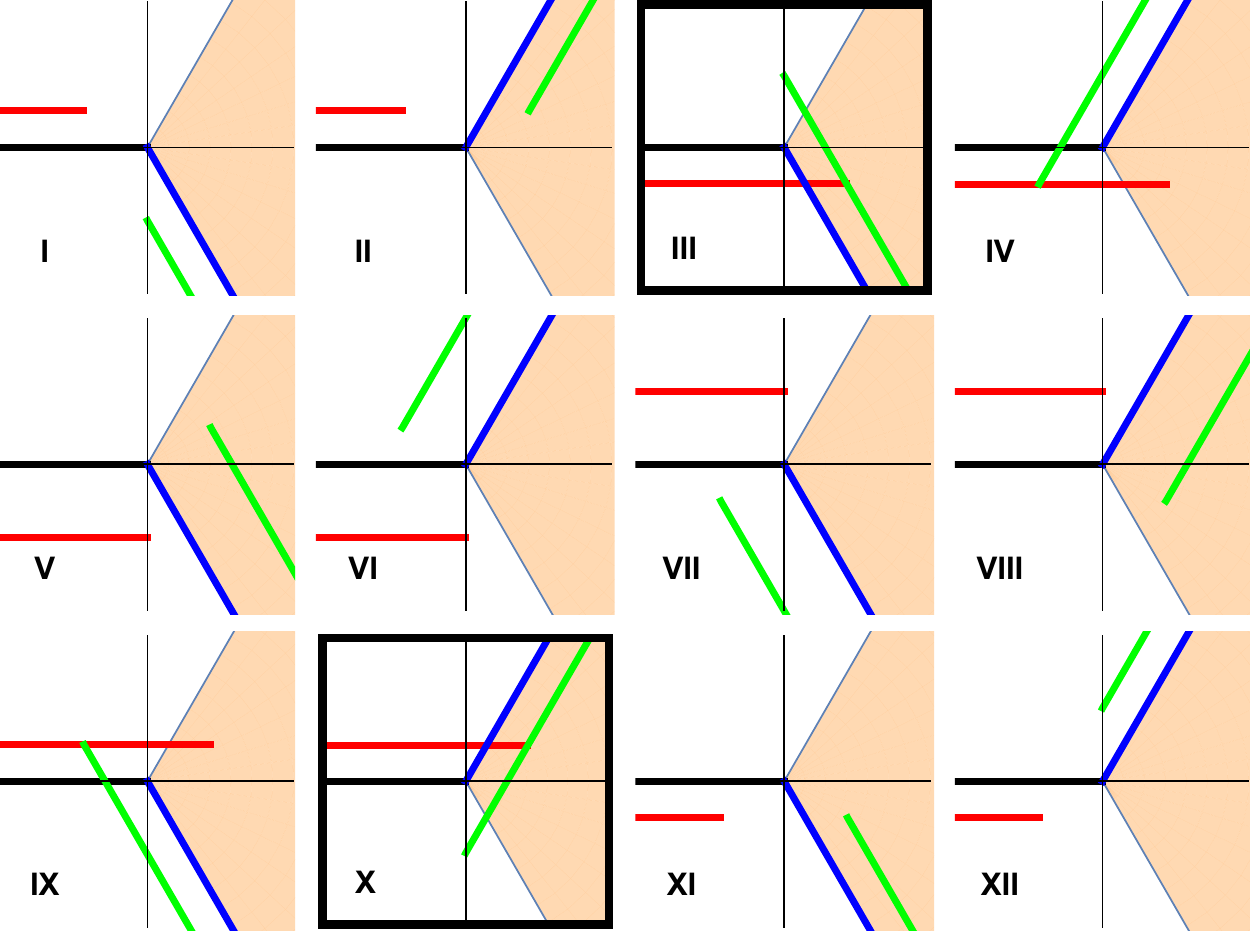}
\caption{\label{Airy-sectors}The 12 types of Airy function pairings $w_{1,2}=\left\{\Ai(z),\Ai(\mu z)\right\}, \left\{\Ai(z),\Ai(\mu^2 z)\right\}, \left\{\Ai(\mu z),\Ai(\mu^2 z)\right\}$ from Table \ref{tab1} and the location of the Airy function arguments with regard to the various asymptotic regions. By substitution into the corresponding SSH equations \rf{s10}, one directly reads off whether the SSH condition $\Ai(\rm{black})=\Ai(\rm{red}) \Ai(\rm{blue})/\Ai(\rm{green})\approx 0$ is fulfilled or not. One finds that the SSH condition, $\Ai(\rm{black})\approx 0$, is met only for configurations III and X (highlighted by black frames).}
\end{center}
\end{figure}

Combining \rf{s3}, \rf{s7}, \rf{s9} with $k=0$ in \rf{s7} we have
\ba{s11}
z_\pm:=z(x=\pm L)&=&e^{-i\frac\pi6}\k (\pm 1 -i\cE),\qquad  \k:=\hbar^{-2/3}g^{1/3}L\in \RR_+,\quad \cE:=E(gL)^{-1}\in\CC\label{s11a}\\
z_+-z_-&=&e^{-i\frac\pi6}2\k\label{s11b}
\ea
what allows us to reduce the SSH check to a simple sectorial convergence analysis. Using the entire-function property \ref{en1a}, $\Ai(e^{2\pi i}z)=\Ai(z)$, and assuming, e.g., the roots of the characteristic equation \rf{s9}, \rf{s10} induced by the roots of $w_1(z_+)=\Ai(z_+)$ one has
\ba{s12}
z_+=-s&\in & -\RR_+=e^{i\pi}\RR_+\label{s12a}\\
z_-&\in & e^{i\pi}\RR_+ + e^{i\frac{5\pi}6}2\k\label{s12b}\\
\mu z_+&\in& e^{-i\frac\pi3}\RR_+\label{s12c}\\
\mu z_-&\in& e^{-i\frac\pi3}\RR_+ - i2\k\label{s12d}\\
\mu^2 z_+&\in& e^{i\frac\pi3}\RR_+\label{s12e}\\
\mu^2 z_-&\in& e^{i\frac\pi3}\RR_+ +e^{i\frac\pi6}2\k\label{s12f}
\ea
for the two Airy function combinations $\left\{w_1(z),w_2(z)\right\}=\left\{\Ai(z),\Ai(\mu z)\right\}, \left\{\Ai(z),\Ai(\mu^2 z)\right\}$.
These $z_\pm$, $\mu z_\pm$, $\mu^2 z_\pm$ combinations are listed as cases I and II in Table \ref{tab1} and are depicted with regard to the corresponding asymptotic Airy function sectors in Fig. \ref{Airy-sectors}.
\begin{table}
\caption{\label{tab1} Half-line (ray) interrelations}
\begin{ruledtabular}
\begin{tabular}{|c|c|c|c|c|c|}\hline
$w_1$&$w_1$, \ black &$w_1$, \ red &$w_2$, \ blue & $w_2$, \ green & \\ \hline
$\Ai(z_+)$&$z_+\in e^{i\pi}\RR_+ $&$z_-\in e^{i\pi}\RR_+ + e^{i\frac{5\pi}6}2\k$&
$\mu z_+\in e^{-i\frac\pi3}\RR_+$&$\mu z_-\in e^{-i\frac\pi3}\RR_+ - i2\k $ & I\\
&&& $\mu^2 z_+\in e^{i\frac\pi3}\RR_+$ & $\mu^2 z_-\in e^{i\frac\pi3}\RR_+ + e^{i\frac\pi6}2\k$ & II\\
\hline
$\Ai(z_-)$&$z_-\in e^{i\pi}\RR_+ $&$z_+\in e^{i\pi}\RR_+ + e^{-i\frac{\pi}6}2\k$&
$\mu z_-\in e^{-i\frac\pi3}\RR_+$&$\mu z_+\in e^{-i\frac\pi3}\RR_+ + i2\k $ & III\\
&&& $\mu^2 z_-\in e^{i\frac\pi3}\RR_+$ & $\mu^2 z_+\in e^{i\frac\pi3}\RR_+ + e^{-i\frac{5\pi}6}2\k$ & IV\\
\hline
$\Ai(\mu z_+)$&$\mu z_+\in e^{i\pi}\RR_+ $&$\mu z_-\in e^{i\pi}\RR_+ -i 2\k$&
$\mu^2 z_+\in e^{-i\frac\pi3}\RR_+$&$\mu^2 z_-\in e^{-i\frac\pi3}\RR_+ +e^{i\frac\pi6}2\k $ & V\\
&&& $ z_+\in e^{i\frac\pi3}\RR_+$ & $ z_-\in e^{i\frac\pi3}\RR_+ + e^{i\frac{5\pi}6}2\k$ & VI\\
\hline
$\Ai(\mu z_-)$&$\mu z_-\in e^{i\pi}\RR_+ $&$\mu z_+\in e^{i\pi}\RR_+ + i2\k$&
$\mu^2 z_-\in e^{-i\frac\pi3}\RR_+$&$\mu^2 z_+\in e^{-i\frac\pi3}\RR_+ + e^{-i\frac{5\pi}6}2\k $ & VII\\
&&& $ z_-\in e^{i\frac\pi3}\RR_+$ & $ z_+\in e^{i\frac\pi3}\RR_+ + e^{-i\frac{\pi}6}2\k$ & VIII\\
\hline
$\Ai(\mu^2 z_+)$&$\mu^2 z_+\in e^{i\pi}\RR_+ $&$\mu^2 z_-\in e^{i\pi}\RR_+ + e^{i\frac{\pi}6}2\k$&
$ z_+\in e^{-i\frac\pi3}\RR_+$&$ z_-\in e^{-i\frac\pi3}\RR_+ + e^{i\frac{5\pi}6}2\k $ & IX\\
&&& $\mu z_+\in e^{i\frac\pi3}\RR_+$ & $\mu z_-\in e^{i\frac\pi3}\RR_+ -i2\k$ & X\\
\hline
$\Ai(\mu^2 z_-)$&$\mu^2 z_-\in e^{i\pi}\RR_+ $&$\mu^2 z_+\in e^{i\pi}\RR_+ + e^{-i\frac{5\pi}6}2\k$&
$ z_-\in e^{-i\frac\pi3}\RR_+$&$ z_+\in e^{-i\frac\pi3}\RR_+ + e^{-i\frac\pi6}2\k $ & XI\\
&&& $\mu z_-\in e^{i\frac\pi3}\RR_+$ & $\mu z_+\in e^{i\frac\pi3}\RR_+ + i2\k$ & XII\\
\hline
\end{tabular}
\end{ruledtabular}
\end{table}
From the graphics I and II in Fig. \ref{Airy-sectors} one reads off the asymptotic behavior of the Airy function combinations in the two representations of the characteristic equation \rf{s9}
\ba{s13}
I:\qquad \qquad \Ai(z_+)\Ai(\mu z_-)-\Ai(z_-)\Ai(\mu z_+)&=&0\qquad z_+\in e^{i\pi}\RR_+\nn\\
\sin \left(\frac23 |z_+|^{3/2}+\pi/4\right)\times \infty - \infty \times \cO(1)&&\nn\\
II:\qquad\qquad \Ai(z_+)\Ai(\mu^2 z_-)-\Ai(z_-)\Ai(\mu^2 z_+)&=&0\qquad z_+\in e^{i\pi}\RR_+\nn\\
\sin \left(\frac23 |z_+|^{3/2}+\pi/4\right)\times 0 - \infty \times \cO(1)&&
\ea
where we symbolically indicated exponential growth of the modulus by $\infty$ and its exponential decay by $0$. Both cases I and II are not of SSH type and we discard them. In total there are 12 cases to check: the characteristic (secular) equation \rf{s9} has 3 representations (induced by the 3 Airy function combinations $\left\{w_1(z),w_2(z)\right\}=\left\{\Ai(z),\Ai(\mu z)\right\}$, $\left\{\Ai(z),\Ai(\mu^2 z)\right\}$,$\left\{\Ai(\mu z),\Ai(\mu^2 z)\right\}$) and for each of these representations there are 4  component functions $w_1(z_\pm)$, $w_2(z_\pm)$ whose arguments should be,  step by step, assumed as contained in the negative real half-axis $e^{i\pi}\RR_+$.  Checking all 12 cases of function sets  (see Table \ref{tab1} and Fig. \ref{Airy-sectors}) one observes that these cases split into 6 groups of pairwise complex conjugate line arrangements. Five of these pairs (I+XII, II+XI, IV+IX, V+VIII and VI+VII) do not fulfill the SSH condition \rf{s10}. For example, in the case of IV+IX, possible line segments would cross the negative semi-axis so that the simple asymptotic behavior \rf{s5a} would break down and \rf{s5a} should be extended via Stokes multipliers to adjacent Riemann sheets. It turns out that to reproduce the observed straight-line-spectrum it suffices to keep within the single Riemann sheet so that the pair IV+IX can be discarded as well. Only the pair III+X shows clear SSH behavior over certain $z_\pm-$line segments in the complex plane
\ba{s14}
III:\qquad \qquad \Ai(z_-) \Ai(\mu z_+)-\Ai(z_+)\Ai(\mu z_-)&=&0\qquad z_-\in e^{i\pi}\RR_+\nn\\
\sin \left(\frac23 |z_-|^{3/2}+\pi/4\right)\times \infty - 0 \times \cO(1)&&\nn\\
\sin \left(\frac23 |z_-|^{3/2}+\pi/4\right)\approx \frac{0\times \cO(1)}\infty&\approx &0\nn\\
X:\qquad\qquad \Ai(\mu^2 z_+) \Ai(\mu z_-)-\Ai(\mu^2 z_-)\Ai(\mu z_+)&=&0\qquad \mu^2 z_+\in e^{i\pi}\RR_+\nn\\
\sin \left(\frac23|\mu^2 z_+|^{3/2}+\pi/4\right)\times \infty - 0 \times \cO(1)&&\nn\\
\sin \left(\frac23|\mu^2 z_+|^{3/2}+\pi/4\right)\approx \frac{0\times \cO(1)}\infty&\approx &0.
\ea
From the III+X-graphics in Figs. \ref{Airy-sectors}, \ref{airy-paths} these line segments can be read off as
\ba{s15}
III:\qquad\qquad z_+&=&-s+e^{-i\frac\pi6}2\k,\qquad \mu z_+= s e^{-i\frac\pi3} +i2\k, \qquad s\in\left(0,s_c\right)\nn\\
X:\qquad\qquad \mu^2 z_-&=& -s + e^{i\frac\pi6}2\k,\qquad
z_-=s e^{i\frac\pi3} -i2\k, \qquad s\in\left(0,s_c\right)
\ea
where their end points  are defined as intersection points of the $z_+,\mu z_+ $ and $z_-,\mu^2 z_-$ line segments with the rays $e^{\pm i\frac\pi3}\RR_+$ of the sectorial boundaries where the SSH behavior breaks down. It turns out that all these segment end points (intersection points) yield the same parameter value
\be{s15a}s=s_c:=\frac{2\k}{\sqrt{3}}.
\ee
Using the parametrization \rf{s15} in \rf{s11}, the line segments $z_\pm(s)$ can be mapped into the complex energy plane \cite{suppl-rem-4}
\ba{s16}
III:\qquad\qquad \cE(s)&=&i+e^{-i\frac\pi3}s\k^{-1},\qquad s\in(0,s_c)\nn\\
X:\qquad\qquad \cE(s)&=&-i+e^{i\frac\pi3}s\k^{-1}
\ea
with end point $s_c=\frac{2\k}{\sqrt{3}}$ mapped into the real energy value
\be{s17}\cE(s_c)=\frac1{\sqrt{3}}.
\ee
A further restriction comes from the SSH condition \rf{s10}, \rf{s14} itself. In the used zeroth-order approximation it restricts the parameter $s$ to the sign-inverted Airy function roots
\ba{s18}
III:\qquad\qquad \Ai(z_-)&=&\Ai(-s)\approx 0,\qquad s\in\RR_+\nn\\
X:\qquad\qquad \Ai(\mu^2 z_+)&=&\Ai(-s)\approx 0
\ea
and, hence, (in full agreement with \cite[9.9.6]{dlmf}) to
\ba{s19}
\Ai(-s)&\approx &0\quad\Longrightarrow\quad \sin \left(\frac23 s^{3/2}+\frac\pi4\right)\approx 0\label{s19a}\\
\frac23 s^{3/2}+\frac\pi4 &\approx& n\pi,\qquad n\in\NN\nn\\
s^{3/2}&\approx& \frac 38 \pi (4n-1)\nn\\
s_n&\approx& \left[\frac 38 \pi (4n-1)\right]^{2/3},\qquad n\in\NN\label{s19b}.
\ea
We note that the zeroth-order WKB approximation is valid for $z_\pm\gg1$, i.e. for $s\gg1$. Nevertheless, comparison of the numerical values of the first (lowest) root $s_1$ given in \rf{s19b} with the high-accuracy value of the actual (sign-inverted) Airy function root $\tilde s_1$ from \cite[9.9(v)]{dlmf} shows that  $s_1=(9\pi/8)^{2/3}\approx 2.32 $ and $\tilde s_1\approx 2.34$. Therefore, within this accuracy, the zeroth-order estimate \rf{s19b} can be used for all $s_n$, $n\in\NN$.

Eqs \rf{s19a} and \rf{s19b} imply that there will be no complex energy eigenvalue as long as the upper bound $s_c$ is smaller than the first Airy function root $s_1$. In this way, the $\cP\cT$ phase transition threshold and the estimate for the number of complex energy states from \cite{lt-czech2004,gsz-jmp2005} is recovered
\ba{s20}
s_c&=&s_n\nn\\
s_c=\frac{2\k}{\sqrt{3}}=\frac2{\sqrt{3}}\hbar^{-2/3}g^{1/3}L&=& \left[\frac 38 \pi (4n-1)\right]^{2/3}\nn\\
\frac2{\sqrt{3}}g^{1/3}L&=&\left[\frac 38 \hbar\pi (4n-1)\right]^{2/3}\nn\\
\left(\frac43\right)^{3/2}gL^3&=&\left[\frac 38 \hbar\pi (4n-1)\right]^2,
\ea
i.e., all eigenvalues $E$ remain real as long as
\ba{s21}
s_c< s_1\qquad \Longrightarrow\qquad gL^3 < \left(\frac34\right)^{3/2}\left(\frac98\hbar\pi\right)^2.
\ea
Moreover, one observes that a given rescaled complex conjugate eigenvalue pair
\be{s22}
\cE(s_n):=\pm i+e^{\mp i\frac\pi3}s_n\k^{-1}=\pm i+e^{\mp i\frac\pi3}s_n \hbar^{2/3}g^{-1/3}L^{-1}
\ee
moves toward the segment end points $\pm i$ when the coupling $g$ and the length $L$ are increased (in full agreement with the numerical observations in \cite{gsz-jmp2005}). This movement (for increasing $L$) can be interpreted as spectral UV$\to$ IR renormalization group flow on the corresponding complex conjugate  scaling graph segments $\cR_{CO}\subset \cR$ (see the discussion in the main text of the Letter). For the unscaled complex conjugate eigenvalue pairs $E(s_n)$ one has from \rf{s11}, \rf{s22}
\be{s23}
E(s_n)=\pm i gL+e^{\mp i\frac\pi3}s_n (g\hbar)^{2/3}
\ee
so that for constant coupling $g$ the real part $\Re E(s_n)$ remains constant when $L$ changes (again in agreement with \cite{gsz-jmp2005}), whereas $|E(s_n)|_{L\to\infty}\to\infty$.
\begin{figure}[htb]
\begin{center}
\includegraphics[scale=0.6]{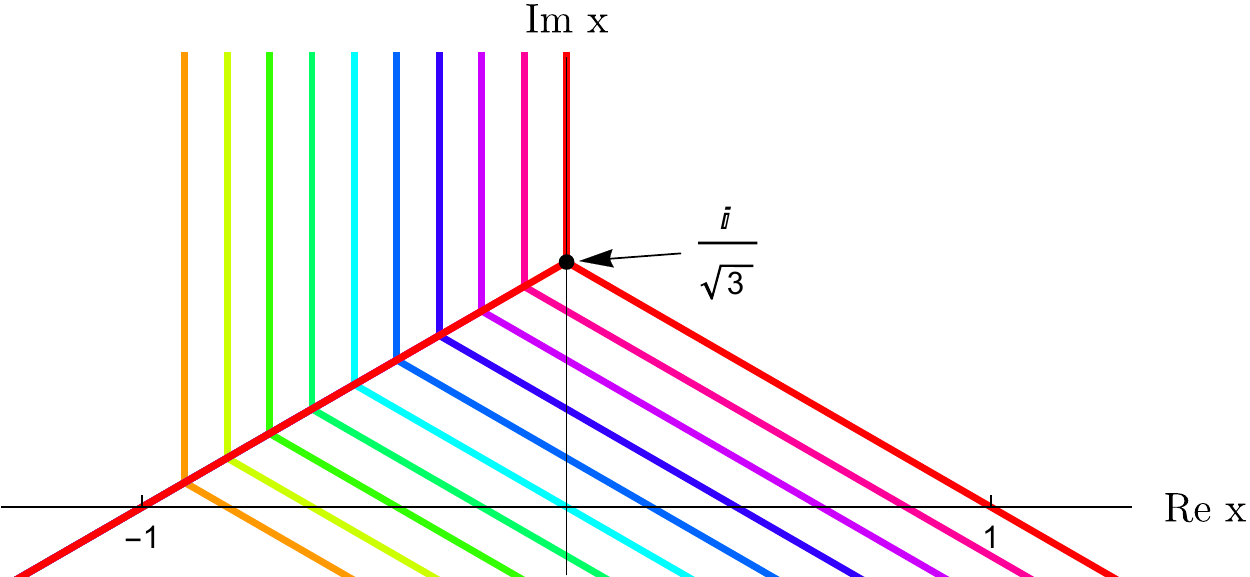}\hspace{2em}
\includegraphics[scale=0.6]{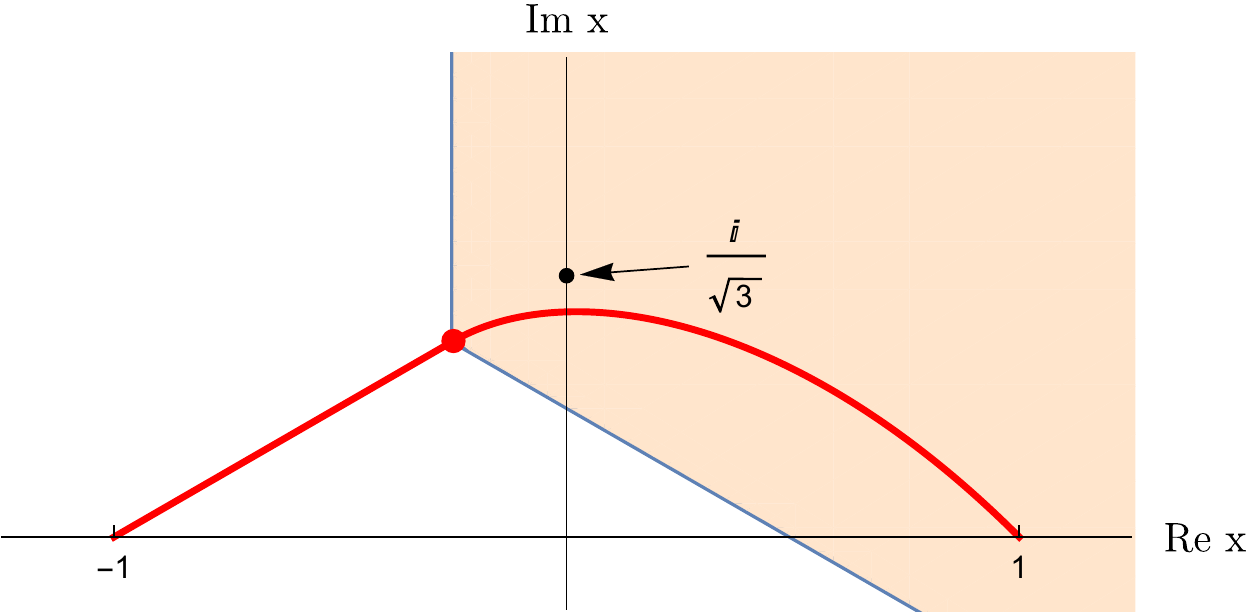}
\caption{\label{ix-energy-region} Anti-Stokes lines of a family of Airy functions (with different complex energy eigenvalues) in the complex $x/L-$plane (left graphics). The end point $x/L=-1$ has to coincide with one of the Airy function zeros, $\Ai[z(x/L=-1)]=0$, so that it has to be located on the negative $z-$axis, $z(x/L=-1)\in e^{i\pi}\RR_+$, of the family of Airy functions. The interval $x/L\in\left[-1,1\right]$ is deformed into a path in the complex $x/L-$plane (red curve, right graphics) such that it follows the negative $z-$axis segment from $z(x/L=-1)$ to the turning point $z=0$ of the corresponding SSH Airy function (marked as red dot) and continues into the subdominant Stokes sector to end in the point $x/L=1$. The Dirichlet BC at $x/L=1$ is satisfied within zeroth-order WKB accuracy due to the exponential decay of solution in the subdominant Stokes sector.}
\end{center}
\end{figure}

The basic mechanism underlying the emergence of the complex eigenvalue branches is easily understood from the location of the end points $x=\pm 1$ of the real interval $\left[-1,1\right]\subset\RR$ with regard to the dominant and subdominant Stokes sectors of the specific SSH type Airy function (see Fig. \ref{ix-energy-region}). Subsequently, we demonstrate this for the SSH configuration III. (The discussion of the second SSH configuration (X) follows that of configuration III just after a $\cP\cT$ transformation, i.e. all the system (with its various paths) has to be complex conjugated in the complex $\cE(s)-$plane and reflected on the imaginary axis in the complex $x-$plane.). For configuration III, the real interval $x\in \left[-1,1\right]\subset\RR$ can be deformed into a path $x\in\cG\subset\CC$ in the complex $x-$plane so that starting from $x=-1$ this path coincides with a segment of the negative axis $z\in e^{i\pi}\RR_+$ for the Airy function $\Ai(z)$ until it reaches the WKB turning point of that Airy function at $x(z=0)$ (marked as red dot in the right graphics of Fig. \ref{ix-energy-region}). At $x=-1$ the Dirichlet BC is satisfied by an Airy function zero, $\Ai[z(x=-1)]=0$. From the turning point $x(z=0)$ the path can be continued into the subdominant Stokes sector (shadowed area) until it reaches the other real end point $x=1$. In the subdominant Stokes sector the Airy function shows exponentially decaying behavior and to accuracy below the chosen zeroth-order WKB approximation the path can be connected to $x=1$, where the Dirichlet BC is satisfied with the same accuracy. This mechanism works as long as the path end point $x=1$ is located sufficiently deep inside the subdominant Stokes sector. When the rescaled complex energy eigenvalue $\cE$ reaches the real limit $\cE\to \cE(s_c)=\frac1{\sqrt{3}}$ the turning point $x(z=0)$ reaches the imaginary $x-$axis at $x[z(s_c)=0]=\frac i{\sqrt{3}}$ and one of the anti-Stokes lines (one of the boundaries of the subdominant Stokes sector) hits the end point $x=1$ (see the left graphic of Fig. \ref{ix-energy-region}). For Airy function turning points in the right complex $x-$half-plane the end point $x=1$ would be located in a dominant Stokes sector so that the exponentially large modulus of the Airy function would make it impossible to satisfy the Dirichlet BC at $x=1$. This implies that, for complex energy eigenvalues $\cE(s)$, the WKB turning points of the corresponding Airy functions have to be located on the finite straight line segment $x[z(s)=0]= -1 +e^{i\frac\pi6} s\k^{-1}\in \left[-1,\frac i{\sqrt{3}}\right]\subset\CC$, $s\in \left[0,s_c\right]\subset \RR_+$ in the complex $x-$plane \cite{suppl-rem-5}.

The setup can be compared to that of a wave function for a particle moving on an interval $\Om\in\RR$ of the real line bounded to the right by a potential wall with finite slope, and to the left by an infinitely high box wall, i.e. a turning point to the right and a Dirichlet BC to the left (see, e.g., \cite[sect. A.2]{child-book}). (The second Dirichtlet BC to the right of the turning point is imposed sufficiently deep in the classically forbidden region so that this BC is satisfied to zeroth order in the WKB approximation.) The corresponding quantization condition formally coincides with the eigenvalue condition \rf{s19a}, i.e., it has the same Maslov term \cite{suppl-rem-6}  $\frac\pi4$.  The difference is in the generalization of the standard textbook problem \cite[sect. A.2]{child-book} living on an interval $\Om\subset\RR$ of the real line to a problem living on a path segment $\cY\subset\cG\subset \CC$ in the complex coordinate plane and for complex energies $E\in\CC$ --- with both path segment $\cY\subset\cG$ and energies $E$ tuned (constrained) in such a way as to keep the effective classical action over $\cY$ purely real and positive. In terms of the coordinate $s\in\RR_+$ this yields anti-Stokes (classically allowed) type behavior along the path segment $\cY$ starting at the turning point $x[z(s=0)=0]$ and ending at the Dirichlet boundary point $x[s_n]/L=-1$. Explicitly one finds from \rf{s7} and $z=e^{i\pi}s$, $s\in\RR_+$
\ba{s24}
s&=&e^{-i\frac{7\pi}6}\hbar^{-2/3}g^{1/3}(x-iEg^{-1})\ge0\nn\\
\frac23 s^{3/2}&=&\int_0^s\sg^{1/2}d\sg\ge0\nn\\
&=&\frac1\hbar\int_{x[z(s=0)=0]}^{x=-L}\left(E+igx\right)^{1/2}dx
\ea
\ba{s25}
0=\sin \left(\frac23 s^{3/2}+\frac\pi4\right)&=&\sin\left(\int_0^s \sg^{1/2}d\sg +\frac\pi4\right)\nn\\
&=&\sin\left(\frac1\hbar\int_\cY \sqrt{E+igx}\,dx+\frac\pi4\right)\nn\\
\int_\cY p\,dx=\int_\cY \sqrt{E+ig x}\,dx&=&\int_{x[z(s=0)=0]}^{-L}\sqrt{E+ig x}\,dx\in\RR_+\,.
\ea
By comparison with known results on quadratic differentials \cite{strebel-book} one identifies $\cY$ as segment of a horizontal trajectory in the sense of Strebel \cite[definition 5.5.3]{strebel-book}
\ba{s26}
d\Phi^2:=(E+ig x)dx^2 &\ge& 0\nn\\
\Phi:=\int_\cY\sqrt{E+ig x}\,dx&\in&\RR_+\nn\\
\arg d\Phi^2=\arg\left[ (E+ig x)dx^2\right]&=&0.
\ea
Moreover, we note that the obtained SSH scheme with its deformation path (Fig. \ref{ix-energy-region}b) can be regarded as  leading to a uniform WKB approximation  \cite{langer-pr-1937}, \cite[sect. 2.3]{child-book} of the wave function $\psi$ in terms of a single Airy function defined over that path $x\in\cG\subset\CC$ (the deformed interval $x\in[-L,L]\subset\RR$) in an extended vicinity of the corresponding single turning point.

Finally, we describe the real part of the energy spectrum related to the real straight-line segment $\cE\in [3^{-1/2},\infty)\subset \RR_+$ (see Fig. \ref{figsa1}). For this purpose we start from the Airy function combinations of configurations III and X, and just assume $\cE\in [3^{-1/2},\infty)=3^{-1/2}+ \RR_+$ as ansatz in \rf{s11a}. With $s_c:=\frac{2\k}{\sqrt{3}}$ from \rf{s15a} this gives the relevant parameterizations
\ba{s27}
III:\qquad\qquad z_-&=&-s_c + e^{-i\frac{2\pi}3}\RR_+\nn\\
z_+&=& e^{-i\frac\pi3} s_c + e^{-i\frac{2\pi}3}\RR_+\nn\\
\mu z_- &=& e^{-i\frac\pi3} s_c +\RR_+\nn\\
\mu z_+&=& e^{i\frac\pi3} s_c +\RR_+\nn\\
X:\qquad\qquad \mu^2 z_+&=&-s_c+ e^{i\frac{2\pi}3}\RR_+\nn\\
\mu^2 z_-&=&e^{i\frac\pi3}s_c + e^{i\frac{2\pi}3}\RR_+\nn\\
\mu z_+&=& e^{i\frac\pi3}s_c +\RR_+\nn\\
\mu z_-&=&e^{-i\frac\pi3}s_c +\RR_+
\ea
depicted as dashed lines in Fig. \ref{airy-paths}.
\begin{figure}[htb]
\begin{center}
\includegraphics[scale=0.5]{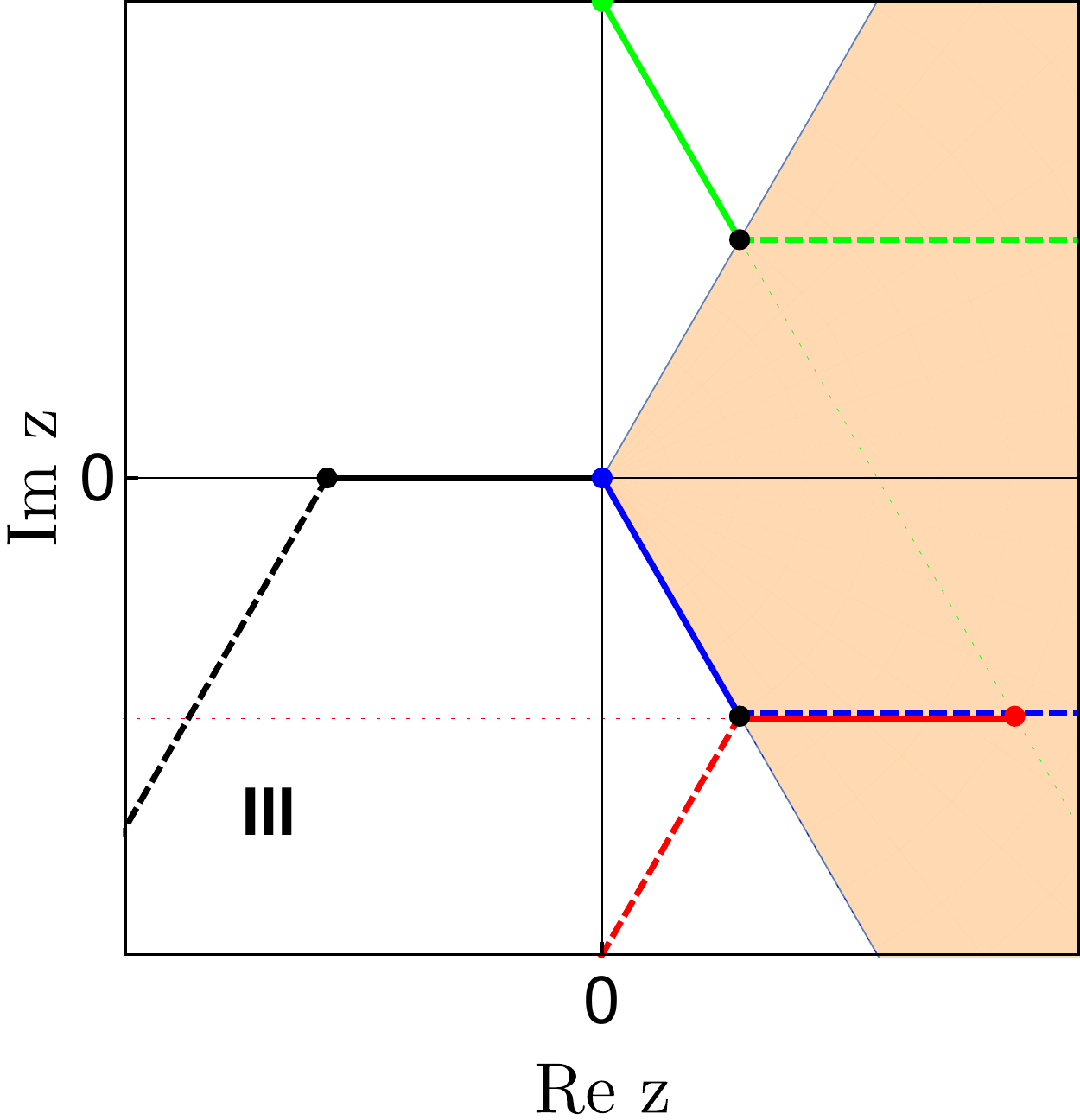}\hspace{2em}
\includegraphics[scale=0.5]{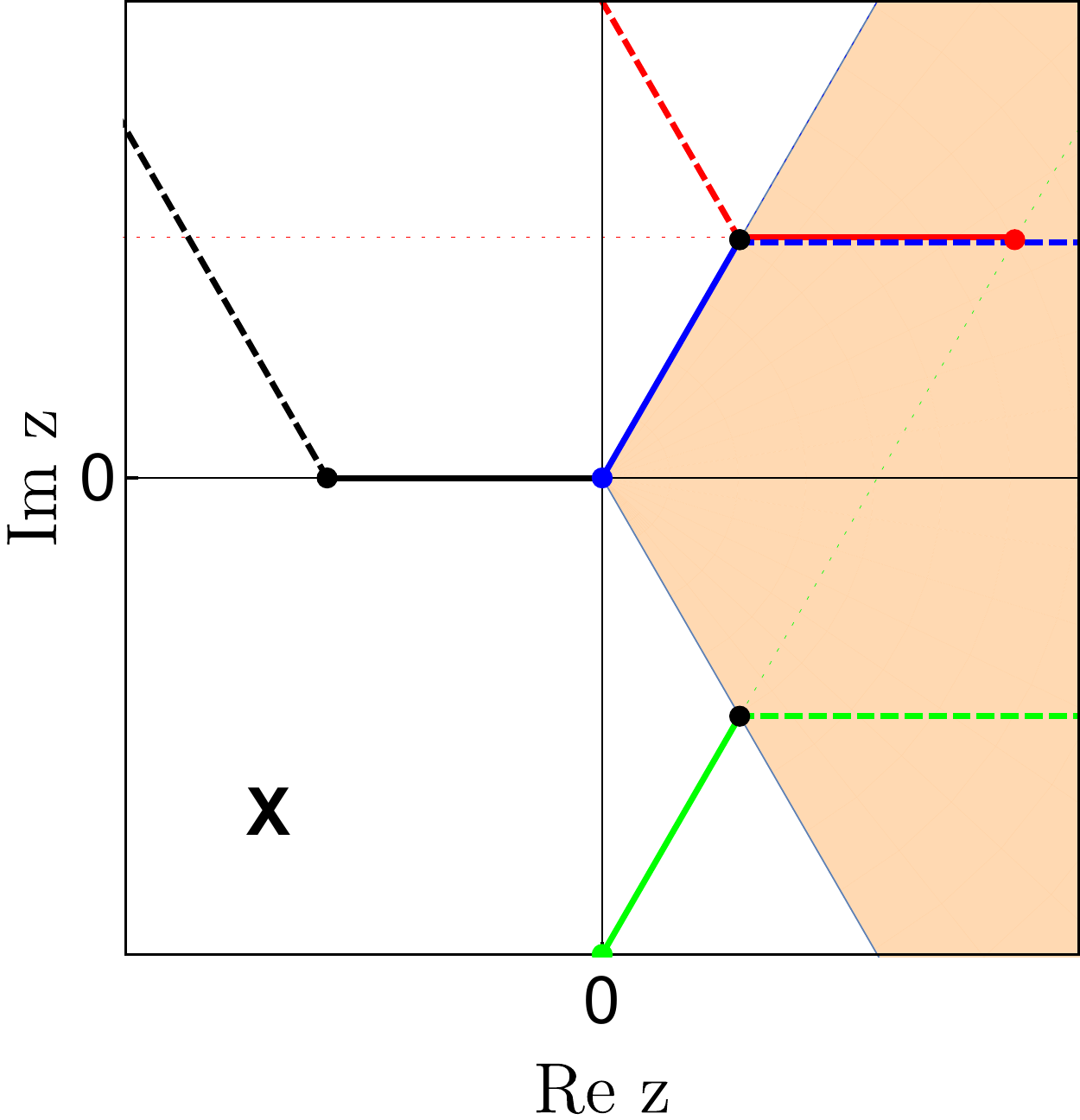}
\caption{\label{airy-paths} Stokes sector picture in the complex $z-$plane for the Airy functions entering the cases III and X. Continuous line segments correspond to complex eigenvalue branches, whereas dashed lines correspond to real eigenvalues.}
\end{center}
\end{figure}
The corresponding rays are located in the single Riemann sheet of the asymptotic expansion \rf{s5a}. From the graphics one sees, e.g., that for real energy values smaller than the lower boundary $\cE< \cE(s_c)=3^{-1/2}$ all the dashed lines should be continued beyond the black dots (corresponding to $\cE(s_c)=3^{-1/2}$). For the black dashed lines this means that they should cross the negative semi-axis (which can be understood here as possible cut) and pass to adjacent Riemann sheets. Due to additional Stokes factors this would change the asymptotic form of the Airy functions and as result destroy the spectrum-generating mechanism based on the ansatz with  real straight-line ray $\cE(s_c)\le\cE\in 3^{-1/2}+\RR_+$.

For the two SSH-type Airy function combinations III and X the corresponding spectrum-defining characteristic equations \rf{s9} yield in zeroth-order WKB approximation
\ba{s28}
III:\qquad\qquad 0&=&\Ai(z_+)\Ai(\mu z_-)-\Ai(z_-)\Ai(\mu z_+)\nn\\
&\approx &\frac1{4\pi}(\mu z_+z_-)^{-1/4}\left[e^{\frac 23\left(z_-^{3/2}-z_+^{3/2}\right)}-e^{-\frac 23\left(z_-^{3/2}-z_+^{3/2}\right)}\right]\nn\\
X:\qquad\qquad 0&=& \Ai(\mu^2 z_+)\Ai(\mu z_-)-\Ai(\mu^2 z_-)\Ai(\mu z_+)\nn\\
&\approx&\frac1{4\pi}(z_+z_-)^{-1/4}\left[e^{\frac 23\left(z_-^{3/2}-z_+^{3/2}\right)}-e^{-\frac 23\left(z_-^{3/2}-z_+^{3/2}\right)}\right]
\ea
and, therefore, the same condition
\ba{s29}
e^{\frac 23\left(z_-^{3/2}-z_+^{3/2}\right)}-e^{-\frac 23\left(z_-^{3/2}-z_+^{3/2}\right)}\approx 0.
\ea
Via \rf{s11}, $z=e^{-i\frac\pi6}\k (\xi -i\cE)$ and
\ba{s30}
\frac 23\left(z_-^{3/2}-z_+^{3/2}\right)&=&\int_0^{z_-}\zeta^{1/2}d\zeta -\int_0^{z_+}\zeta^{1/2}d\zeta=-\int_{z_-}^{z_+}\zeta^{1/2}d\zeta \nn\\
&=&-e^{-i\frac\pi4}\k^{3/2}\int_{-1}^1\sqrt{\xi -i\cE}d\xi\nn\\
&=& i \k^{3/2}\int_{-1}^1\sqrt{\cE +i\xi}d\xi
\ea
this reduces to the $\cP\cT-$symmetric generalization of the standard quantization condition for a particle in a box
\ba{s31}
\sin\left[\k^{3/2}\int_{-1}^1\sqrt{\cE +i\xi}\,d\xi\right]&=&0\nn\\
\frac1\hbar \int_{-L}^L\sqrt{E +igx}\,dx&=& \pi n, \qquad n\in\NN,
\ea
where the $\cP\cT$ symmetry of the integrand and the integration interval imply
\ba{s32}
\sqrt{E +igx}&=:&f_+(x)+i f_-(x),\qquad f_\pm(-x)=\pm f_\pm(x)\in\RR \nn\\
\int_{-L}^L\sqrt{E +igx}\,dx&=&\int_{-L}^L f_+(x)\, dx\in\RR.
\ea

Finally, we note that, due to the invariance of the spectral scaling graph $\cR$ (Fig. \ref{figsa1}) under changes of the IR-cutoff $L$ and the coupling $g$, in the IR completion limit $L\to\infty$ the non-scaled energies $E_n$ according to Eq. \rf{s23} move to infinity, $|E_n(L\to\infty)|\to\infty$, leaving the finite spectral plane empty. In this way, Herbst's empty-spectrum result \cite{Herbst-1} for the $V=-ix$ model for $x\in\RR$ is recovered (see, also \cite{gsz-jmp2005}).

\vspace{3ex}
{\bf Essential conclusion:} For the model with potential $V=-ix$ and Dirichlet BCs at the end-points of the interval of definition, $\Omega\subset\RR$, complex eigenvalues are related to a deformation of the  interval $\Omega$ into a path $x\in\cG\subset\CC$ in the complex $x-$plane. One of the BCs is fulfilled exactly as a zero of the relevant SSH Airy function, whereas the Dirichlet BC at the other end-point of the interval should be located deep in the subdominant Stokes sector of this Airy function so that the latter BC is fulfilled to accuracy of the WKB approximation. The path $\cG\subset\CC$ itself has to be chosen in such a way that it follows the negative real axis of the Airy function (with anti-Stokes type/horizontal behavior of the WKB approximation) pass through the turning point and connect to the other interval end-point in the subdominant Stokes sector of the Airy function.
This observation serves as main heuristic test ansatz and working hypothesis for the conjectured quantization condition for complex conjugate eigenvalue-pairs in general $\left[-g(ix)^{2n+1},x\in[-L,L]\right]$ models.

\section{E: $\left[-g(ix)^{2n+1},x\in[-L,L]\right]$ models: resolution of the anti-Stokes-line constraints}

There is strong evidence that the spectral behavior of the $\left[-g(ix)^{2n+1},x\in[-L,L]\right]$  models, including that of the paradigmatic $(igx^3,x\in[-L,L])$ model, is defined by the same basic underlying mechanism which can be qualitatively described within zeroth-order WKB.

The remarkably good reproduction of the eigenvalue-solver based results by $0$th-order WKB (see sects. F,G of the present Supplement) allow for the following explanation. Regardless of the increasing complexity of the Stokes graphs with increasing powers $2n+1$ (sect. G) and in agreement with \cite{bender-jones-2}, the real energies are defined by the same secular equation pair \rf{p8}  ---  keeping $Q(x)=E-V(x)=E+g(ix)^{2n+1}$ and deforming the interval $x\in[-L,L]\subset\RR$ into a path $\cG\subset\CC$ passing through those pairs of nearest turning points which for suitable $\cE_c$ allow for direct anti-Stokes-line connections between them and to the path end points $x=\pm L$, and which for even (odd) $n$ are located in the upper (lower) complex $x-$half-plane.

We start from the working hypothesis that the complex scaled/mapped eigenvalues (MEs) $\cE_j=\frac{E_j}{gL^{2n+1}}$ are located on a single curve in the complex $\cE-$plane (and its complex conjugate $\cP\cT-$symmetric mirror image) defined by the reality (horizontality \cite{strebel-book}) constraint for the anti-Stokes-line segment $\cY$ connecting the turning $a_k$ closest to the interval end point $x=-L$ with $x=-L$ itself ($x=:Ly, \ a_k=:L\a_k$)
\ba{c1}
\int_\cY\sqrt{Q(x')}dx'&=&\int_{a_k}^{-L}\sqrt{E+g(ix')^{2n+1}}dx'\in\RR_+\nn\\
&=&g^{1/2}L^{(2n+3)/2}\int_{\a_k}^{-1}\sqrt{\cE+(-1)^niy'^{2n+1}}dy'\nn\\
&=:&g^{1/2}L^{(2n+3)/2}\tau(\cE)\in\RR_+,\qquad y'=\a_k \xi\nn\\
\tau&=&\cE^{1/2}\a_k\int_1^{-\a_k^{-1}}\sqrt{1-\xi^{2n+1}}d\xi\in\RR_+.
\ea
The parameters $\cE$ and $\a_k$ have been scaled out from the integrand of $\tau$ to keep the integral of single type under differentiation. Making use of $\cE=-i(-1)^n\a_k^{2n+1}$ and its implications $\frac{d\a_k}{d\cE}=\frac{\a_k}{(2n+1)\cE}$, \ $\a_k\frac{d(-\a_k^{-1})}{d\cE}=\frac1{(2n+1)\cE}$ the differential version of the reality constraint can be obtained as
\ba{c2}
d\tau=\frac{d\tau}{d\cE}d\cE&=&\left[\frac12 \tau +\frac1{2n+1}\tau+\frac1{2n+1}\cE^{1/2}\sqrt{1-(-\a_k^{-1})^{2n+1}}\right]\frac{d\cE}{\cE}\in\RR\nn\\
&=&\left[\frac{2n+3}2 \tau +\sqrt{\cE-(-1)^n i}\right]\frac{d\cE}{(2n+1)\cE}\in\RR
\ea
with resulting ODE
\be{c3}
\frac{d\cE}{d\tau}=\frac{(2n+1)\cE}{\frac{2n+3}2\tau+\sqrt{\cE-(-1)^n i}}
\ee
as defining equation for one of the complex branches of the scaling graph [Eq. \rf{8} in the main text of the Letter]. The initial condition is most conveniently chosen close to $\cE(\tau=0)=(-1)^{n}i$. This point corresponds to a vanishing integral in \rf{c1}, i.e. a configuration with one of the turning points coinciding with the interval end point $y=-1$. Once three anti-Stokes line segments are starting in a turning point and the ODE \rf{c3} itself is singular at $\cE(\tau=0)=(-1)^{n}i$, the relevant initial condition has to be fixed at a point slightly shifted off $\cE(\tau=0)$ but located on the complex solution curve $\cE(\tau)$. Substituting for small $\tau\ll 1$ a Puiseux expansion ansatz
\ba{c4}
\cE(\tau)= (-1)^{n}i +A_1\tau^{\mu_1}+A_2\tau^{\mu_2}+o(\tau^{\mu_3}),\qquad 0<\mu_1<\mu_2<\mu_3<\ldots,\qquad A_k=\const
\ea
into \rf{c3} yields the initial condition  as
\be{c5}
\tau=\d \tau\ll1, \qquad \cE(\tau=\d\tau)=(-1)^n i+\left[\frac{3(2n+1)}2\d \tau\right]^{2/3}\exp\left[-(-1)^n i\frac{7\pi}3\right]+o(\d\tau^{4/3}).
\ee
This initial condition selects the correct integration path and is used to obtain numerical solutions and the graphical input for the comparison with the shooting-method results (see section G, below).

Next, we note that for general $\left[-g(ix)^{2n+1},x\in[-L,L]\right]$ models with $n\ge1$ the integrals of type \rf{c1} lead to expressions in terms of incomplete Beta-functions $B_z(b,c)=\int_0^z t^{b-1}(1-t)^{c-1}dt $ with $z=i(-1)^n\cE$, $b=1/(2n+1)$, $c=1/2$ what makes an explicit resolution of the reality constraints highly complicated compared to the equivalent complex ODE \rf{c3}. In the simple $(-ix,x\in[-L,L])$ case, i.e. for $n=0$,  one obtains from \rf{s24}, \rf{s25} the relation $\tau=\frac23 \hbar (gL^3)^{-1/2} s^{3/2}$ which together with \rf{c3} and \rf{s11a}, \rf{s15a}, \rf{s16} yields the solution branch
\be{c6}
\cE(\tau)=i+e^{-i\frac\pi3}\left(3\tau/2\right)^{2/3},\qquad \tau\in(0,\tau_c),\quad \tau_c=2^{5/2}3^{-7/4}.
\ee

Furthermore, within $0$th-order WKB, the $\cP\cT$ phase transition point $\cE_c$ can be roughly estimated as end-point of the complex branch $\cE_c\approx\cE(\tau_c)\in\RR_+$ and with obvious fitting of break-up rule \rf{p10} and quantization rule [Eq. \rf{5} in the main text of the Letter]. The $\cY-$identification $-\hbar I_L=\int_{a_k}^{-L}\sqrt{Q(x')}dx'=\int_\cY\sqrt{Q(x')}dx'$ at $\cE_c$ indicates an underlying selection rule for the $\cP\cT$ phase transition and Stokes graph break-up toward complex energies: $ \pm I_L+\frac\pi4=N_3\pi \quad\longrightarrow\quad \pm\Im \cE>0$, an effect which should find an explanation within an explicit WKB-Riemann-surface based technique (still to be developed).
\newpage

\section{F: The $(igx^3,x\in[-L,L])$ model: graphical supplement}

\subsection{Spectral data  (shooting-method)}
The 20 lowest eigenvalues $\left\{E_j(L)\right\}_{j=1}^{20}$ have been calculated numerically to high precision by a shooting-method-based eigenvalue solver for complex Sturm-Liouville problems. The corresponding single data set $\left\{E_j(L)\right\}_{j=1}^{20}$ for the $(ix^3,x\in [-L,L])$ model is analyzed under various scalings to visualize the hidden simple structure laws and patterns contained in these spectral data.
\begin{figure}[htb]
\begin{center}
\includegraphics[scale=0.68]{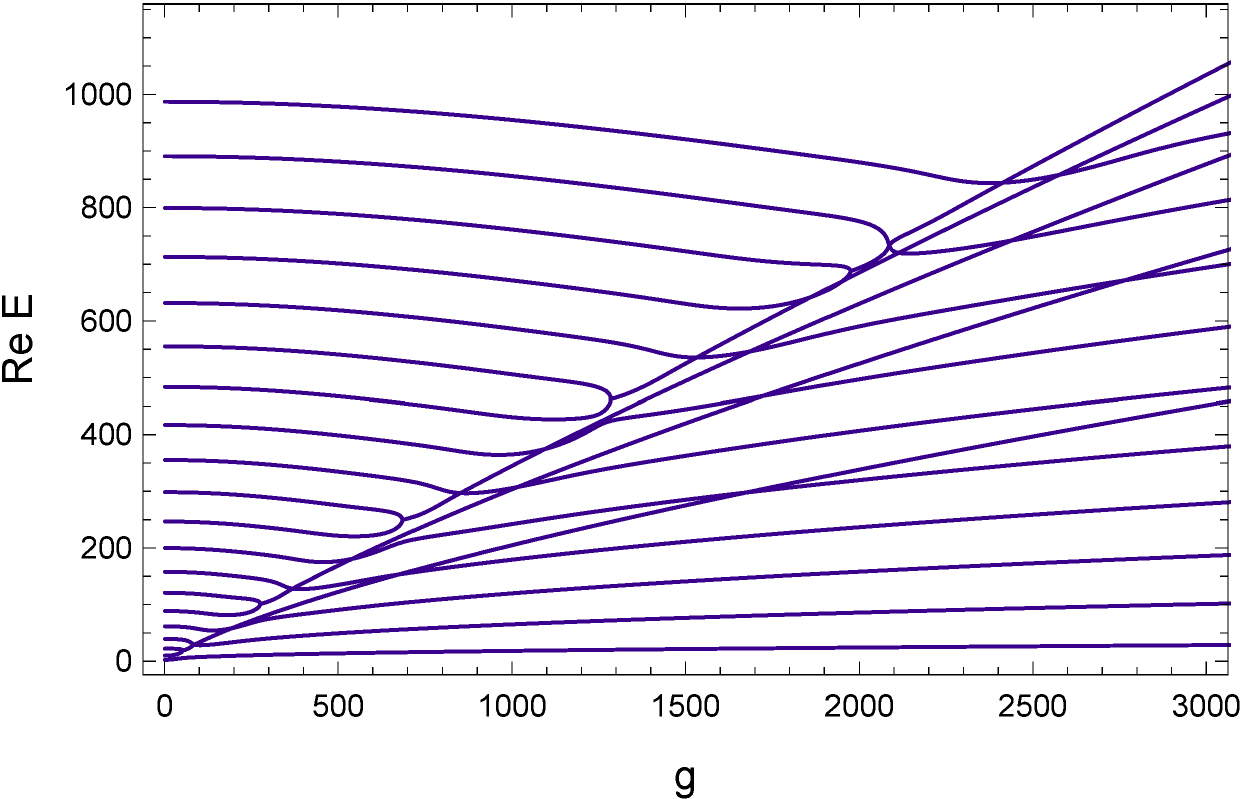}\hspace{2em}
\includegraphics[scale=0.68]{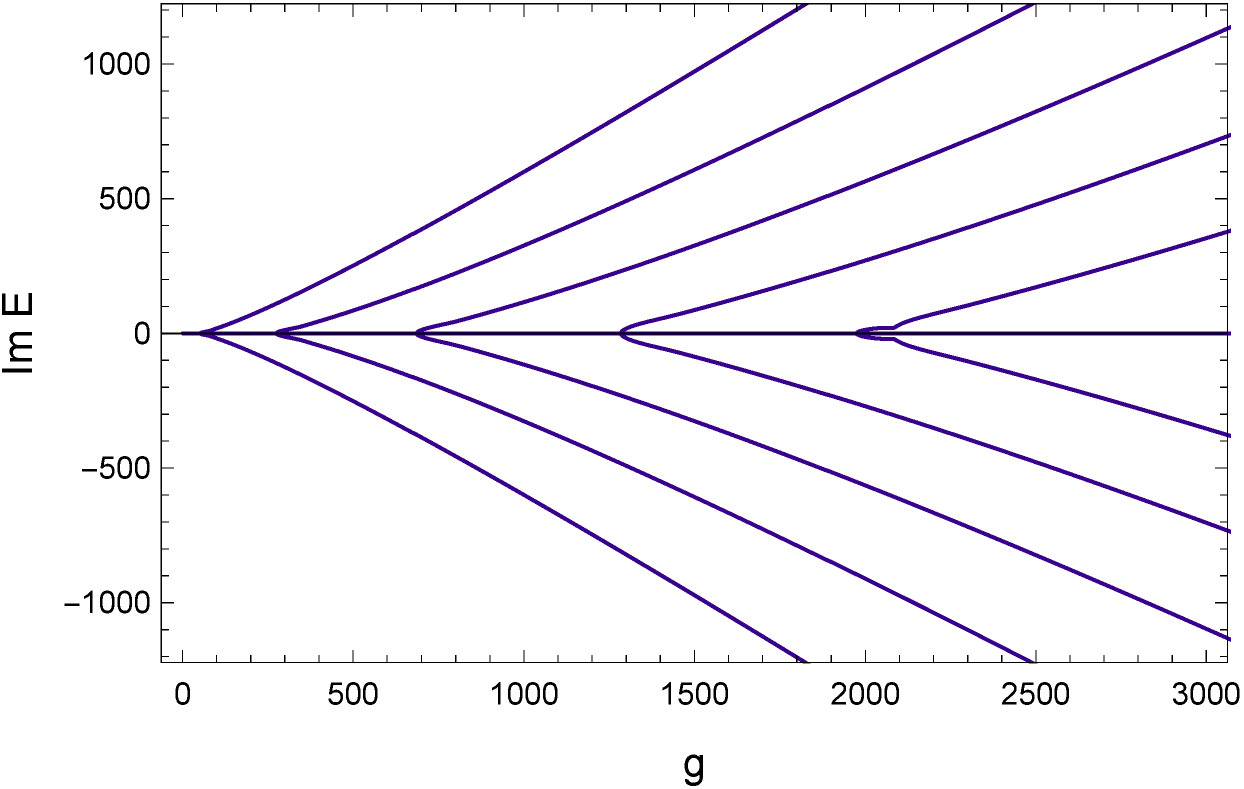}
\caption{\label{ix3-over-g} Eigenvalue behavior of the $(igy^3,y\in[-1,1])$ model. This spectral behavior, known from \cite{bender-jones-2}, is reproduced from the $(ix^3, x\in [-L,L])$ model data $\left\{E_j(L)\right\}_{j=1}^{20}$ by a rescaling $x=Ly$, \ $g=L^5$, \ $\tilde E_j=L^2 E_j(L)$. Such a spectral behavior with its transitions from purely real to pairwise complex-conjugate branches is a typical pattern for various $\cP\cT-$symmetric setups \cite{bender-jones-2}.}
\end{center}
\end{figure}
\begin{figure}[htb]
\begin{center}
\includegraphics[scale=0.67]{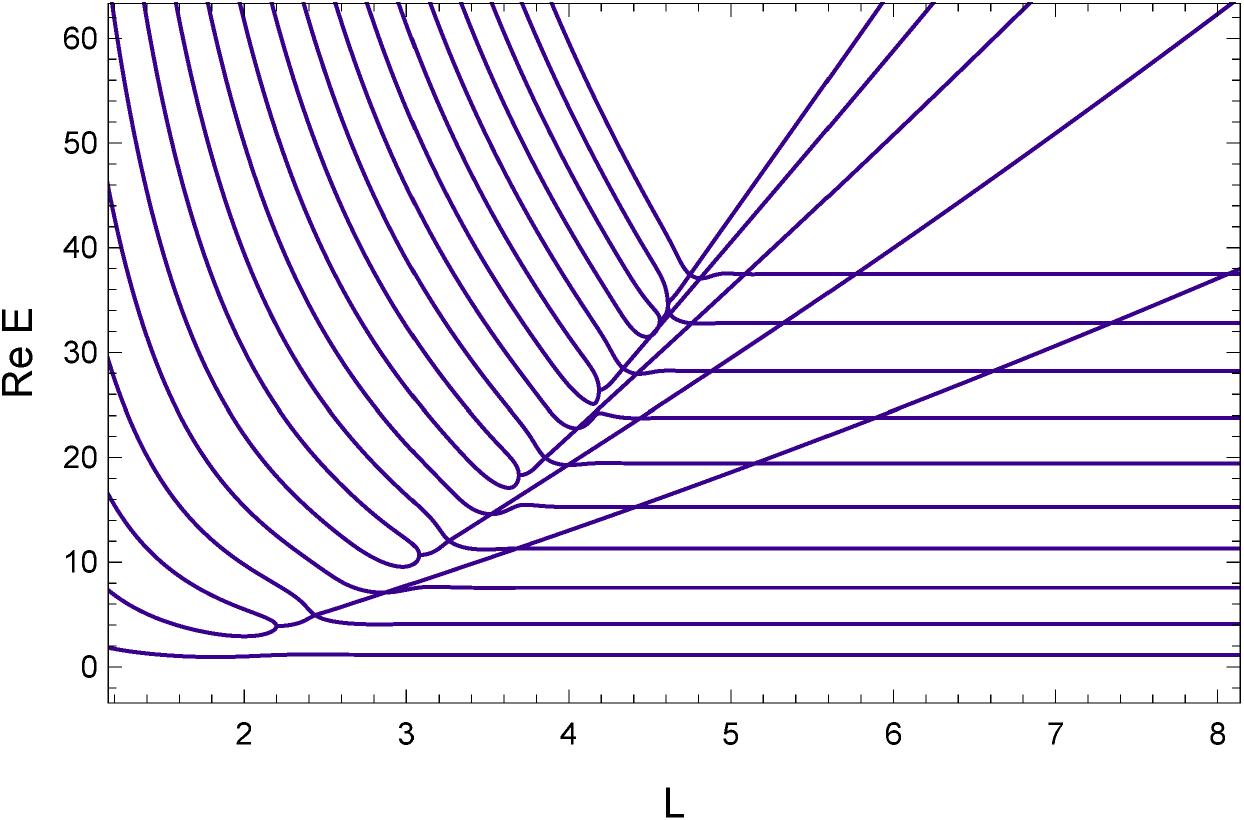}\hspace{2em}
\includegraphics[scale=0.68]{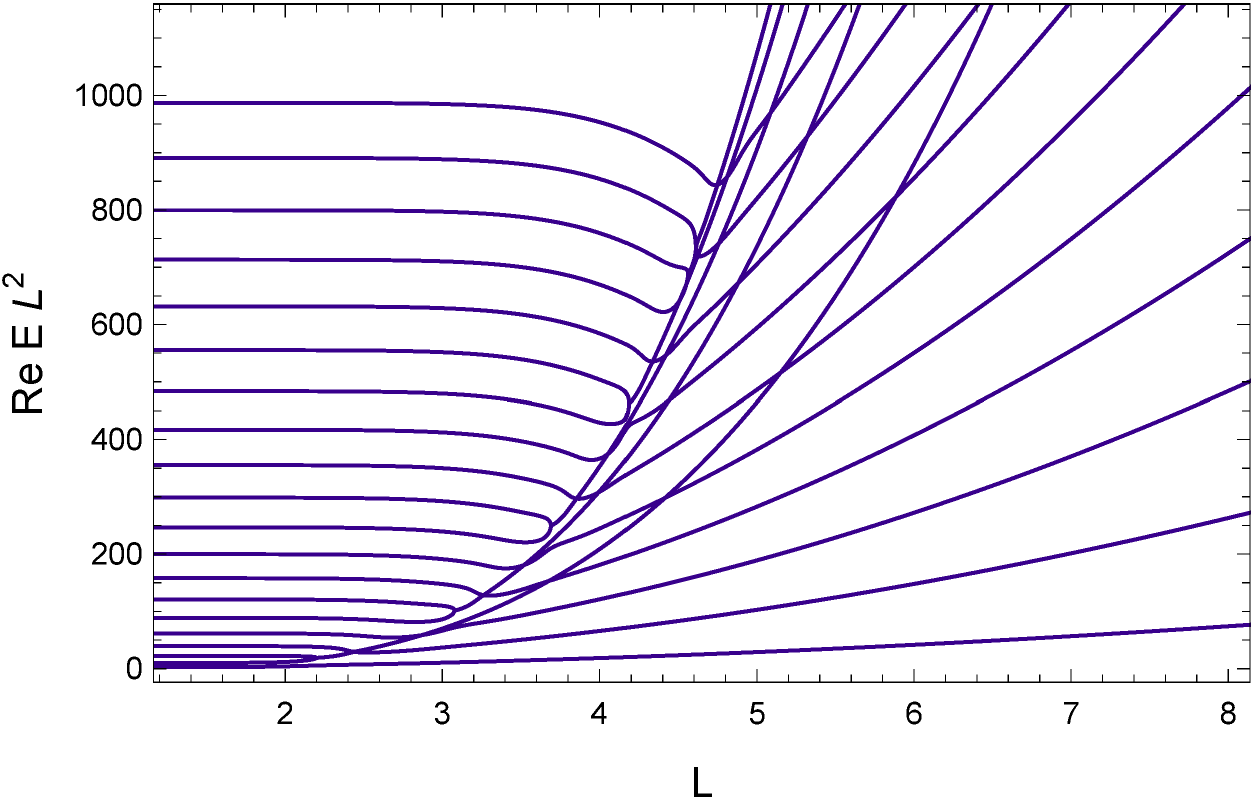}\\
\caption{\label{ix3-real-energy-over-L} Eigenvalue behavior of the $(ix^3,x\in[-L,L])$ model. Clearly visible is the $L-$independence of the low-lying Bohr-Sommerfeld (BS) type real eigenvalues for large $L$ (left graphic) which reproduce the corresponding low-lying eigenvalues of the IR-completed $(ix^3,x\in\RR)$ model. Clearly pronounced is also the $L-$independence of the high-lying $L^2-$multiplied box-type (BT) real eigenvalues with $E_j L^2\approx \pi^2 j^2/4$ for small $L$ (right graphic). This specific behavior of the real eigenvalues, including the $BT\rightleftarrows BS$ transition between the two spectral regimes, is described by the secular equation pair \rf{p8}.}
\end{center}
\end{figure}
%\newpage
\begin{figure}[htb]
\begin{center}
\includegraphics[scale=0.67]{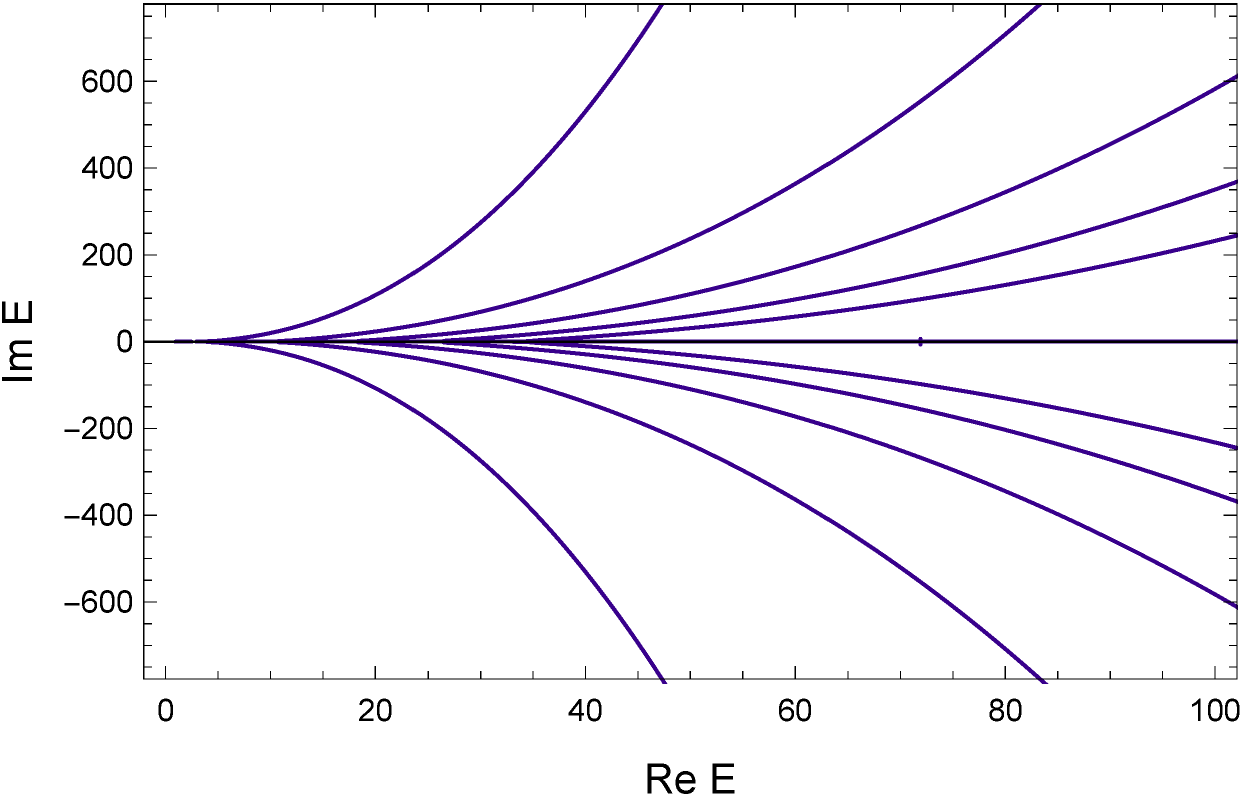}\hspace{2em}
\includegraphics[scale=0.68]{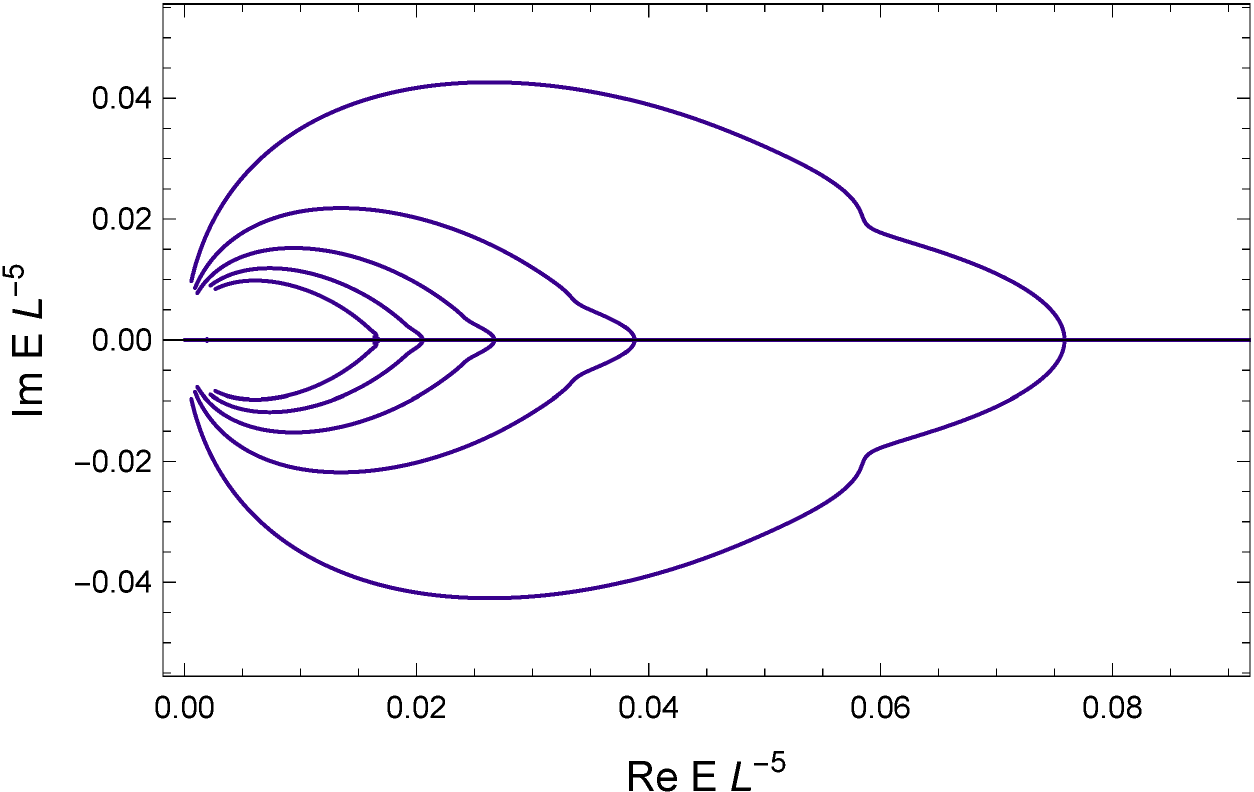}\\
\caption{\label{ix3-real-imaginary-energy-general} Eigenvalue behavior of the $(ix^3,[-L,L])$ model. The pairwise complex conjugate eigenvalue branches are clearly separated for unscaled eigenvalues (left graphic) and for arbitrarily scaled eigenvalues (here sampled for $L^{-5}E_j(L)$, right graphic). An exceptional role play the $L^{-3}-$scaled eigenvalues $L^{-3}E_j(L)$ for which the complex branches coincide outside a well-defined $\cP\cT$ phase transition region (next figure).}
\end{center}
\end{figure}
\begin{figure}[htb]
\begin{center}
\includegraphics[scale=0.665]{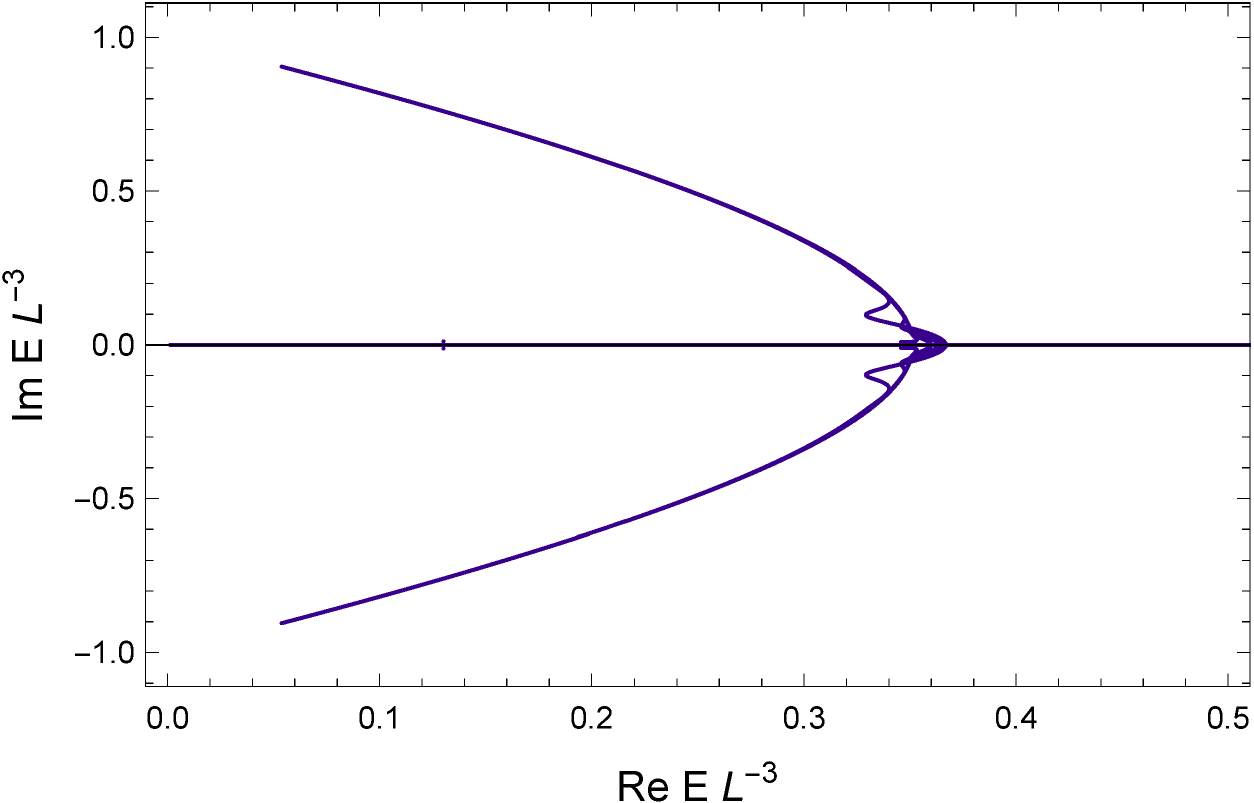}\hspace{2em}
\includegraphics[scale=0.665]{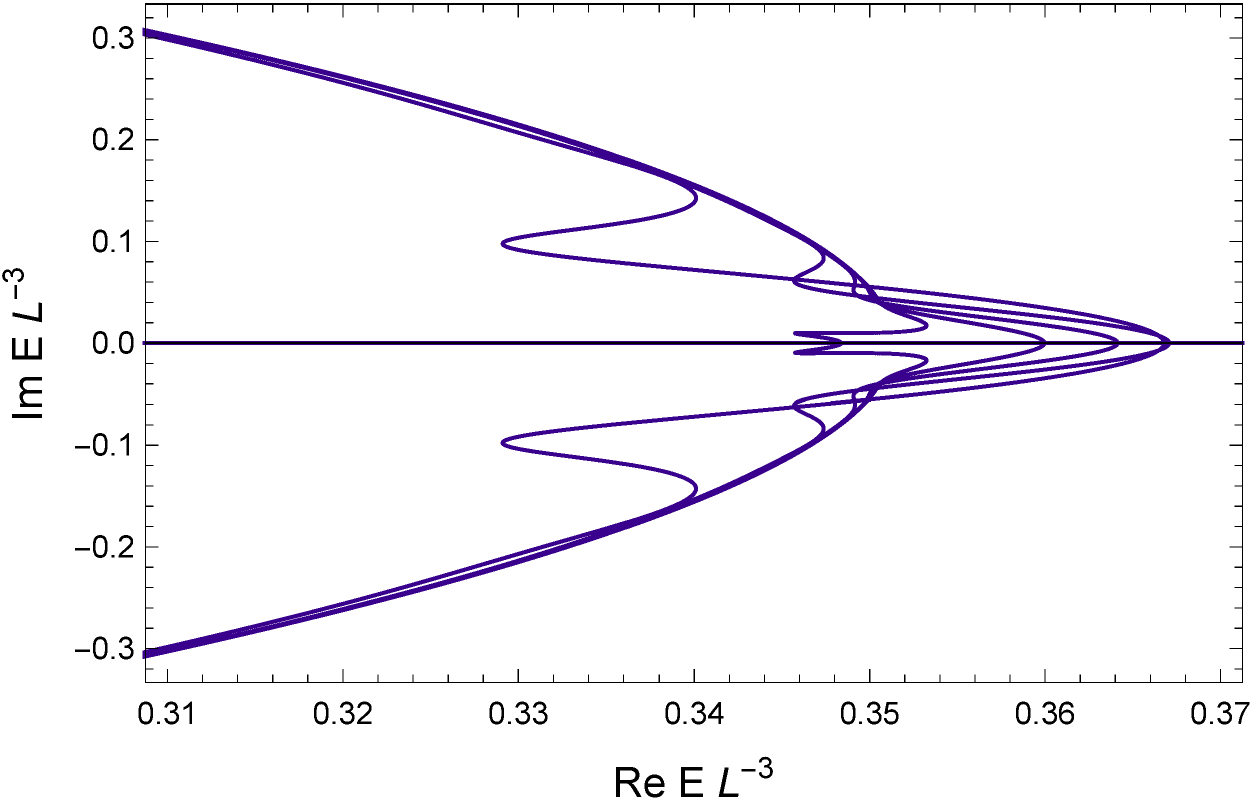}\\
\caption{\label{ix3-real-imaginary-energy-general} Eigenvalue behavior of the $(ix^3,[-L,L])$ model. The $L^{-3}-$scaled complex eigenvalue branches $\cE_j=L^{-3}E_j(L)$ coincide outside a well-defined $\cP\cT$ phase transition region (left graphic). The coincidence curve forms the complex part $\cR_{CO}$ of the spectral scaling graph $\cR= \cR_{CO}\cup \RR_+$ and can be described analytically within zeroth-order WKB (for a comparison see Fig. \ref{ix3-real-imaginary-energy-comparison} below). There exists a $\cP\cT$ phase transition region in the $L^{-3}-$scaled spectral plane (the spectral $\cE-$plane) where the complex branches do not coincide (right graphic) and where the zeroth-order WKB approximation breaks down. From the numerical data one observes that the strongest deviations from zeroth-order WKB are produced by the eigenvalue branches with the lowest mode numbers $j$, whereas an increasing $j$ is associated with shrinking deviations. More sophisticated techniques will be required for a comprehensive analytical description in this $\cP\cT$ transition region and are still to be developed.}
\end{center}
\end{figure}

\clearpage

\subsection{Stokes graphs}
\mbox{}\hfill\mbox{}\vspace{0.1cm}\mbox{}
\begin{figure}[htb]
\begin{center}
\includegraphics[scale=0.45]{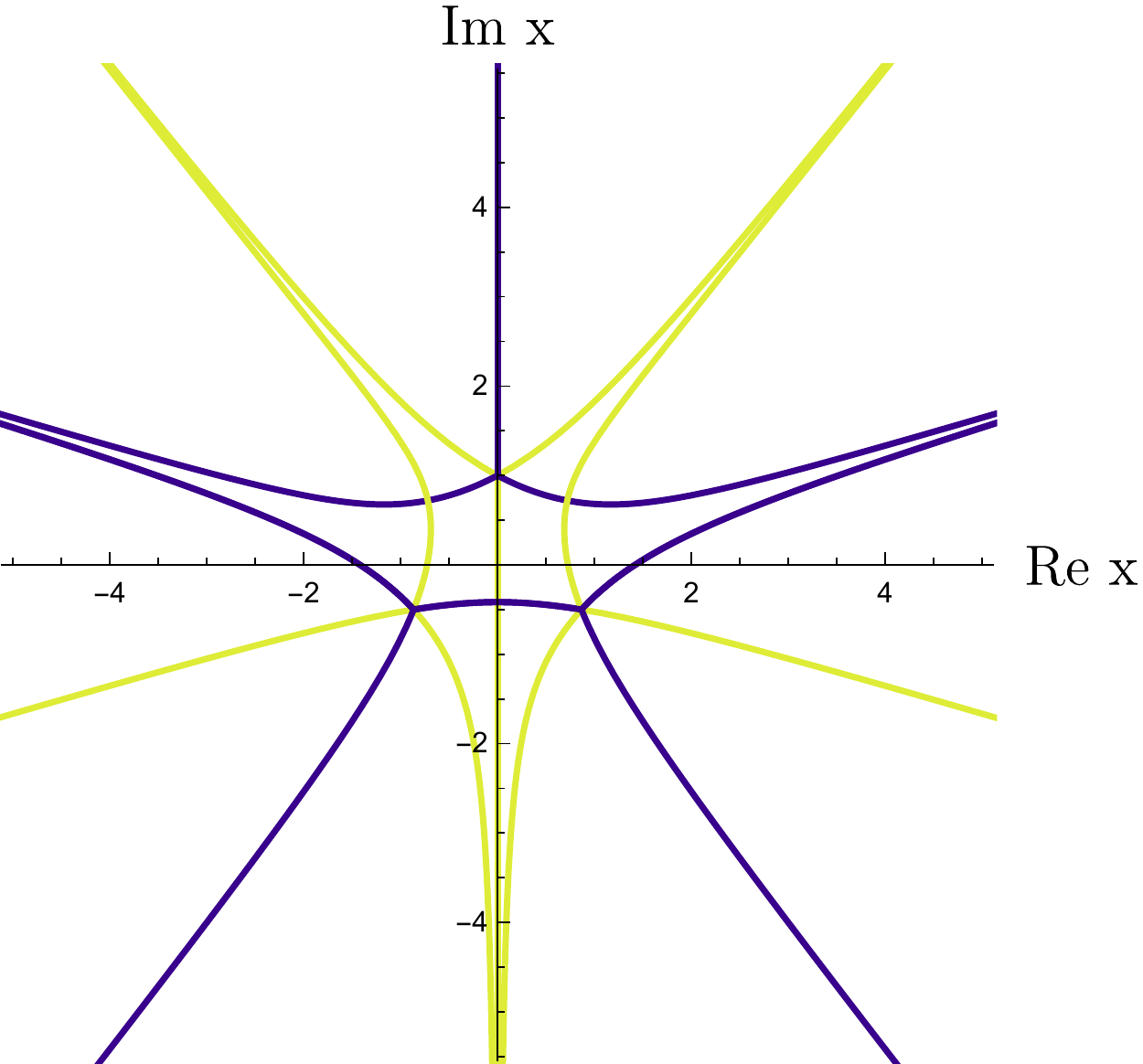}\hspace{2em}\\[4ex]
\includegraphics[scale=0.45]{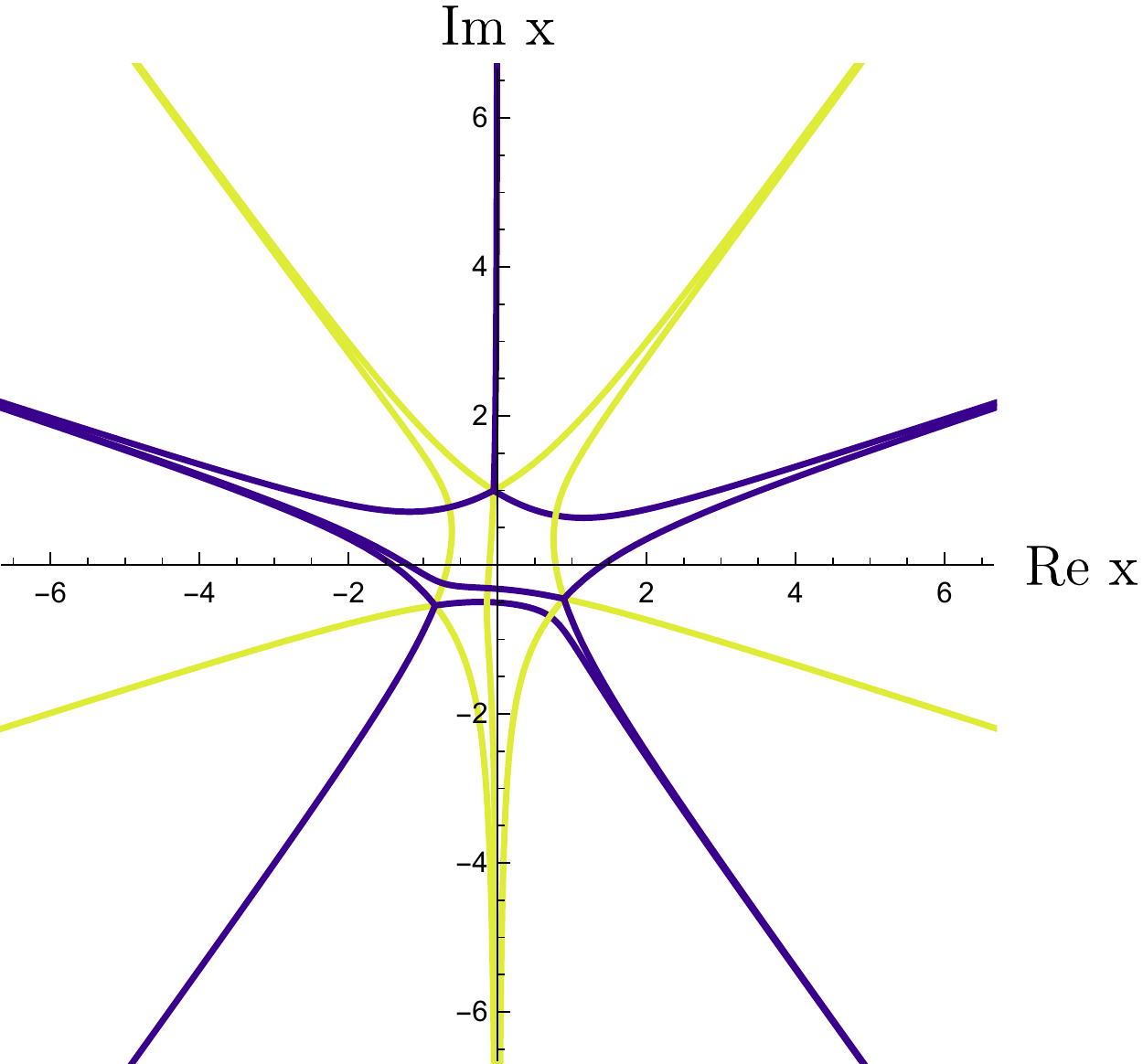}\hspace{2em}
\includegraphics[scale=0.45]{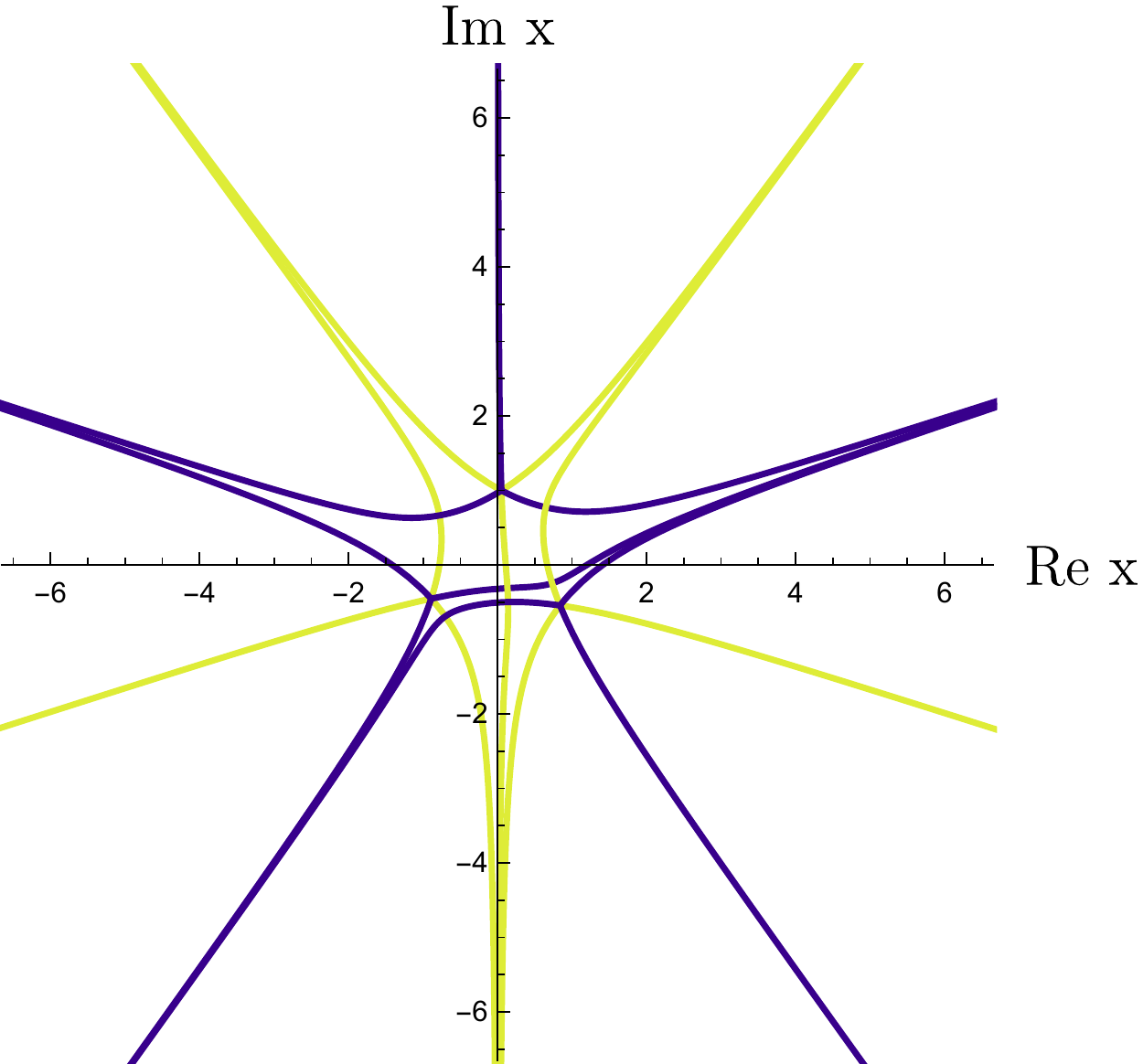}\\[4ex]
\includegraphics[scale=0.45]{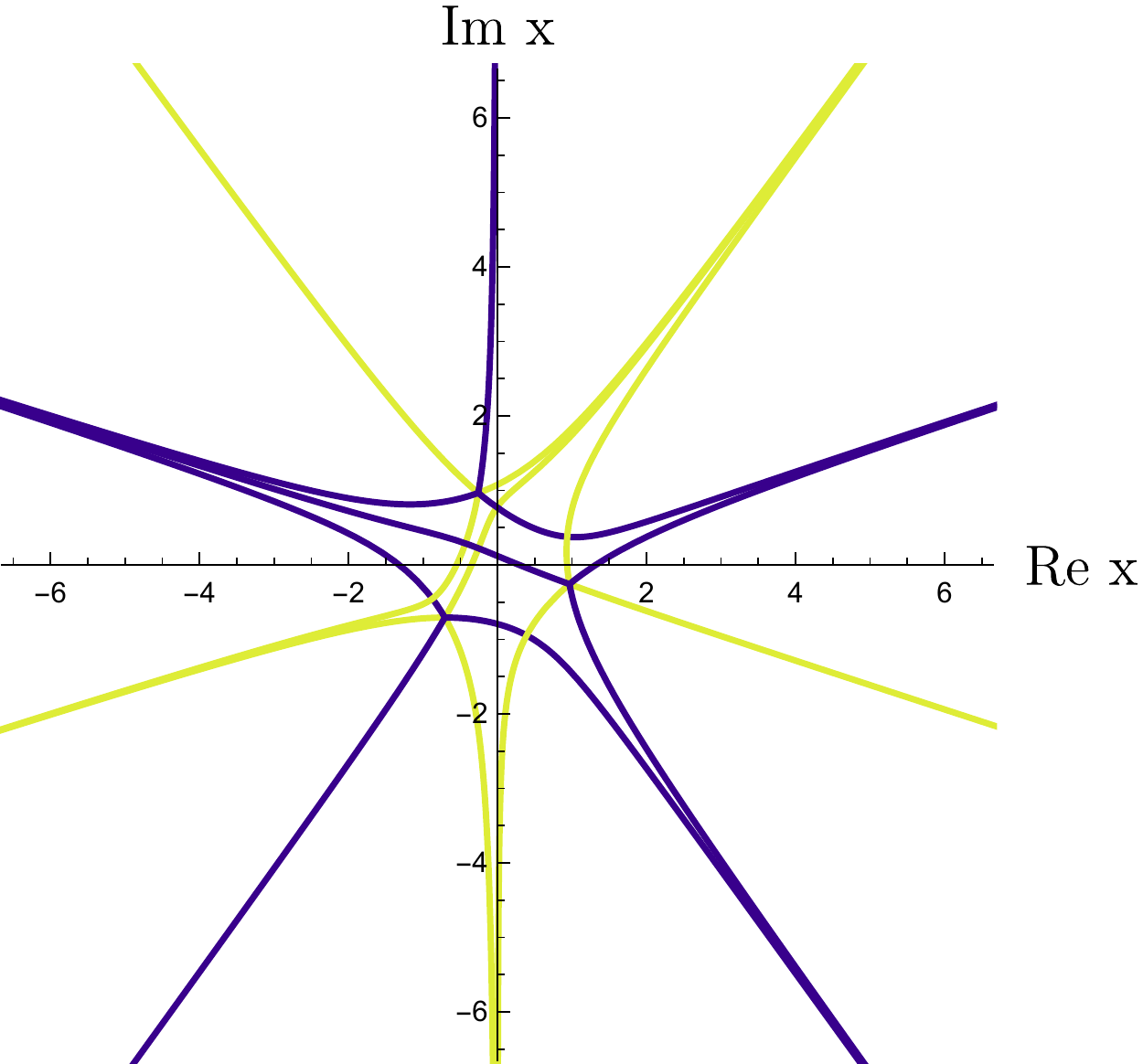}\hspace{2em}
\includegraphics[scale=0.45]{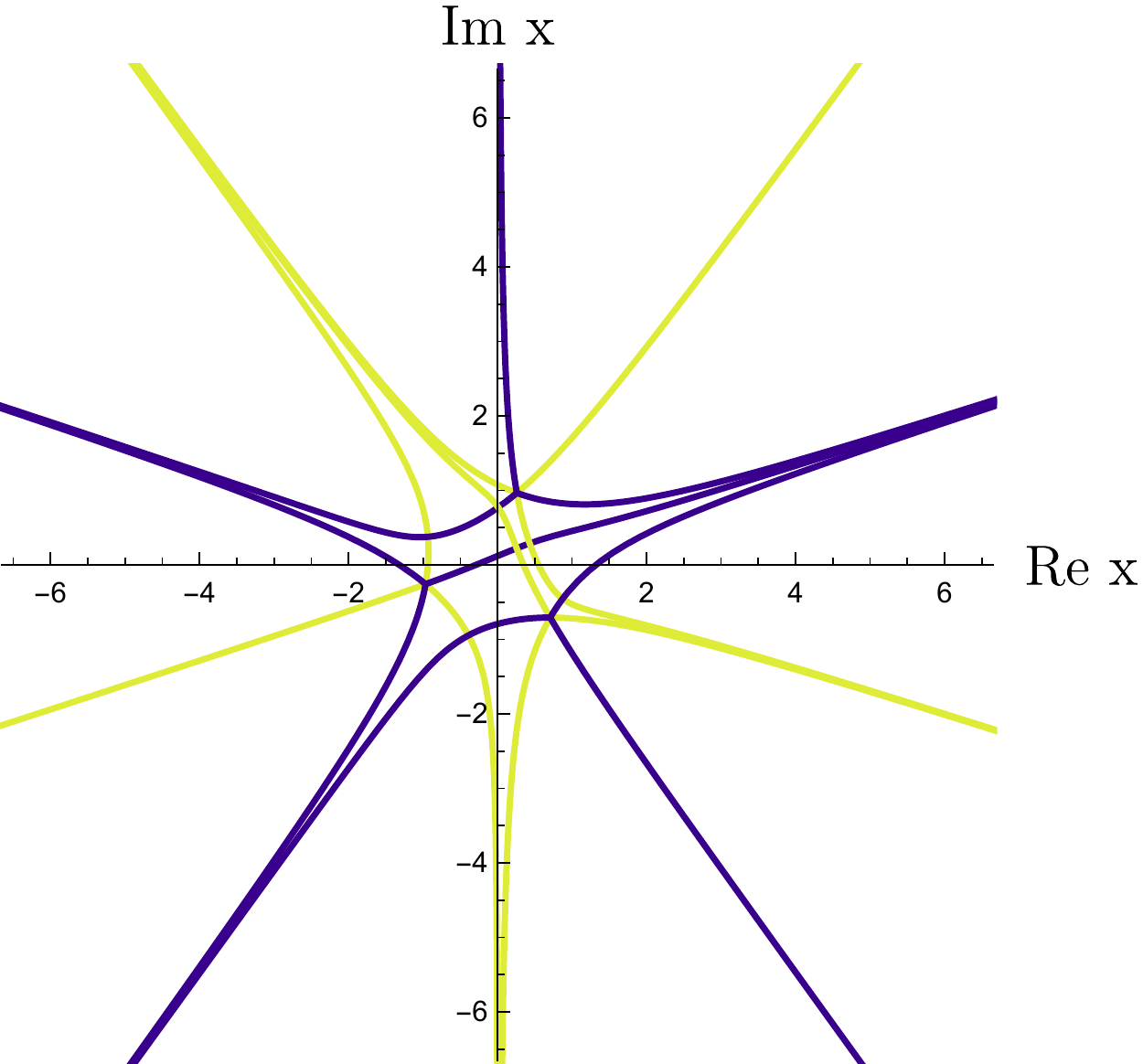}\\[4ex]
\caption{\label{sample-stokes} Stokes graphs of WKB wave functions $\psi_\pm(x)=Q^{-1/4}(x)\exp\left[\pm i\hbar^{-1}\int_{x_0}^x \sqrt{E-ix'^3}dx'\right]$ for sample energies $E=1$ (top), $E=e^{i\pi/20}$ (middle left), $E=e^{-i\pi/20}$ (middle right), $E=e^{i\pi/4}$ (bottom left), $E=e^{-i\pi/4}$ (bottom right). Anti-Stokes lines (blue curves) correspond to oscillatory behavior of $\psi_\pm$, whereas Stokes lines (green curves) indicate steepest exponential decay or blow-up. Clearly visible is the $\cP\cT$ symmetry relation between graphs for complex conjugate energies which shows up as mutual reflection symmetry with regard to the imaginary axis in the complex $x-$plane.}
\end{center}
\end{figure}
%\newpage
\begin{figure}[htb]
\begin{center}
\includegraphics[scale=0.6]{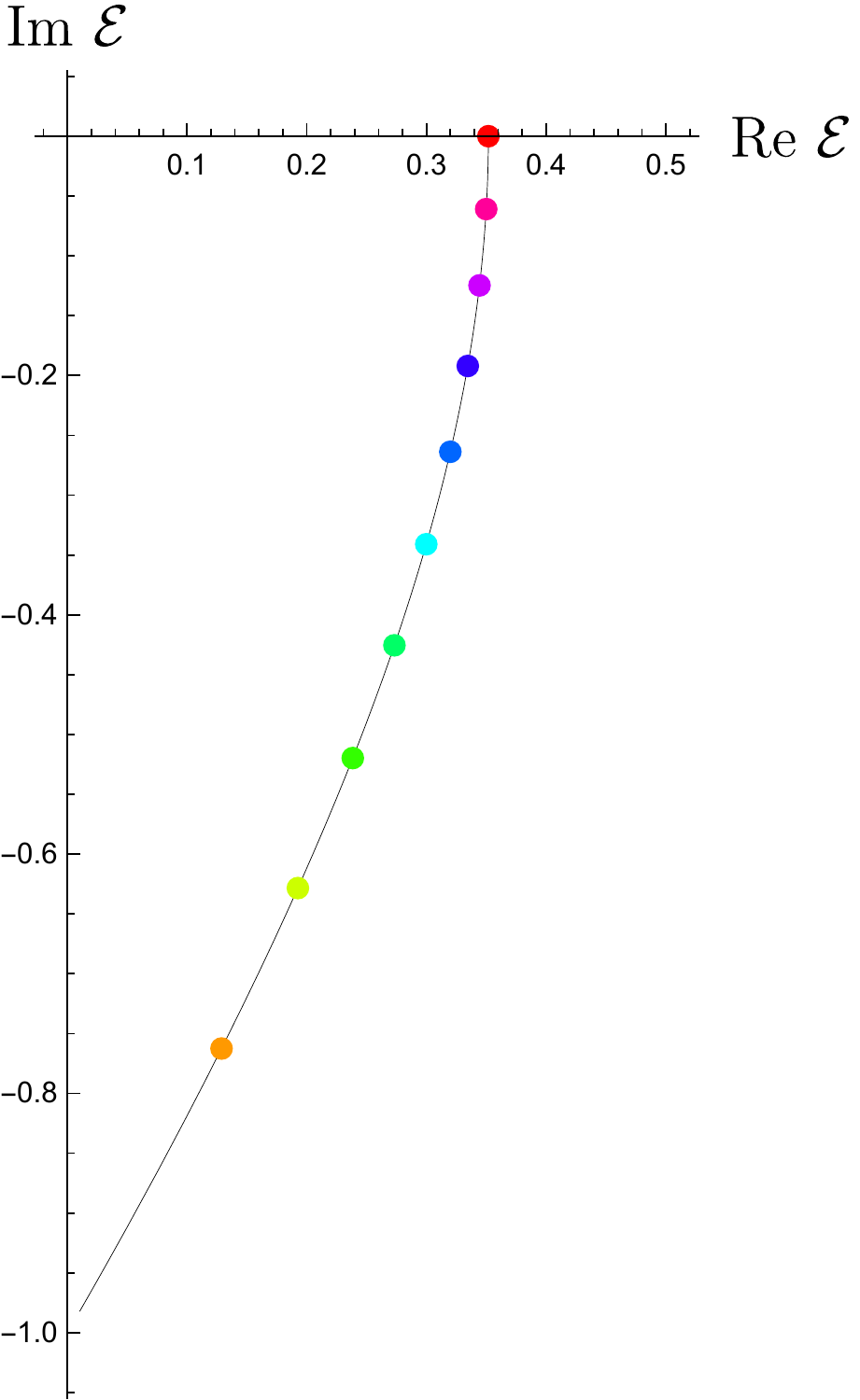}\hspace{4em}
\includegraphics[scale=0.8]{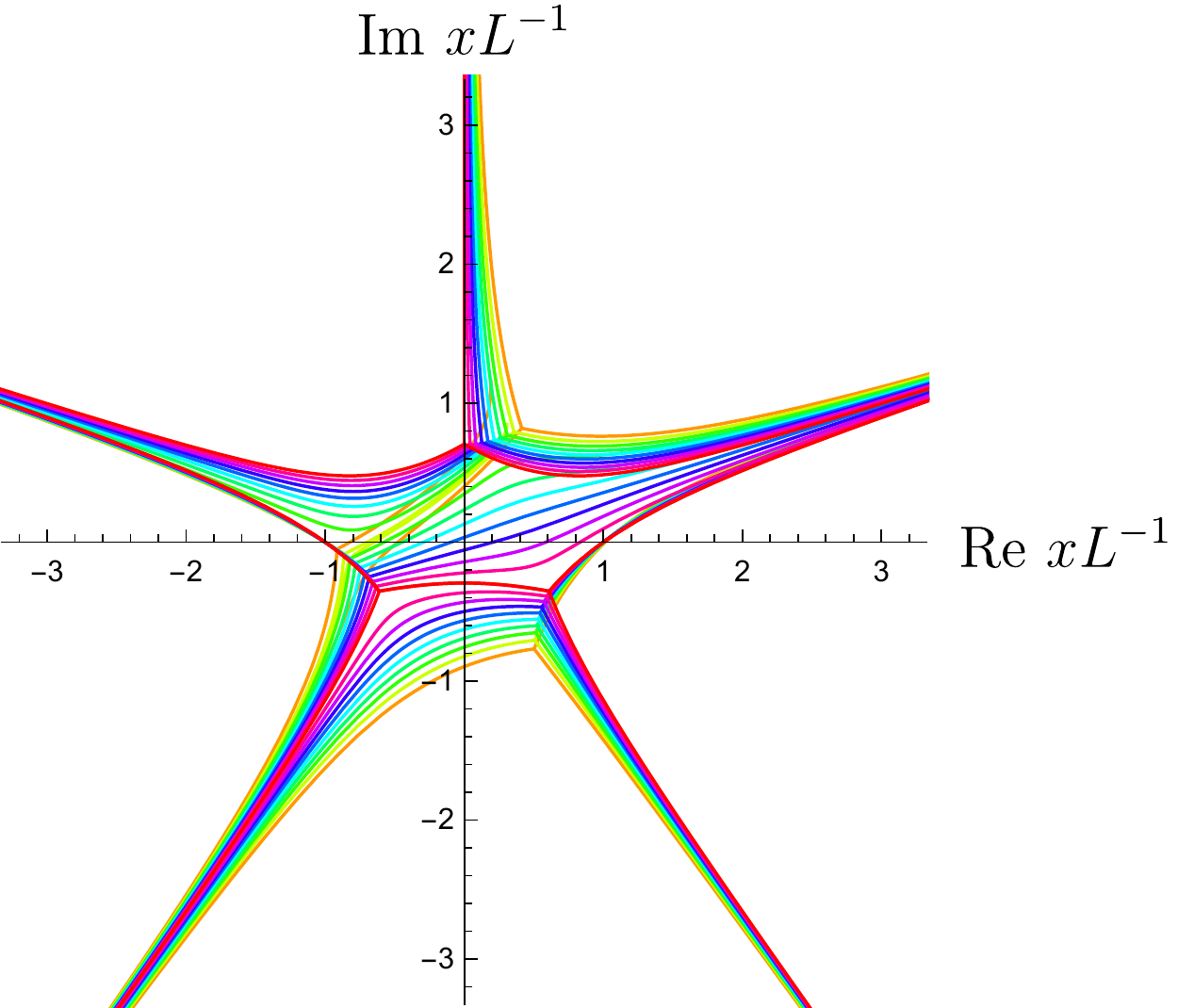}\\[6ex]
\includegraphics[scale=0.65]{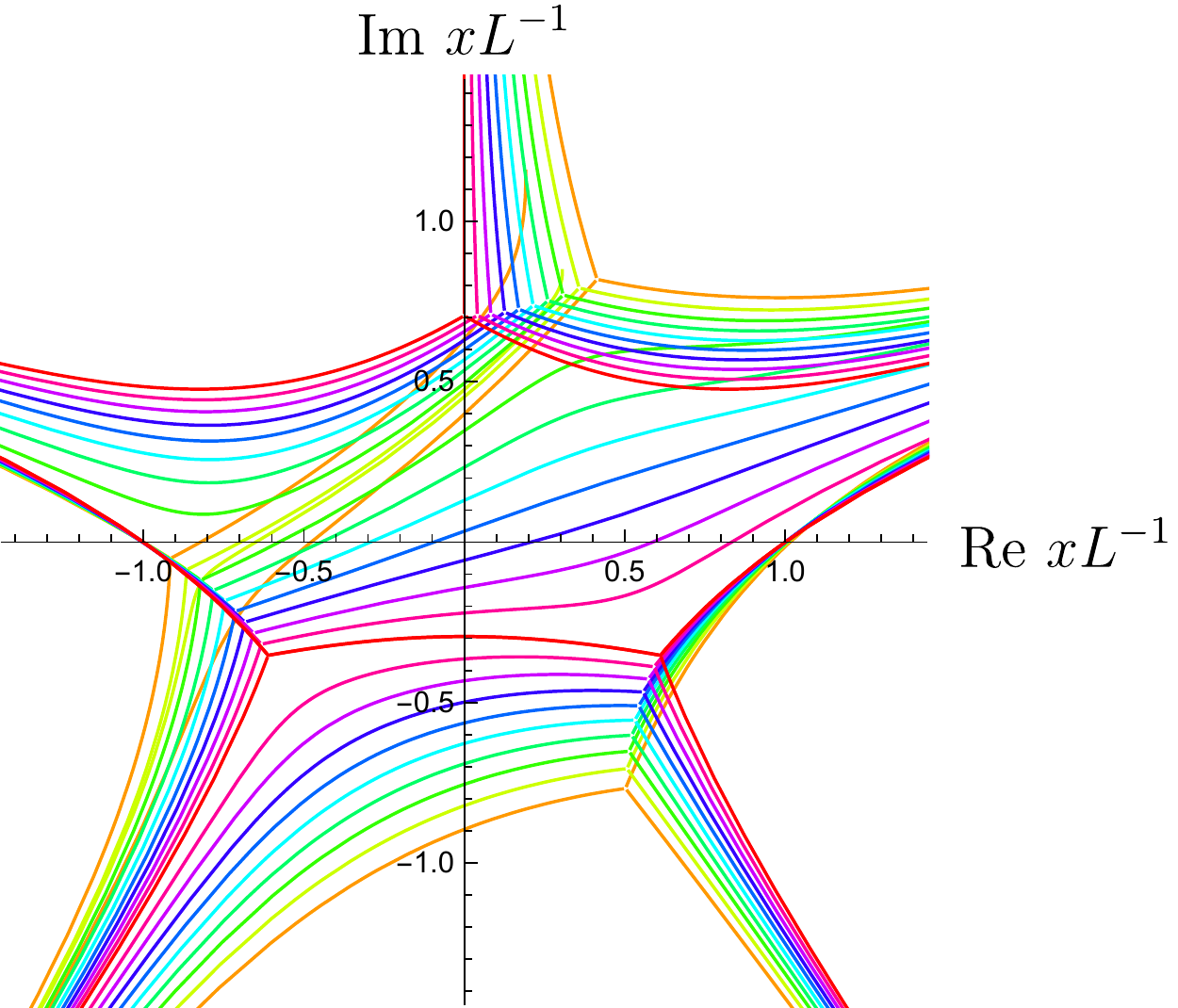}\hspace{4em}
\includegraphics[scale=0.65]{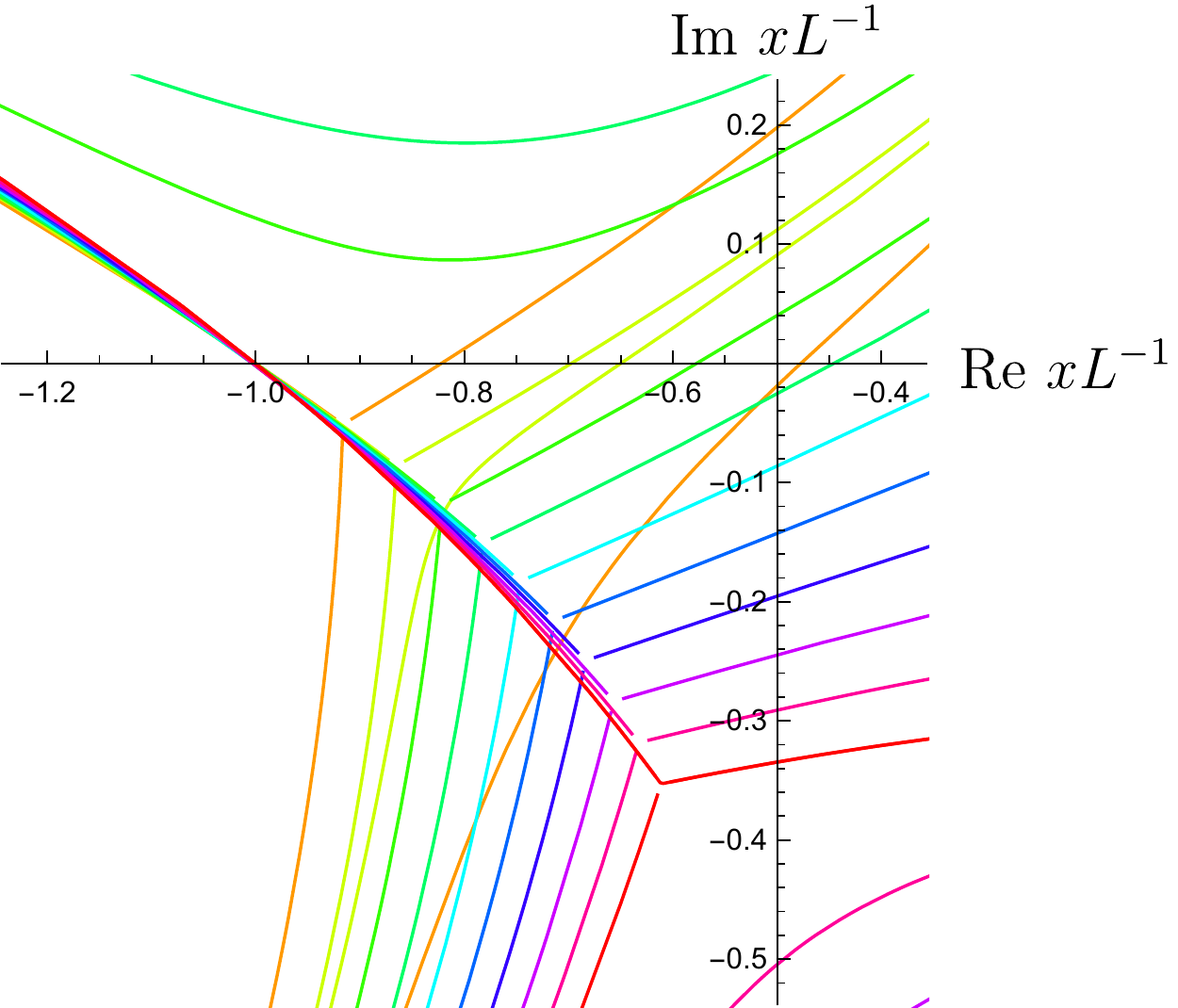} \\[6ex]
\caption{\label{sample-stokes-deformation} The $(ix^3,x\in[-L,L])$ model. Spectral scaling graph segment $\cR_{CO}(\Im \cE< 0)\cup \cE_c$ with 10 marked sample rescaled/mapped energies (MEs) $\cE_j=L^{-3}E_j=\cE\left( j\tau_c/10\right)\in \cR_{CO}$, $j=1,\ldots,9$ $\cup$ $\cE_{10}=\cE_c\in\RR_+$ (top, left) and the corresponding Stokes graphs (anti-Stokes lines) in various resolutions (top, right; bottom). For all MEs $\cE_j\in\cR_{CO}$, one of the Dirichlet BCs for the wave function has to be satisfied at one of the interval end points exactly (here at $y=xL^{-1}=-1$ with $\psi(x=-L)=0$) and an anti-Stokes line has to pass through that point,  $y=-1\in \cS$ (as clearly visible; bottom right). For $\cE_c\in\RR_+$, the condition $y=\pm 1\in \cS$ holds at both end points with exact Dirichlet BCs satisfied simultaneously $\psi(x=\pm L)=0$.}
\end{center}
\end{figure}

\begin{figure}[htb]
\begin{center}
\includegraphics[scale=0.8]{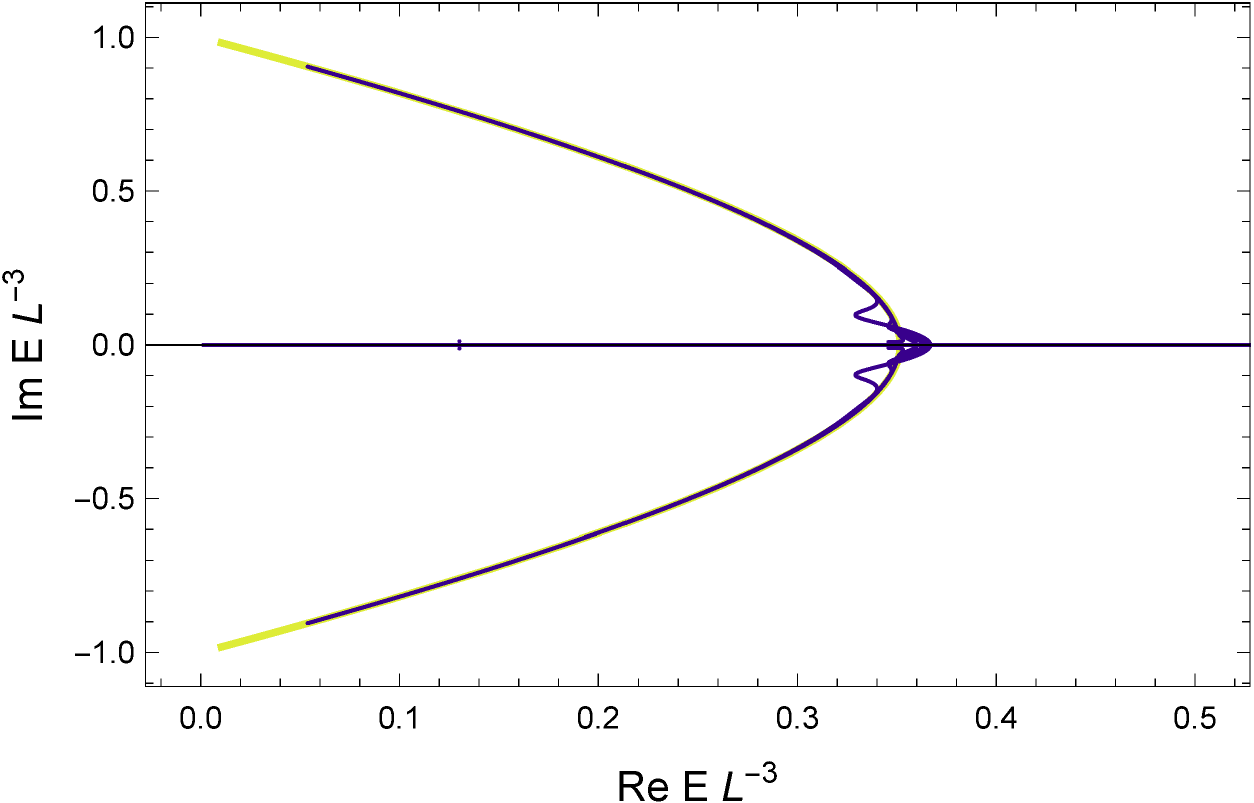}\\[6ex]
\includegraphics[scale=0.8]{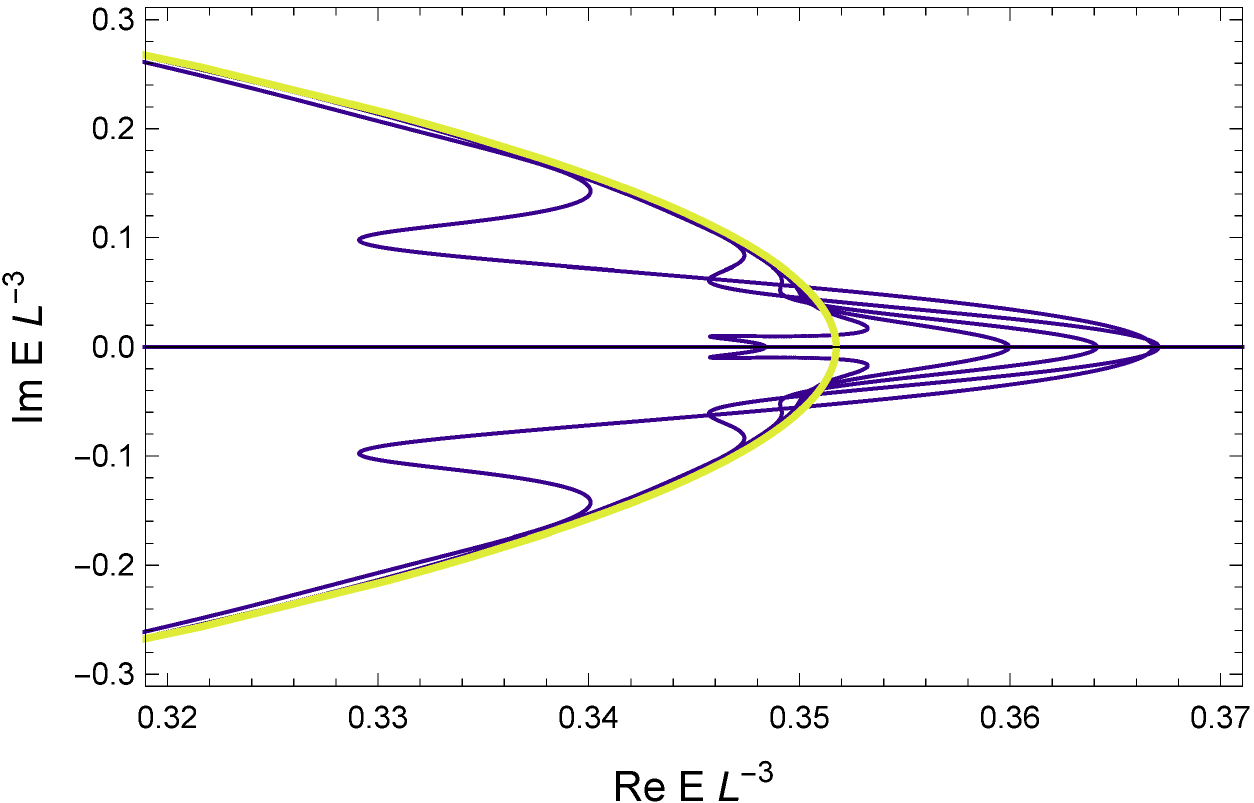}\\[6ex]
\caption{\label{ix3-real-imaginary-energy-comparison} Eigenvalue behavior of the $(ix^3,[-L,L])$ model. Comparison of the $L^{-3}-$scaled complex eigenvalue branches $L^{-3}E_j(L)$ (blue curves) obtained by a shooting method with the
zeroth-order WKB approximation results (green curve). A remarkably good coincidence of the two curves holds outside a certain $\cP\cT$ phase transition region (upper graphic). From the numerical data one observes that the strongest deviations from zeroth-order WKB are produced by the eigenvalue branches with the lowest mode numbers $j$, whereas an increasing $j$ is associated with shrinking deviations. More sophisticated (higher-order) techniques will be required for a comprehensive analytical description in this $\cP\cT$ transition region (lower graphic) and are still to be developed.\mbox{}\vspace{3cm}\mbox{}}
\end{center}
\end{figure}

\clearpage

\section{G: $\left[-g(ix)^{2n+1},x\in[-L,L]\right]$ models and their asymptotic spectral scaling graphs:\\ comparison to shooting-method results}
\begin{figure}[htb]
\begin{center}
\includegraphics[scale=0.59]{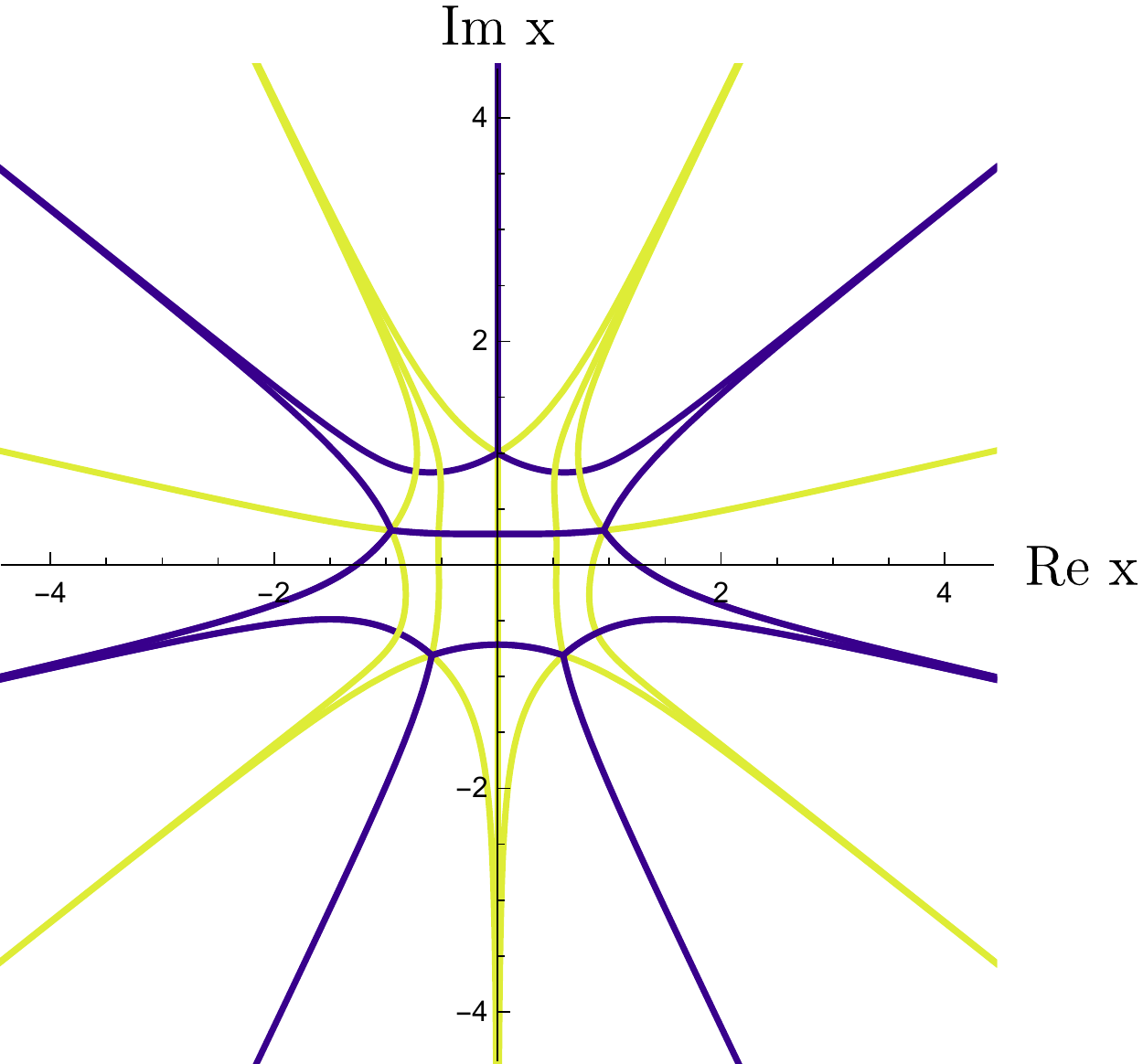}\hspace{2em}
\includegraphics[scale=0.59]{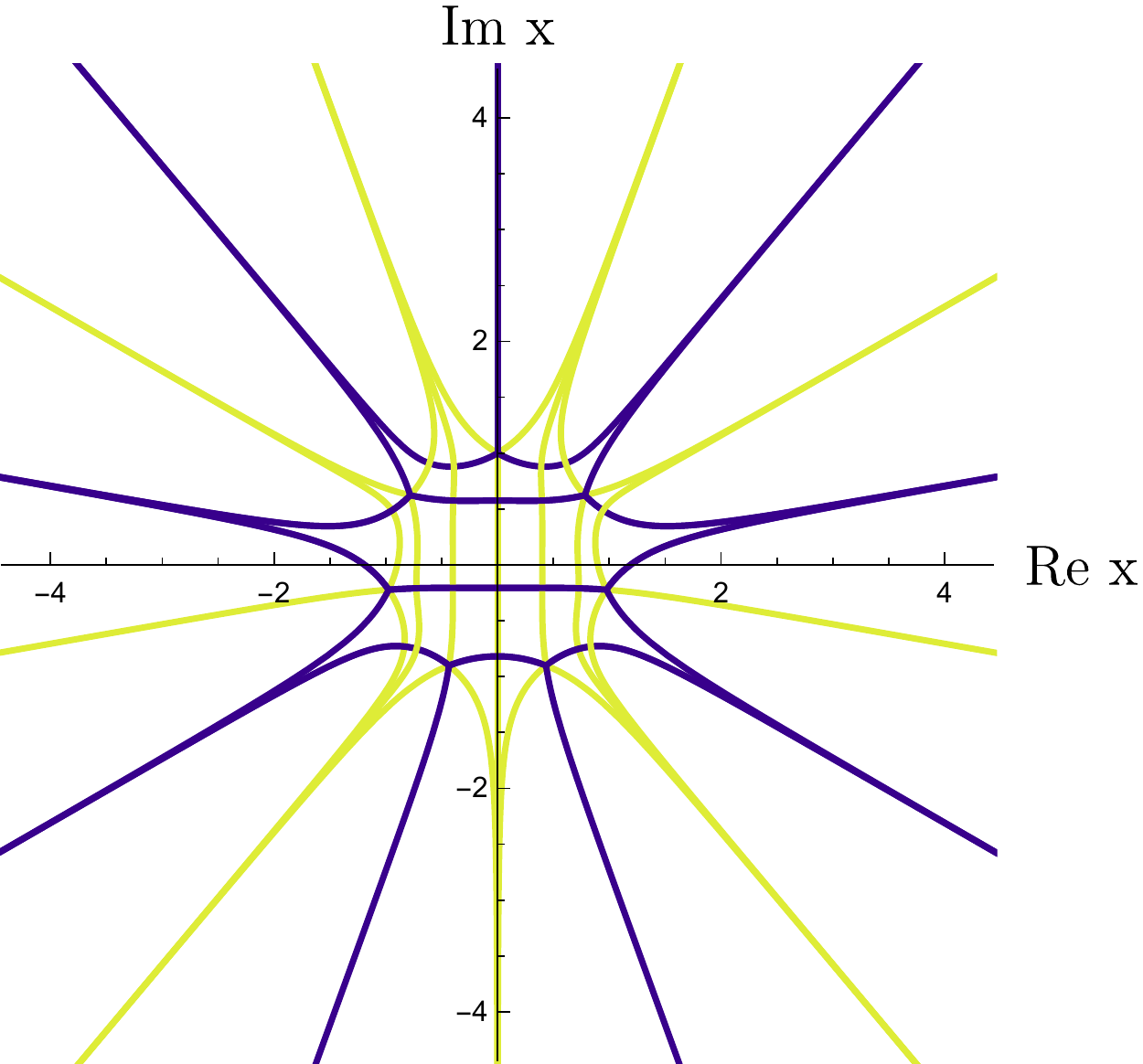}\\[2ex]
$V=-ix^5$\hspace{20em}$V=ix^7$\\[4ex]
\includegraphics[scale=0.59]{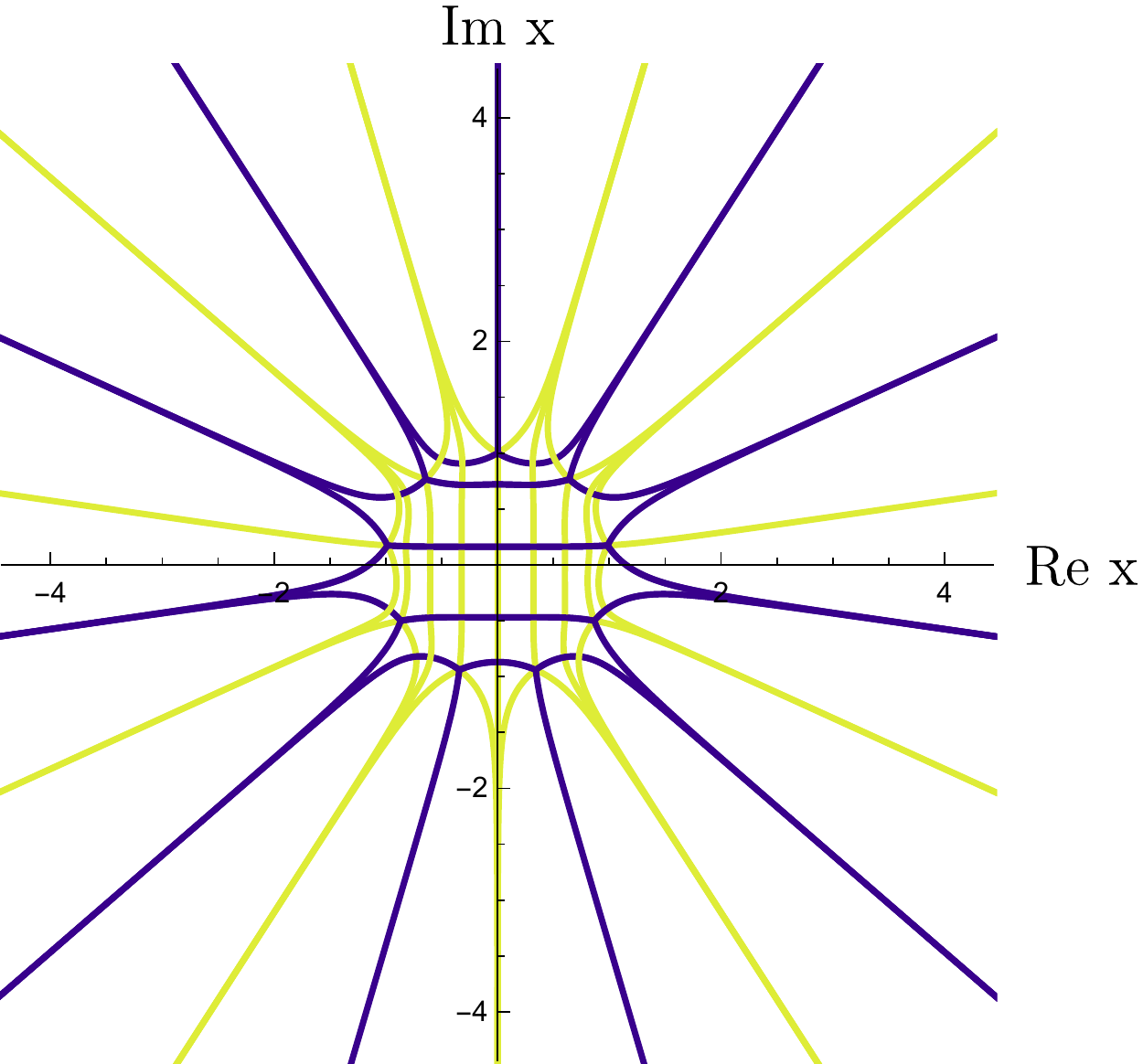}\hspace{2em}
\includegraphics[scale=0.59]{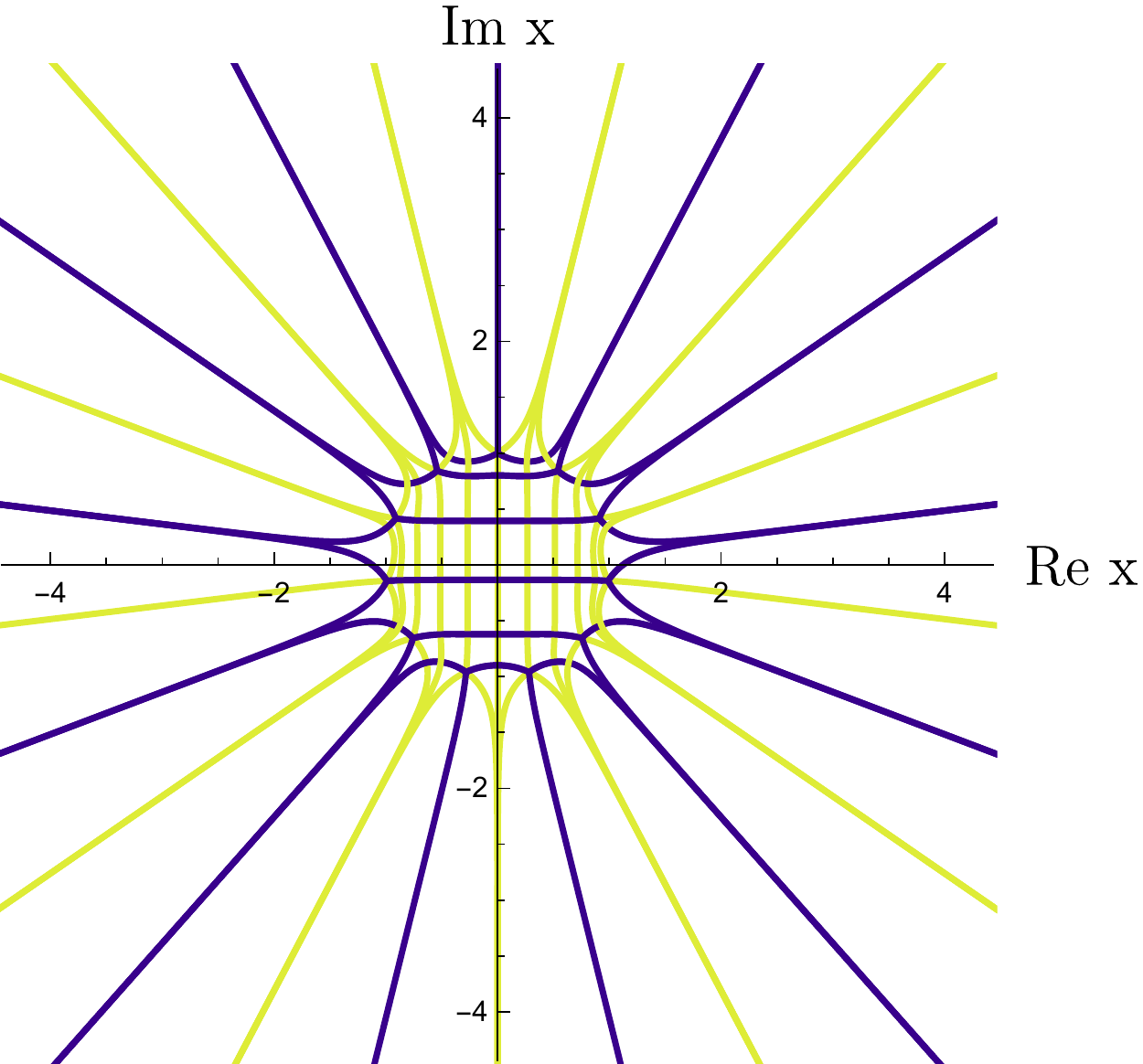}\\[2ex]
$V=-ix^9$\hspace{20em}$V=ix^{11}$\\[4ex]
\caption{\label{sample-stokes-deformation} $\left[-(ix)^{2n+1},x\in[-L,L]\right]$ models. Stokes graphs of the models with $n=2,\ldots,5$, i.e.  with $V=-ix^5$ $\ldots$ $V=ix^{11}$ for the same fixed real sample energy $E=1$. Anti-Stokes lines (blue curves) correspond to oscillatory behavior, whereas Stokes lines (green curves) indicate steepest exponential decay or blow-up. Regardless of the increasing complexity of the Stokes graphs with increasing $n$, the basic mechanism responsible for the location of the complex conjugate eigenvalue pairs remains the same kind of zeroth-order WKB approximation with an anti-Stokes line segment connecting one of the interval end points $x=\pm L$ (with exact Dirichlet BC at this point) with a nearest turning point defining the corresponding complex eigenvalue via uniform Airy-function approximation \cite{langer-pr-1937}, \cite[sect. 2.3]{child-book} of the wave function. From the structure of the Stokes graphs one reads off that these relevant nearest turning points are located in the lower (upper) complex $x-$plane for $n$ odd, i.e. $V=ix^7, ix^{11}$ ($n$ even, i.e. $V=-ix^5,-ix^9$).}
\end{center}
\end{figure}

\begin{figure}[htb]
\begin{center}
\includegraphics[scale=1]{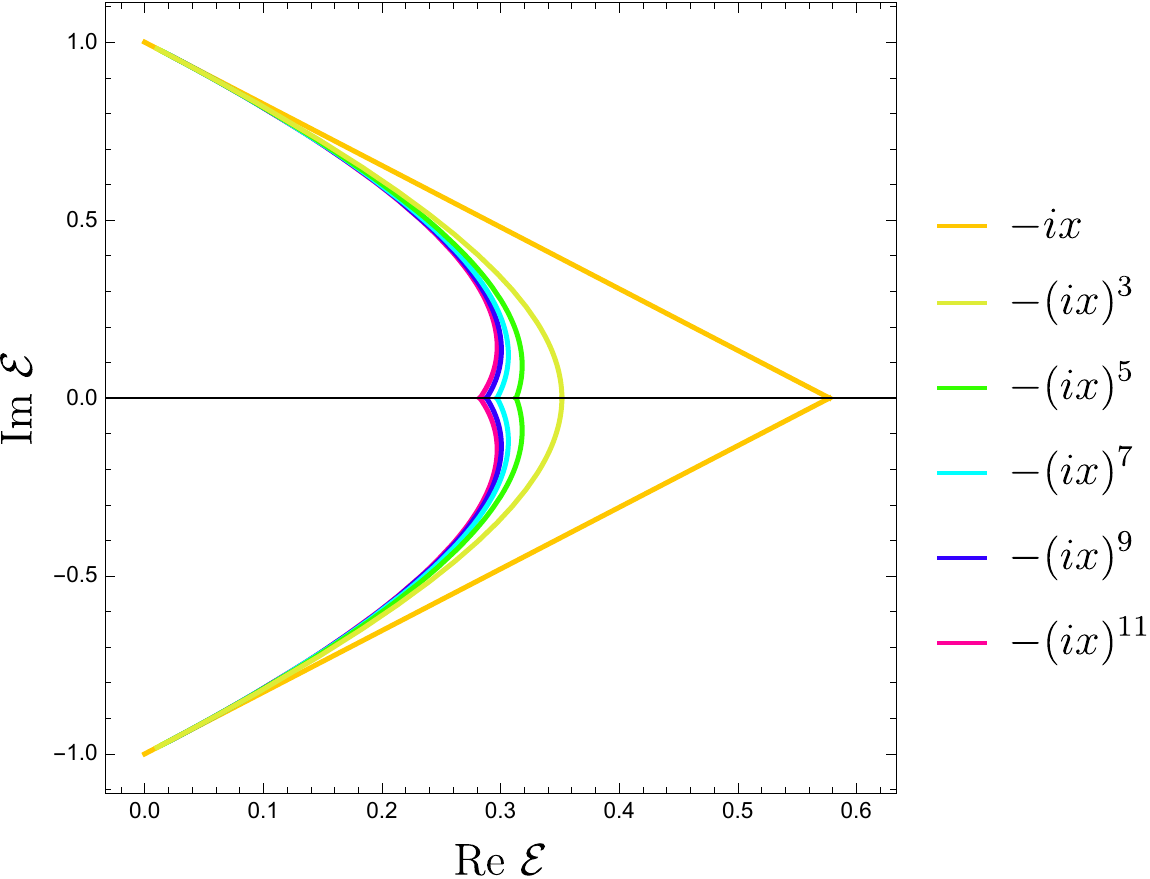}\\[6ex] %\hspace{2em}
\mbox{}\hspace{-8em}
\includegraphics[scale=0.72]{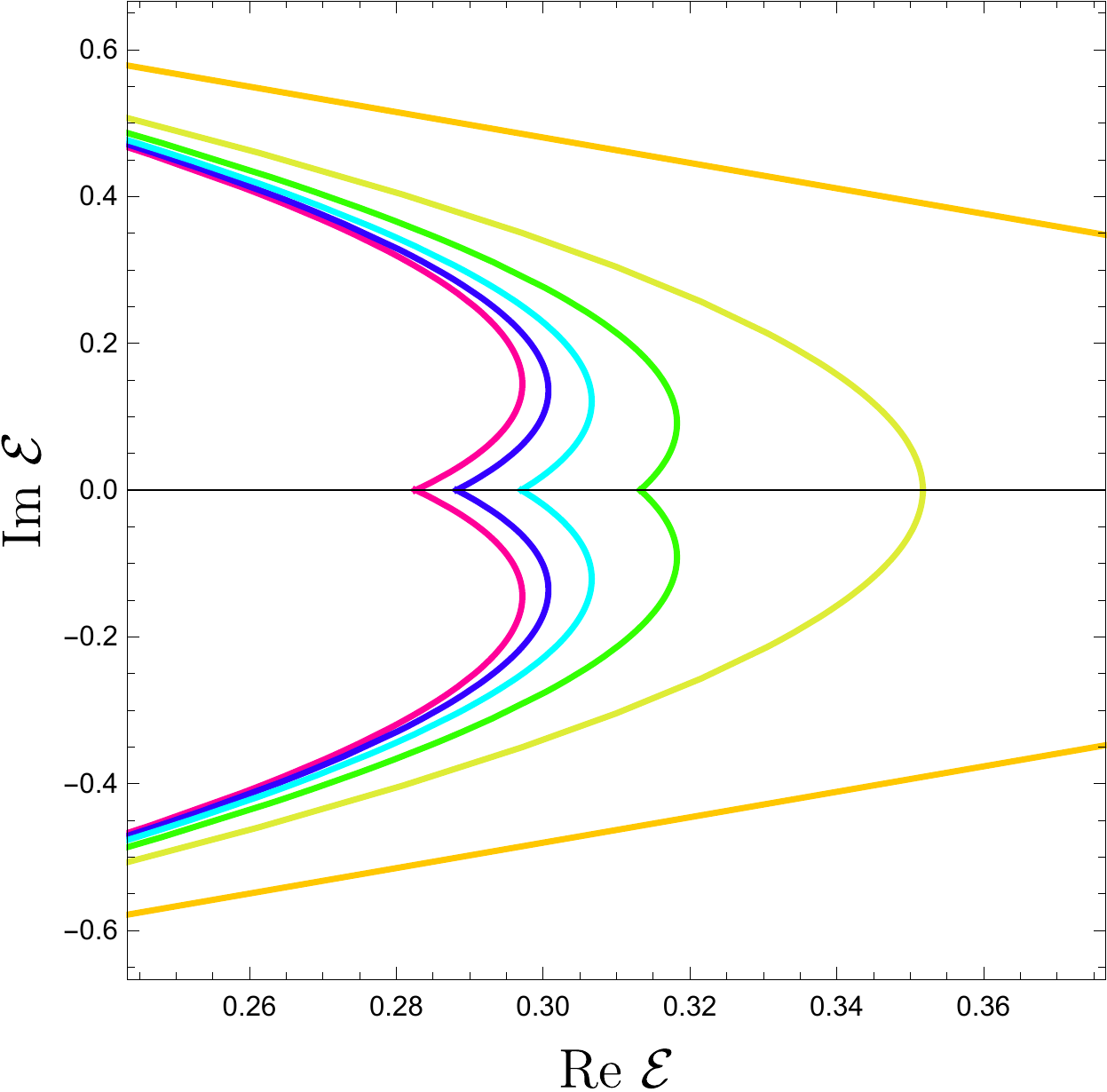}\\[4ex]
\caption{\label{complex-scaling-graph-comparison} The complex components $\cR_{CO}$ of the asymptotic scaling graphs $\cR=\cR_{CO}\cup\RR_+$ for scaled/mapped energies (MEs) $\cE=\frac{E}{gL^{2n+1}}$ of models with polynomial potentials $V=-(ix)^{2n+1}$, $n=0,\ldots 5$  obtained as solution curves of the corresponding differential reality-constraints over anti-Stokes-line segments $\cY$ defined within the zeroth-order WKB technique. This technique is a so called uniform WKB approximation of the wave functions by suitably chosen Airy-functions \cite{langer-pr-1937}, \cite[sect. 2.3]{child-book}.}
\end{center}
\end{figure}

\begin{figure}[htb]
\begin{center}
\includegraphics[scale=0.5]{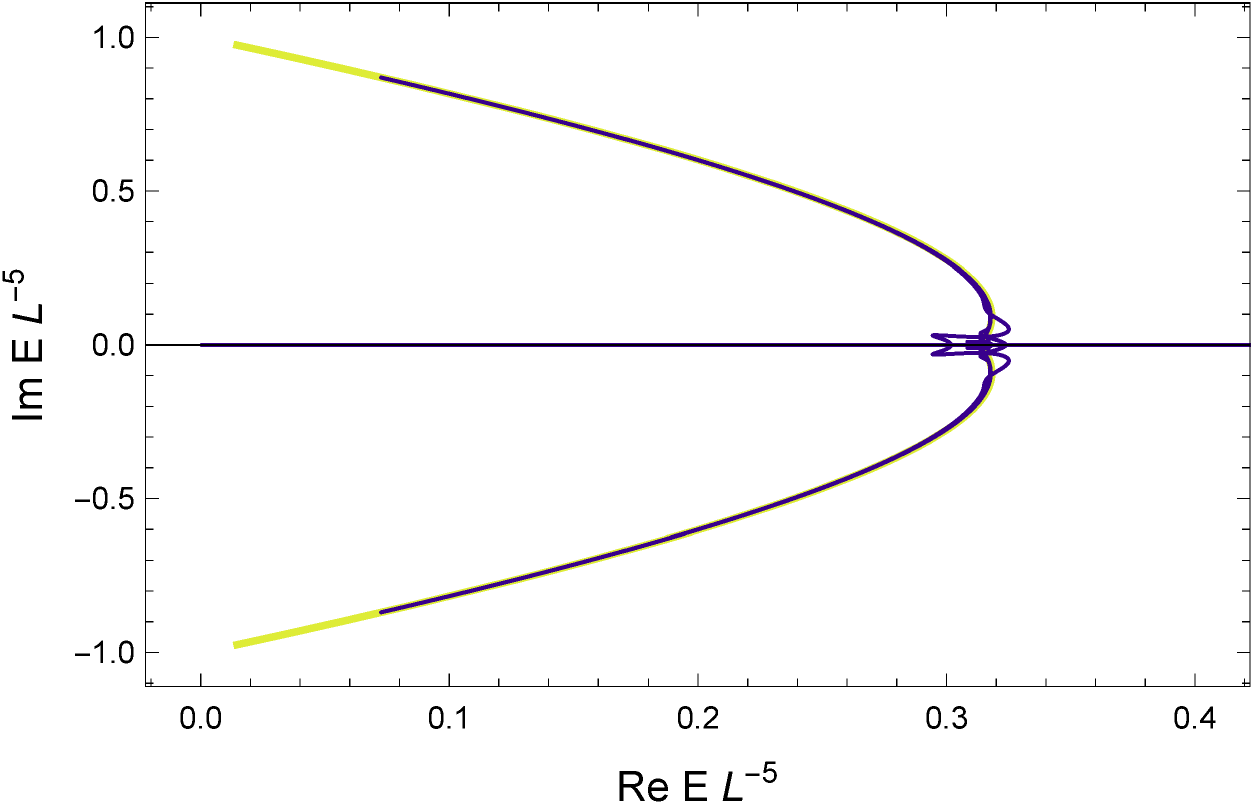}\hspace{2em}
\includegraphics[scale=0.5]{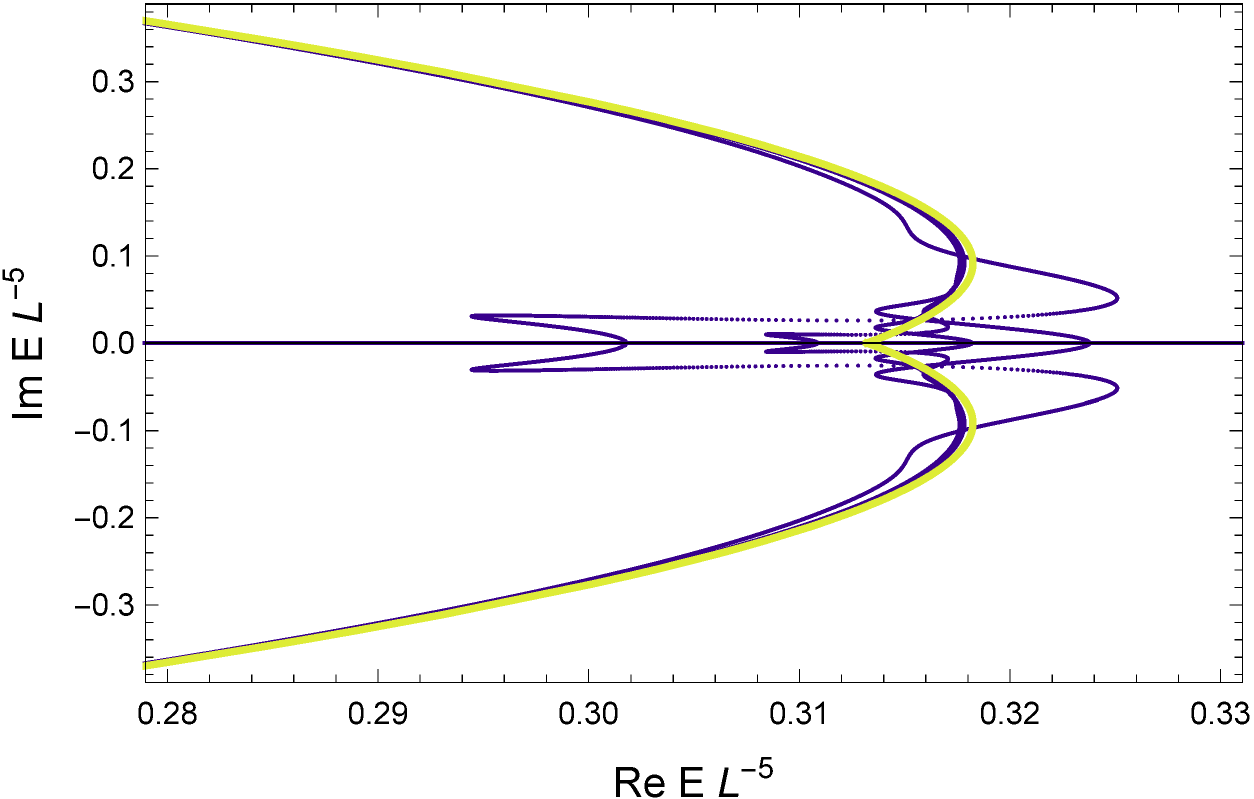}\\[2ex]
$V=-ix^5$\\[4ex]
\includegraphics[scale=0.5]{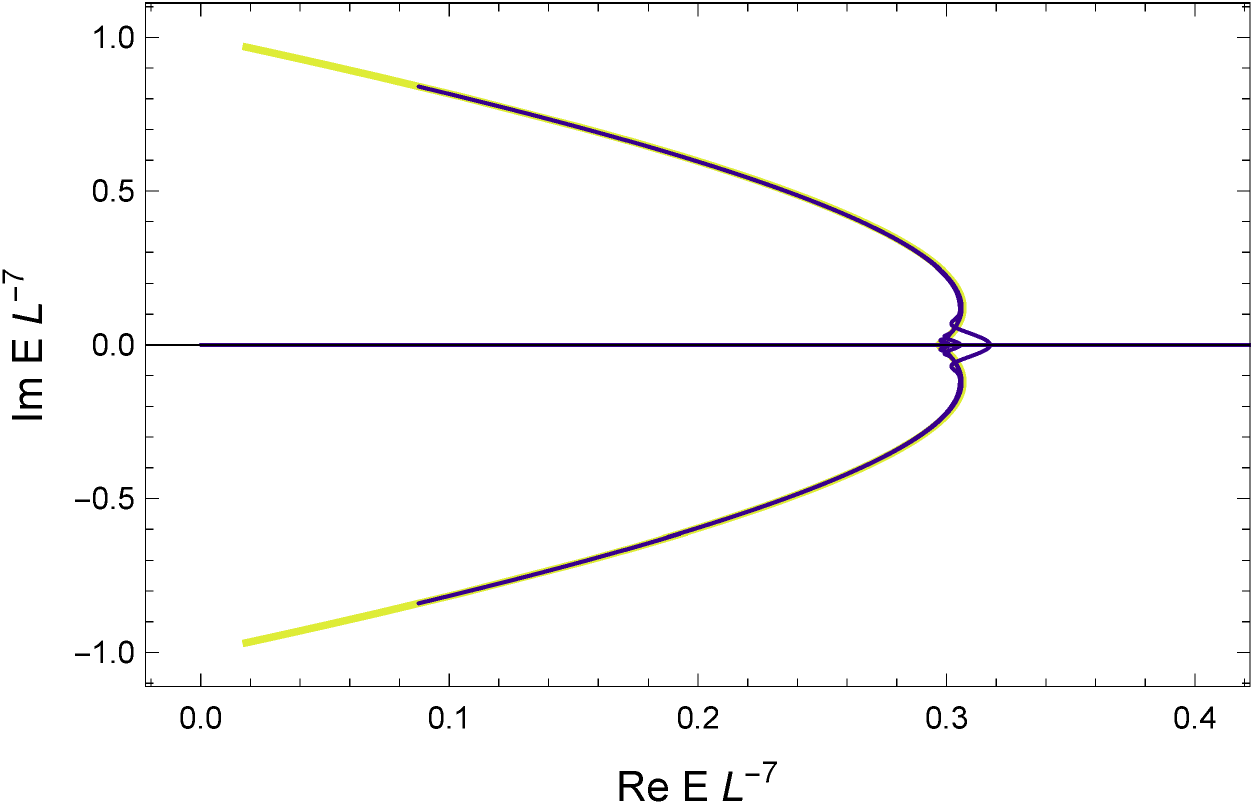}\hspace{2em}
\includegraphics[scale=0.5]{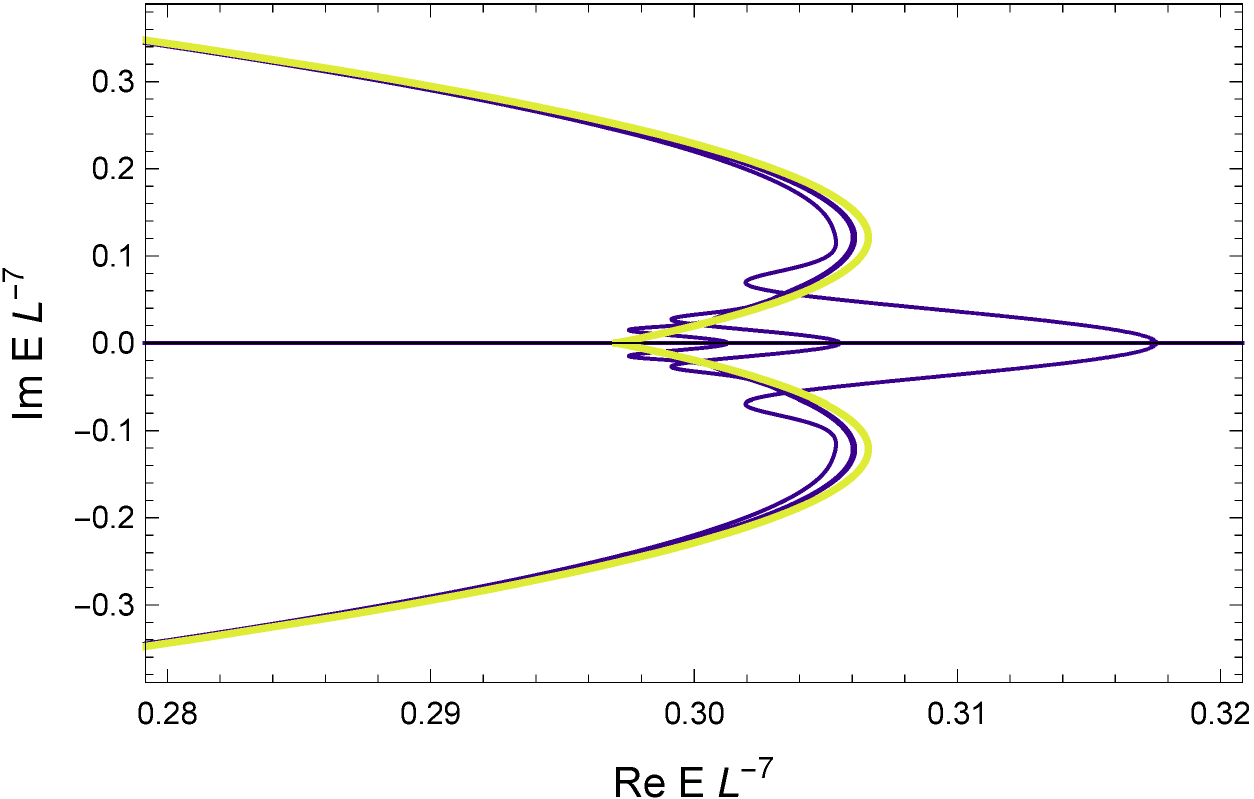}\\[2ex]
$V=ix^7$\\[4ex]
\includegraphics[scale=0.5]{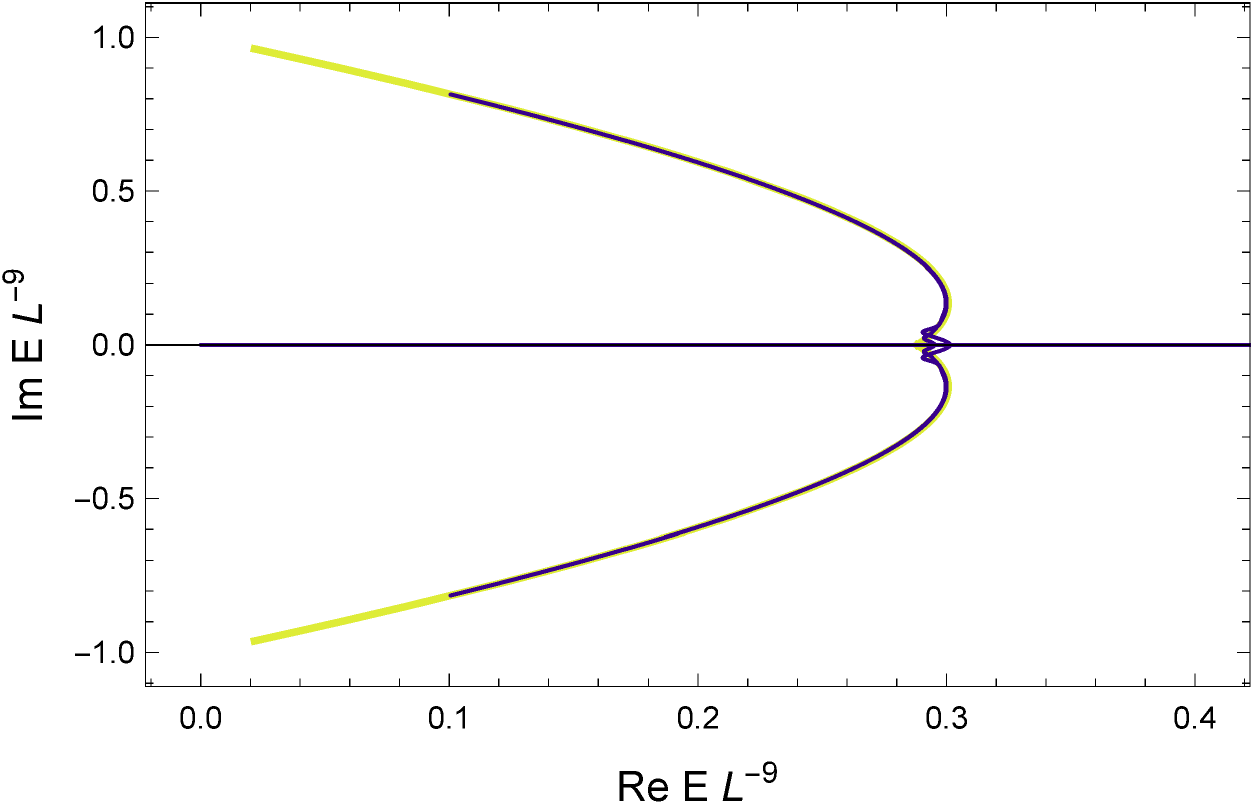}\hspace{2em}
\includegraphics[scale=0.5]{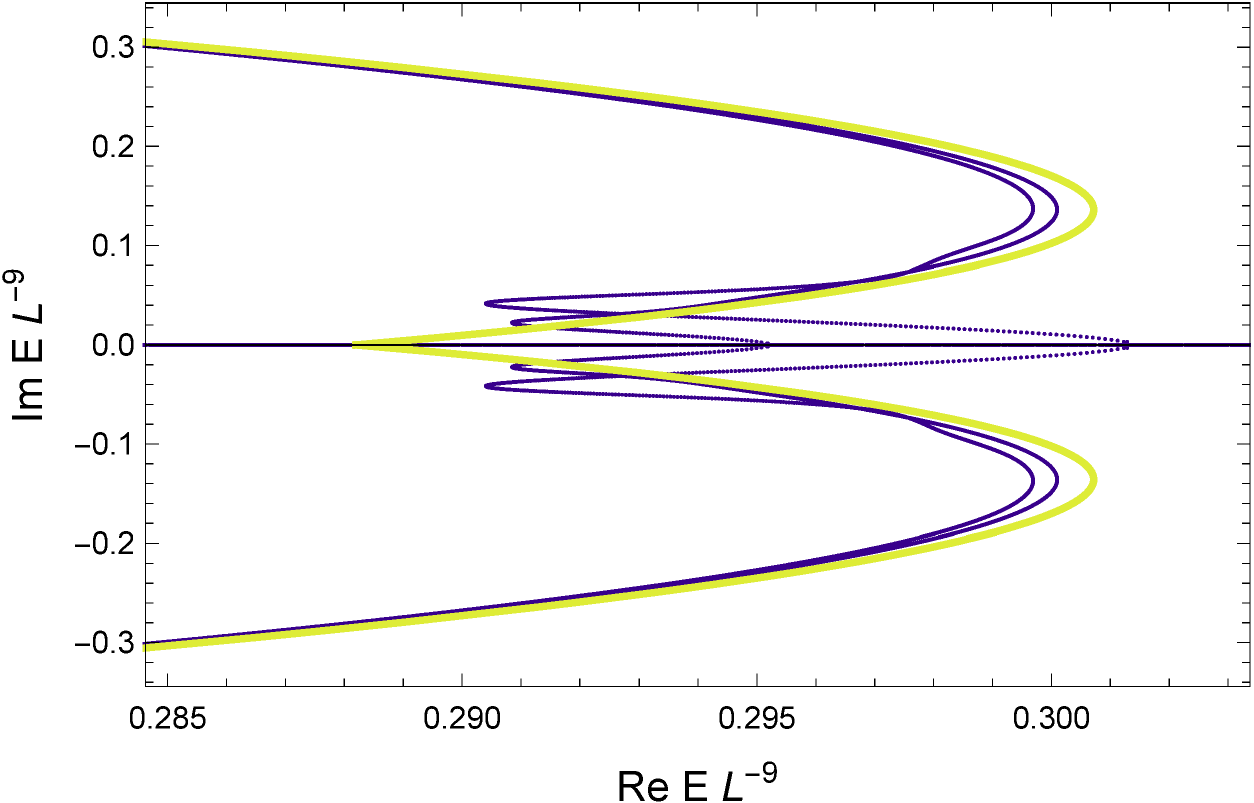}\\[2ex]
$V=-ix^9$\\[4ex]
\includegraphics[scale=0.5]{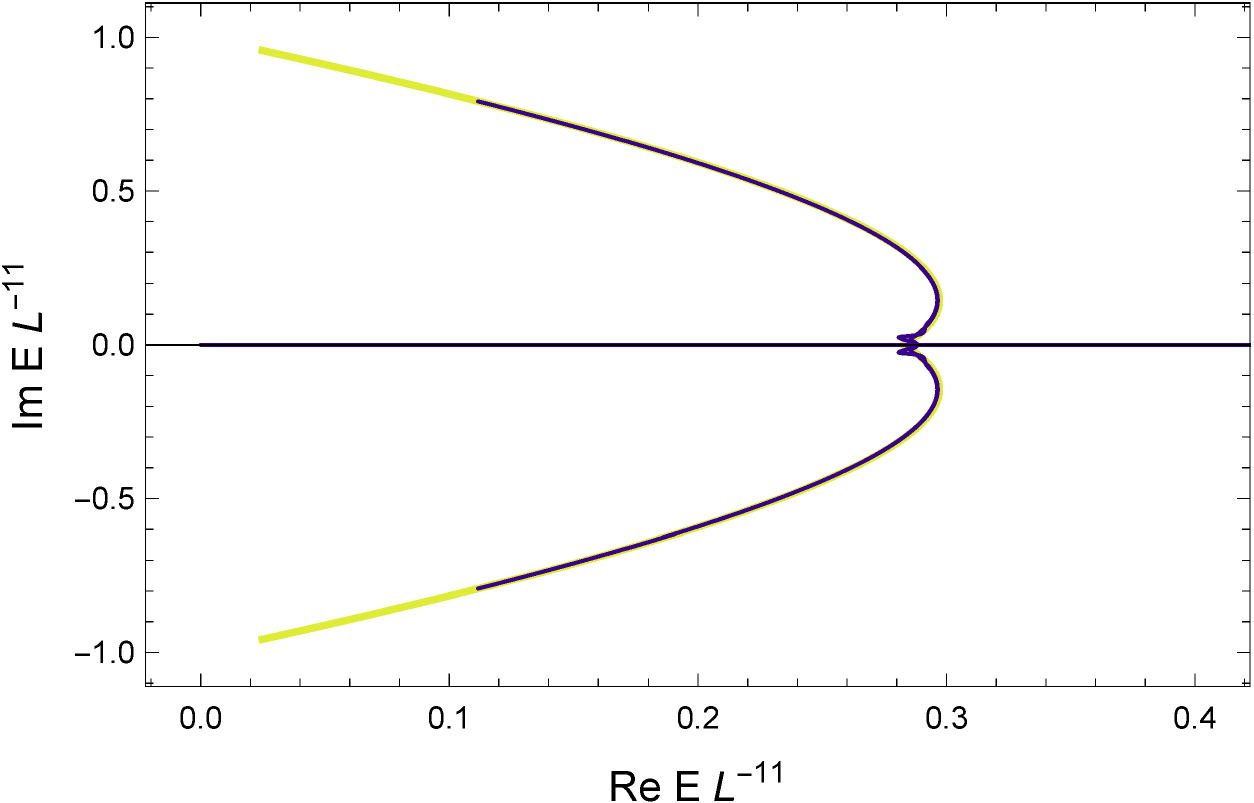}\hspace{2em}
\includegraphics[scale=0.5]{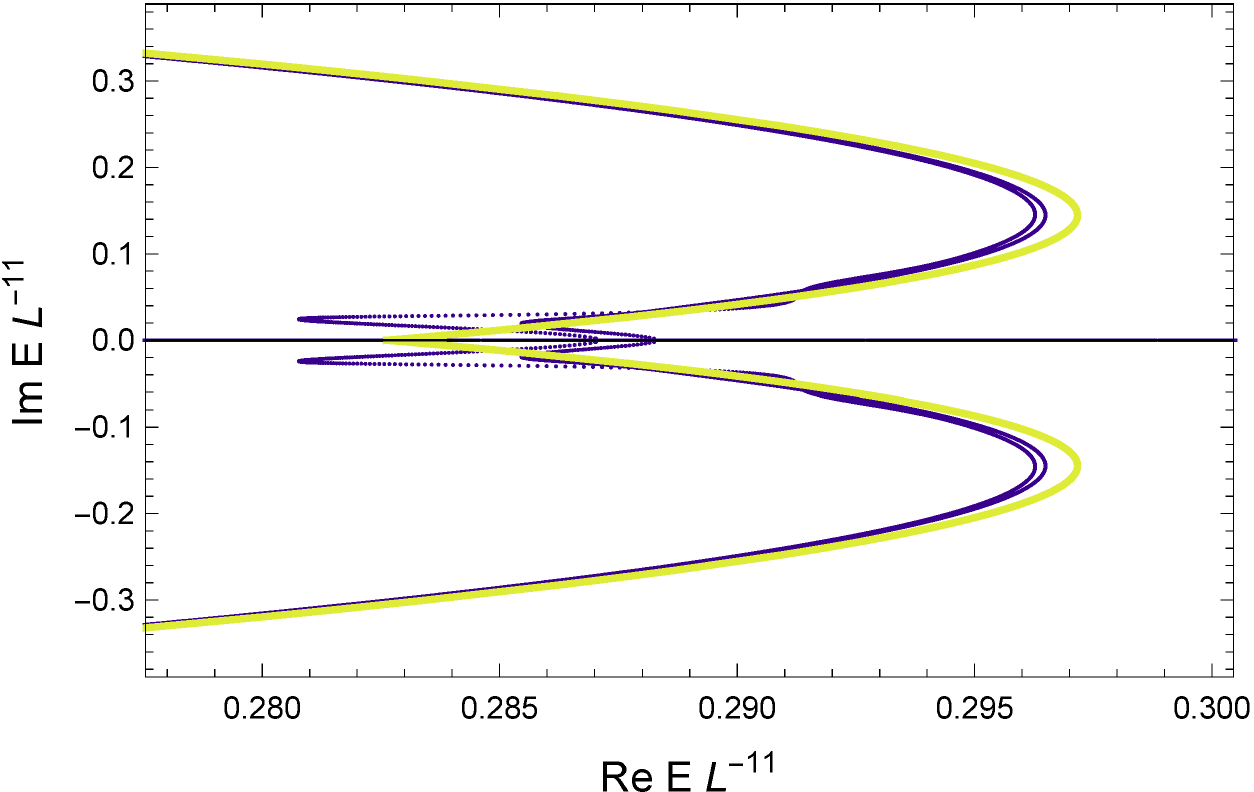}\\[2ex]
$V=ix^{11}$\\[4ex]
\caption{\label{multiple-wkb-scaling-graph-comparison} $\left[-(ix)^{2n+1},x\in[-L,L]\right]$ models with $n=2,\ldots,5$, i.e.  with $V=-ix^5$ $\ldots$ $V=ix^{11}$. Comparison of the scaled complex eigenvalue branches $\cE_j=\frac{E_j(L)}{L^{2n+1}}$ obtained by a shooting method (blue curves) with the results of the zeroth-order WKB approximation (green curves). The remarkably good (graphical) coincidence of these results outside the $\cP\cT$ phase transition regions provides strong numerical evidence that the hypothesized anti-Stokes-line approach (the uniform zeroth-order WKB approximation \cite{langer-pr-1937}, \cite[sect. 2.3]{child-book} in terms of single Airy functions) with its differentially resolved reality constraint holds for models with higher-polynomial potentials as well. As for the $(ix^3,x\in[-L,L])$ model, the uniform zeroth-order approximation breaks down in the $\cP\cT$ phase transition region. From the numerical data one observes that the strongest deviations from zeroth-order WKB are produced by the eigenvalue branches with the lowest mode numbers $j$, whereas an increasing $j$ is associated with shrinking deviations. More sophisticated (higher-order WKB) techniques will be required for a comprehensive  analytical description in this region which are still to be developed.}
\end{center}
\end{figure}

\clearpage

\section{H: $\cP\cT$ phase transitions and asymptotic spectral scaling graphs: some details}
The flow behavior of the mapped eigenvalues (MEs) $\cE_j=\frac{E_j}{gL^{2n+1}}$ on the corresponding asymptotic spectral scaling graph $\cR=\cR_{CO}\cup \RR_+=\cR_{CO}\cup\cR_{BS}\cup\cR_{BT}\cup \cE_c$ with $\cR_{CO}:=\left\{\cE(\tau)\cup[\cE(\tau)]^*|\,\tau\in[0,\tau_c)\right\}$, $\cR_{BS}:=(0,\cE_c)\subset\RR_+$, $\cR_{BT}:=(\cE_c,\infty)\subset\RR_+$ can be qualitatively described as follows. As shown in \cite{lt-czech2004}, considering $V(x)$ as perturbation of an empty-box eigenvalue problem over $x\in[-L,L]$ with eigenvalues $E_j=\frac{\pi^2}{4L^2}j^2$, the large $E_j$ with
\begin{equation}\label{11}
\frac{\pi^2}{8L^2}(2j+1)>||V||_\infty=\sup_{x\in[-L,L]}|V(x)|=gL^{2n+1}
\end{equation}
are not involved in $\cP\cT$ phase transitions and remain real. In combination with Fig.~\ref{fig3}(b) in the main text of the Letter, this leads to the following scaling behavior for fixed $g$ and varying $L$: for sufficiently small $L$, inequality \rf{11} holds for all $j$, the system shows essentially BT behavior with $E_j\approx \frac{\pi^2}{4L^2}j^2$, and the corresponding mapped eigenvalues (MEs) $\cE_j\in \cR_{BT}:=(\cE_c,\infty)$. With increasing $L$, the $E_j$ (for fixed $j$) decrease and \rf{11} becomes violated for an increasing number $j\sim L^{2n+3}$ of the lowest eigenvalues. At $\cE_j\approx \cE_c$, part of the BT MEs $\cE_j\in\cR_{BT}$ undergoes pairwise transitions to pairs of complex-conjugate eigenvalues, another part passes from real BT to real BS (with details of transition mechanism and selection rules still to be clarified). For further increasing $L$ the BS MEs $\cE_j=E_j/(gL^{2n+1})$ (due to $L-$indepenent $E_j$) flow from $\cE_c$ toward $\cE=0$ densifying on this BS segment $\cR_{BS}:=(0,\cE_c)$, whereas the complex MEs flow from $\cE_c$ toward $\cE=\pm i$ densifying on $\cR_{CO}:=\left\{\cE(\tau),[\cE(\tau)]^*| \ \tau\in[0,\tau_c)\right\}$. From the quantization condition [Eq. \rf{5} in the main text of the Letter] follows that the distance $\tau(\cE_j)$ between $\cE_j\in\cR_{CO}$ and the end point $\cE(\tau=0)=\pm i$ on the corresponding $\cR_{CO}-$branch for increasing $L$ shrinks as $ \tau(\cE_j)=\frac14 (4j-1)\pi\frac{\hbar}{\sqrt{g L^{2n+3}}}$.


\begin{thebibliography}{99}
\bibitem{remark-1-airy-paths} For $E/g\neq 0$ the turning points are simple roots of $Q(x)=0$ so that in the vicinities of $a_k$ linear approximations $Q(x)\approx Q'(a_k)(x-a_k)$ can be used. Substitution into the Schr\"odinger equation yields Airy function approximations of the wave function in the vicinities of the turning  points (where the WKB approximations \rf{3} break down).  Matching the Airy function approximations in the turning point vicinities of $a_{1,2}$ with standard lowest-order WKB wave functions over the path segments connecting the turning point vicinities with each other and with the path end points $x=\pm 1$, a secular equation was derived  \cite{bender-jones-2}.
\bibitem{remark-1a-stokes-lines}   For simple turning points $a_k$ (as for all models studied in this Letter) three anti-Stokes lines are starting in each turning point \cite{strebel-book}. They either end in another turning point or tend toward infinity $|x|\to\infty$ \cite[sect.~C]{suppl}.
\bibitem{remark-3-end-comment-complex-branches}That $\cE(\tau)$ correctly describes the rescaled location of the complex eigenvalues is independently verified by graphical comparison to the eigenvalues obtained by an eigenvalue-solver for complex Sturm-Liouville problems based on a shooting method \cite[sects.~F,G]{suppl}.
\bibitem{remark-3a-end-comment-complex-branches}
     Within $0$th-order WKB, the $\cP\cT$ phase transition point $\cE_c$ can be roughly estimated as end-point of the complex branch $\cE_c\approx\cE(\tau_c)\in\RR_+$ with obvious fitting of break-up \rf{4q} and quantization \rf{5} rules.
\bibitem{remark-5-condensation}Complementary insight into the hypothesized condensation mechanism may be expected from the close relation between  $(igx^3,x\in\RR)$  and a field theory describing polymer percolation in 2D (inverting the dimensional reduction \cite{DDT-pramana,cardy-poly}) and analyzing the 2D-field-theoretic equivalent of the 1D eigenfunction limit toward collinearity.
\bibitem{suppl-rem-1} These are good approximations, but not asymptotic approximations, because the zeros of the approximations do not coincide with the exact zeros of $\Ai(z)$.
\bibitem{suppl-rem-2} This follows from the  $\cP\cT-$symmetry, i.e. the reflection symmetry with regard to the imaginary axis in the complex $x-$plane, which implies $\cP\cT a_1=a_2$ so that the integration paths can be chosen $\cP\cT-$symmetric as well.
\bibitem{suppl-rem-3} Via Stokes multipliers induced by the fractional powers in the exponent and the pre-factor $z^{-1/4}$ in \rf{s5a}, (see, e.g., \cite[sect. 3.3]{heading-book}, \cite[App. A]{child-book}). In general, a cut can be chosen along any ray $z=e^{i\phi}r$, $r\in \RR_+$ with different argument $\phi\neq \pm \pi$. In this case, the concrete representation of the function will change, but not its value in the dominant and subdominant Stokes sectors. (For corresponding details see, e.g., \cite[sect. 3.3]{heading-book}.
\bibitem{suppl-rem-4}  We note that in both cases III and X, the two segments $z_\pm(s)\in\CC$ in each of the these cases produce the same line segment $\cE(s)\in\CC$ in the complex energy plane. This double cover is a clear indication of the underlying Riemann surface structure in which the Airy function Sturm-Liouville problem lives. A detailed analysis of these structures will be presented elsewhere.
\bibitem{suppl-rem-5} In the close vicinity of the real end point $\cE(s_c)=\frac1{\sqrt{3}}$ (and with end point $x=1$ close to the anti-Stokes line) the zeroth-order approximation loses its accuracy and higher-order corrections should be taken into account, possibly including techniques of super- and hyper-asymptotical approximations of the relevant SSH Airy functions \cite{hyper-berry} entering \rf{s9}. Moreover, the SSH assumption \rf{s10} itself will break down and the full characteristic equation \rf{s9} should be used instead.
\end{thebibliography}
\end{document}